\documentclass[aps,nofootinbib,,superscriptaddress,groupedaddress, preprintnumbers]{revtex4}
\pdfoutput=1
\usepackage{amsmath,amsfonts,amssymb,amsthm,amssymb,bbm,bm}
\usepackage{pdfsync}
\usepackage{verbatim}

\usepackage{subcaption}
\usepackage{graphicx,import}
\usepackage[latin1]{inputenc}
\usepackage{hyperref}
\usepackage{empheq}
\usepackage[all]{xy}
\usepackage{stmaryrd}
\usepackage{rotating}
\usepackage{color}  
\usepackage{slashed}
\usepackage{cancel}
\usepackage{array, makecell} %
\usepackage{comment}
\usepackage{tikz-cd}
\usepackage{hhline}
\usepackage{comment}
\usepackage{cancel}
\usepackage{url}
\makeatletter

\DeclareFontFamily{U}{matha}{\hyphenchar\font45}
\DeclareFontShape{U}{matha}{m}{n}{
    <5> <6> <7> <8> <9> <10> gen * matha
    <10.95> matha10 <12> <14.4> <17.28> <20.74> <24.88> matha12
     }{}
\DeclareSymbolFont{matha}{U}{matha}{m}{n}
\DeclareMathSymbol{\oright}       {2}{matha}{"69}

\newcommand{\doublehat}[1]{%
\begingroup%
  \let\macc@kerna\z@%
  \let\macc@kernb\z@%
  \let\macc@nucleus\@empty%
  \hat{\raisebox{.55ex}{\vphantom{\ensuremath{#1}}}\smash{\hat{#1}}}%
\endgroup%
}

\newcommand{\p}{\partial}

\newcommand{\bit}{\begin{itemize}}
\newcommand{\eit}{\end{itemize}}
\newcommand{\bd}{\begin{description}}
\newcommand{\ed}{\end{description}}

\newcommand{\bc}{\begin{center}}
\newcommand{\ec}{\end{center}}

\newcommand{\C}{{\mathbb C}}
\newcommand{\N}{{\mathbb N}}
\newcommand{\R}{{\mathbb R}}
\newcommand{\Z}{{\mathbb Z}}

\newcommand{\cM}{{\cal M}}

\newcommand{\cJ}{{\cal J}}

\newcommand{\bT}{{\bar T}}

\newcommand{\cS}{{\mathcal S}}

\newcommand{\bN}{\bar{N}}
\newcommand{\cQ}{{\mathcal{Q}}}

\newcommand{\SL}{\mathrm{SL}}

\renewcommand{\sl}{{\mathfrak{sl}}}

\def\be#1\ee{\begin{align}#1\end{align}}
\newcommand{\bea}{\begin{eqnarray}}
\newcommand{\eea}{\end{eqnarray}}
\newcommand{\bs}{\begin{subequations}}
\newcommand{\es}{\end{subequations}}

\newcommand{\la}{\label}

\newcommand{\tR}{{\rm Tr}}
\newcommand{\f}{\frac}

\newcommand{\bz}{{\bar{z}}}

\def\p{\partial}

\def\bF{\bar {F}}

\def\d{\delta}

\def\rd{\mathrm{d}}
\def\pa{\partial }

\def\k{{\kappa^2} }

\newcommand{\tq}{{\tilde q }}
\newcommand{\tF}{{\tilde F }}
\newcommand{\tr}{{\tilde r }}

\newcommand{\tN}{{\tilde N }}

\newcommand{\Ag}{\mathfrak{g}}
\newcommand{\Res}{\text{Res}}

\usepackage[mathscr]{euscript}

\graphicspath{{Figures/}}
\begin{document}

\title{ Double-soft limit and celestial shadow OPE from charge bracket}

\author{Daniele Pranzetti}
\email{daniele.pranzetti@uniud.it}
\author{Domenico Giuseppe Salluce}
\email{domenicogiuseppe.salluce@uniud.it}

\affiliation{Universit\`a degli Studi di Udine,
via Palladio 8,  I-33100 Udine, Italy}

\begin{abstract}
The dual formulations of an infinite tower of tree-level soft theorems in asymptotically flat spacetimes for scattering amplitudes in the standard  energy-momentum basis and for correlators of a 2D celestial  conformal field theory imply a correspondence between the celestial operator product expansion (OPE) and the higher spin charge bracket. We apply such correspondence to provide first a prescription to solve the double-soft limit ambiguity  in the mixed-helicity sector of celestial OPEs. Furthermore, demanding the {\it charge OPE/bracket correspondence} to remain valid when operators are shadow transformed, we  construct an algorithm to compute shadow celestial OPEs. We first test the algorithm by recovering results in the previous literature involving the celestial energy-momentum tensor;  we then  apply it  to both gravity and Yang--Mills theory and generalize the OPE derivation to arbitrary spins.
 \end{abstract}

\maketitle

\newpage
\tableofcontents

\section{Introduction}

Celestial holography is an implementation of the holographic principle in asymptotically flat spacetimes where $n$-particle scattering amplitudes are recast as   correlators of a celestial two-dimensional conformal field theory (2D CFT)  on the celestial sphere. For massless particles, this rewriting is achieved by switching from the standard basis of energy-momentum eigenstates to a basis of asymptotic boost  eigenstates through a Mellin transform in the energies \cite{Pasterski:2016qvg}. This reformulation strongly relies on the fact that, in the large-$r$ mode expansion of the bulk fields near null infinity, 
the spatial direction of the massless 4-momenta  is identified with the position of the dual CFT operator on the celestial sphere. Such identification allows to relate the collinear limit of amplitudes in momentum space  to the operator product expansion (OPE) of  primary operators in the celestial CFT (CCFT); in particular the leading term in the OPE given by a conformal primary is uniquely determined by  the leading collinear behavior of scattering amplitudes, while demanding global $\SL(2,\C)$  covariance of the OPE determines the contribution from all its descendants \cite{Pate:2019lpp, Guevara:2021tvr,Himwich:2021dau}. 

The second key ingredient of the celestial holography proposal to reformulate the scattering problem is the range of boost weights $\Delta$ labeling, in addition to a  point $(z,\bar z)$ on the celestial sphere, a complete basis of boost eigenstates. It was shown in
\cite{Pasterski:2017kqt} that the continuum of conformal primary wavefunctions with $\Delta=1+i\lambda,\lambda\in \R$  span the complete set of normalizable solutions to the Klein--Gordon equation in $\R^{1,3}$.  Conformal primary wavefunctions of spin $J$ can be constructed out of these and the massive case can be included as well. At the same time, the study of the infrared  sector of the theory, where most of the physically interesting effects have been studied so far through the manifestation of BMS symmetries, requires the introduction of a conformally soft limit \cite{Donnay:2018neh,Fan:2019emx} and the inclusion of
celestial operators with discrete $\Delta$  \cite{Donnay:2020guq,Pasterski:2021fjn,Pasterski:2021dqe,Donnay:2022sdg}. The celestial OPE involving these operators has led to the discovery of an infinite tower of conformally soft theorems governed by a $w_{1+\infty}$ symmetry in gravity and an $S$-algebra symmetry in Yang--Mills (YM) theory, when restricting to the same helicity sector \cite{Guevara:2021abz, Strominger:2021mtt, Himwich:2021dau}. Treatment of conformally soft celestial operators in terms of the  principal continuous series of irreducible
unitary representations of Lorentz seems unpractical; alternatively, inclusion of  the tower of conformally soft operators in a  basis of conformal primary wavefunctions is possible if one works with the  discrete basis $\Delta\in \Z$ introduced in 
 \cite{Freidel:2022skz} for fields in a functional space of Schwartzian type. 

 Moreover, while the $\Delta$-basis is very natural when working with CFT structures, the energy $\omega$-basis and the retarded time $u$-basis can provide a more transparent treatment of different aspects of soft physics \cite{Donnay:2022sdg}; in order to move freely from one basis to another \cite{Freidel:2022skz}, one wants to avoid putting apriori restrictions on the boost weight and study celestial OPEs in the full complex-$\Delta$ plane.

 Among the CFT structures, an important role is played by the {\it shadow transform} 
\cite{Ferrara:1972uq,Simmons-Duffin:2012juh}. This is an operation that sends an operator $O_{(\Delta, J)}(z,\bar z)$ to $S[O_{(\Delta, J)}](z',\bar z')$, where $(z,\bar z),(z',\bar z') $ are two different points on the celestial sphere; the shadow operator transforms as a  conformal primary operator of boost weight and spin $(2-\Delta, -J)$, explicitly\footnote{The normalization constant $K_{\Delta,J}$ coincides with the one in \cite{Osborn:2012vt}, up to a sign.} 
\be\la{Shadow}
S[O_{(\Delta, J)}](z',\bar z'):= K_{\Delta,J}\int\f{d^2z}{2\pi}\f{O_{(\Delta, J)}(z,\bar z)}{(z'-z)^{2-\Delta-J}(\bar z'-\bar z)^{2-\Delta+J}}\quad
{\rm with}\quad K_{\Delta,J}= (-)^{1-\Delta-J}\f{\Gamma(2-\Delta-J)}{\Gamma(\Delta-J-1)}\,.
\ee
The shadow transform is thus a non-local operation that appears in at least three important ways in celestial holography. First and foremost, a {\it local} CCFT energy-momentum tensor  can be constructed as the shadow transform of the sub-leading soft graviton operator
\cite{Kapec:2016jld,Fotopoulos:2019tpe,Fotopoulos:2019vac}, reproducing the standard OPE with a second  conformal primary operator; reproducing the expected $TT$ OPE is however a thorny issue \cite{Banerjee:2022wht} to which we will come back in Section \ref{sec:SET}. The second appearance is due to the realization that, in addition to the local  highest-weight basis for $\Delta\in 1+i\R$ mentioned above, the shadow transform of this family provides a second non-local family of conformal primary wave-functions  spanning the space of plane wave states \cite{Pasterski:2017kqt}. The third appearance has been preliminarily revealed in \cite{GGIseminar}  and it concerns the role of shadow higher spin charges in providing a closed bracket for the mixed-helicity sector of celestial symmetry algebras; this will be the main topic of our upcoming companion paper \cite{MH-bracket}.

These three pieces of evidence, as well as several other applications of the shadow transform \cite{Nande:2017dba,Kapec:2021eug,Kapec:2022hih,Kapec:2022axw,deGioia:2023cbd,Sleight:2023ojm,Banerjee:2024yir,Narayanan:2024qgb,Fan:2021isc,Chang:2022jut,Surubaru:2025qhs,Bhattacharyya:2025nfp}, hint towards an important role played by shadow operators in the dual CCFT; this raises the interesting question of how to derive  celestial OPEs involving at least one shadow celestial operator and how to take their double-soft limit. This last operation is characterized by a degree of ambiguity already in the case of standard (local) operators (see, e.g.,  
\cite{Distler:2018rwu,Anupam:2018vyu,Fotopoulos:2019vac,Campiglia:2021bap,Banerjee:2022wht,Ball:2022bgg,Ball:2023sdz}). 

The main goal of this manuscript is to provide a consistent algorithm to take the double-soft limit of celestial OPEs, including the case when at least one of the two entries is a shadow operator, both in the case of gravity and YM theory. The key tool of our analysis is the equivalence between the celestial OPE with a first soft insertion and the celestial hard charge bracket  action on the same general operator appearing in the second entry, as revealed in \cite{Freidel:2021ytz}. More explicitly, this {\it charge OPE/bracket correspondence} can be written as
\be\la{OPE-bra}
q^1_s(z, \bar z) O_{(\Delta', J')}(z',\bar z')
&\leftrightarrow 
 \frac{1}{i}
\{q^2_s(z, \bar z),  O_{(\Delta', J')}(z',\bar z')\} 
 \,,
\ee
where $q^1_s, q^2_s$ are respectively the soft and the hard charge contributions \cite{Freidel:2021ytz,Freidel:2022skz,Freidel:2023gue}, with $s\in\Z_+$, and $ O_{(\Delta', J')}$ is a conformal primary operator. The correspondence \eqref{OPE-bra} represents a key entry in the bulk/boundary dictionary for flat holography and it is rooted in the equivalent formulations of tree-level soft theorems for scattering amplitudes in the energy-momentum basis (LHS of \eqref{OPE-bra}) and in the boost basis (RHS of \eqref{OPE-bra}). More precisely, in their standard formulation, soft theorems  can be understood as  Ward identities associated to an infinite  number of conserved charges, with one  conservation law for each point of the celestial sphere \cite{Strominger:2013jfa, Kapec:2014opa, Campiglia:2016efb, Freidel:2021dfs}. Leaving antipodal matching  implicit, these
truncated\footnote{For $s\geq2$ in the case of gravity and $s\geq1$ in YM higher order contributions to the charges appear, giving rise to collinear terms in the soft factors \cite{Freidel:2021dfs,Jorstad:2025qxu}.}  Ward identities take the form
\be 
\la{spin-s-cons}
\langle {\rm out}|[q_s^1,\mathcal{S}]|{\rm in}\rangle = -\langle{ \rm out}| [q_s^2, \mathcal{S}]|{\rm in}\rangle.
\ee
The action of the soft charge on the LHS of \eqref{spin-s-cons} generates the insertion of a soft graviton/gluon at  $\mathcal{O}(\omega^s)$  in the  $\mathcal{S}$-matrix element; replacing the Poisson bracket of the quadratic charge in a Fourier mode expansion with the quantum commutator, the RHS of \eqref{spin-s-cons} generates the corresponding soft factor. On the other hand, when expressing scattering observables   in a basis of asymptotic boost, these take the form of conformal correlators, as recalled above; in this formulation then classical soft theorems can be expressed as OPEs of conformal primary operators with a conformally soft limit taken on the first entry \cite{Pate:2019mfs, Guevara:2021tvr}, namely the RHS of \eqref{OPE-bra}. It follows that the celestial OPE with one soft insertion where $\Delta\to 1-s$ is given by the  (sub)$^s$-leading soft factor \cite{Freidel:2021ytz}. 

While the correspondence \eqref{OPE-bra} is thus simply a statement about the validity of tree-level soft theorems when expressed in two different basis, its realization is a non-trivial fact nevertheless, given the different ingredients that enter the derivation of its two sides. On the one hand, the charge bracket action is computed working within the asymptotic covariant phase space  formalism; starting with the conserved charge at spatial infinity and using the asymptotic Einstein/YM equations of motion in a large-$r$ expansion about null infinity, one re-expresses the charge as an integral over scri of powers of the conjugate fields parametrizing the asymptotic phase space as well as  their spatial and (inverse) time derivatives. The charge bracket action is then computed from the symplectic form of the radiative phase space of the theory and it is an on-shell result derived from solving the  recursion relation encoding a truncation (beyond a certain value of the spin)  of the evolution equations for the higher spin charges \cite{Freidel:2021ytz,Freidel:2022skz,Freidel:2023gue}.

On the other hand, celestial conformal primary OPEs have been derived starting from the leading collinear singularities of three-point vertices in momentum space computed with Feynman diagrams and  applying  Mellin transforms; asymptotic symmetry constraints, together with some analyticity assumptions,  are then implemented to determine the coefficients of the leading (primary) as well as subleading (descendants) contributions in terms of the conformal dimensions of the primaries  \cite{Pate:2019mfs,Ebert:2020nqf,Himwich:2021dau}. The construction is purely kinematical, with no dynamical evolution equation involved. 

While very powerful and elegant, the celestial OPE derivation becomes fuzzy when shadow operator are involved due to the non-local nature of the shadow transform, which spoils the coincident collinear limit. In this case then, the leading term in the collinear limit of the three-point function in momentum space does not necessarily dominate over the other terms and both primary and descendants contributions should be included at the same time when computing the OPE  coefficients. A preliminary treatment of this issue by means of the OPE block technique in CCFT \cite{Guevara:2021tvr,Kulp:2024scx} has recently appeared in \cite{Himwich:2025bza} (see also \cite{Banerjee:2025oyu}).

In this work, we will employ the RHS of  \eqref{OPE-bra} to derive OPEs involving at least one soft insertion and where either one or both entries are shadow operators. The advantages of this phase-space-based approach are twofold. Firstly, demanding the OPE of the soft charges to correctly reproduce the first order charge bracket  fixes the double-soft limit prescription to {\it ``first entry goes soft first''}, as we will show in Section \ref{sec:double}. 
Secondly, since the collinear limit does not really play any role when computing the charge bracket action, there is no conceptual obstacle to applying the shadow transform to it, as the quadratic charge expression {\it already} includes all the primary and descendant contributions. Therefore, {\it demanding} the correspondence \eqref{OPE-bra} to remain valid in the case where shadowed operators are included provides a direct algorithm to compute shadow celestial OPEs, which we will apply in Section \ref{sec:garvity} to the case of gravity and in Section \ref{sec:YM} to YM theory. We will first test our proposal for the case of OPEs involving one as well as two copies of the energy-momentum tensor (corresponding to a shadowed subleading soft graviton operator)  and show that we recover results appeared in previous literature. As an extra bonus, the phase-space-based approach allows us to clarify another ambiguity maybe not always explicitly emphasized, namely the commutativity or not of the shadow transform and the conformally soft limit. We will show that the two operations can be performed in any of the two orders, giving the same result. 

Beyond the key observation \eqref{OPE-bra}, our analysis exploits another useful tool represented by the organization of conformally soft primary operators and charges provided by
celestial diamonds  \cite{Pasterski:2021fjn, Pasterski:2021dqe, Guevara:2021abz,Pano:2023slc}. We will review their structure in Section \ref{sec:pre}, together with other preliminary material, and clarify some aspects of the relation between different corners along the way. Concluding remarks are presented in Section \ref{sec:conc} and detailed derivations are extensively provided in the series of Appendices \ref{App:YM}, \ref{App:GR-OPE}, \ref{App:YM-OPE}.

\section{Preliminaries}\la{sec:pre}

We start by reviewing the construction of higher spin charges in gravity and YM theory, their bracket action on fields parametrizing the asymptotic phase space and their relation to other relevant (to our analysis) operators, as captured by celestial diamonds.

\subsection{Higher spin charges and their action}\la{sec:charges-gravity}

\subsubsection{Gravity}\la{sec:gravity}

We describe the gravitational radiation near $\mathscr{I}^+$ through the shear helicity scalar $C(u,z,\bar z)$ and the news $N(u,z,\bar z)=\pa_u C^*(u,z,\bar z)$, being functions of the retarded time $u$ and the transverse coordinates $(z,\bar z)$ on the celestial sphere $S$. 
We define positive and negative energy modes of the shear as
\be
\widetilde C_+(\omega)=\int_{-\infty}^\infty du e^{i\omega u}C(u), \quad \widetilde C_-(\omega)=\int_{-\infty}^\infty du e^{i\omega u}C^*(u), \quad \omega>0
\ee
and the Mellin transforms
\be\la{CDelta}
\widehat C_\pm(\Delta)=\int_0^\infty d\omega \omega^{\Delta-1} \widetilde C_\pm(\omega).
\ee
One can decompose $C(u)$ into positive and negative energy sectors as
\be
C(u)=C_+(u)+C^*_-(u)
\ee
where 
\be
C_{\pm}(u)=\f{1}{2\pi}\int_{0}^\infty d\omega e^{-i\omega u}\widetilde C_{\pm}(\omega).
\ee
Analogous objects can be constructed also for the news $N(u)$.

If we restrict to signals belonging to the Schwartz space, an infinite tower of higher spin charges \cite{Freidel:2021ytz} can be extracted from the asymptotic expansion of truncated (for $s> 3$) Einstein's equations near $\mathscr{I}^+$. These charges have conformal dimension/helicity\footnote{We use interchangeably the tensor notion of `spin' with the scalar notion of `helicity'; see \cite{Freidel:2021dfs} for more details.} $(\Delta, J)=(3,s)$, with $s\in\mathbb{Z}, s\ge-2$, and satisfy the recursive equations
\be\la{rec}
\partial_u\mathcal{Q}_s=D_z\mathcal{Q}_{s-1}+\f{(s+1)}{2}C\mathcal{Q}_{s-2},
\ee
where $\mathcal{Q}_{-2}=\partial_u{N}/2$ is the derivative of the news, and $\mathcal{Q}_{-1}=D_z{N}/2$ is the energy current. Moreover $\cQ_0=\cM_\C$ is the complex mass aspect, and $\cQ_1=\cJ$ is the angular momentum aspect \cite{Freidel:2021qpz}. Here $D_z$ denotes a covariant derivative with respect to $z$ on the celestial sphere $S$; it will also be denoted as $D$ in the following, and similarly for $D_{\bar z}$ and $\bar D$.

Since the charge aspects $\mathcal{Q}_s$ diverge in the limit $u\rightarrow-\infty$, one defines the renormalized higher spin generators (see also \cite{Geiller:2024bgf} for a generalized renormalization procedure) 
\be
\hat{q}_s(u,z,\bar z):=\sum_{n=0}^s\f{(-u)^{s-n}}{(s-n)!}D^{s-n}\mathcal{Q}_s(u,z,\bar z)\,,
\ee
which parametrize the non-radiative phase space $\Gamma_{NR}$ corresponding to the conditions $\partial_u{N}=0=D{N}$. 
The limit  $u\rightarrow-\infty$ of these generators is finite and defines the higher spin charge aspects
\be
q_s(z,\bar z):=\lim_{u\rightarrow-\infty}\hat{q}_s(u,z,\bar z).
\ee
While these charge aspects have helicity $s$, their hermitian conjugates $\bar q_s:=q_s^\dagger$ have helicity $-s$.
We can expand $q_s$ in powers of radiation fields as $q_s=q_s^1+q_s^2+\cdots$, where the first two contributions are the soft charge aspect (linear in radiation fields) and the hard charge aspect (quadratic in radiation fields). Explicitly, by solving the recursion relation \eqref{rec} at linear order, we have
\be\la{q1}
q^1_s(z,\bar z)=D^{s+2}N_s(z,\bar z), \quad \text{where} \quad N_s(z,\bar z)=\f{(-)^{s+1}}{2 s!}\int_{-\infty}^\infty du u^s N(u,z,\bar z)\,,
\ee
is the (sub)$^s$-leading soft graviton.
At quadratic order, the renormalized charge aspect take the form
\be\la{q2}
q^2_s(z,\bar z)
&=-\frac14\int_{-\infty}^{\infty}  du\left[\sum_{\ell=0}^s \frac{(\ell+1) (-u)^{s-\ell}}{(s-\ell)!} 
  D^{s-\ell}\left[C  D^{\ell} \p_u^{-\ell+1}  N  \right](u,z,\bar z) \right]\,,
\ee
where
\be\label{iiint}
 \pa_u^{-n}  := \int_{+\infty}^{u}\rd u_1 \int_{+\infty}^{u_1} \rd u_2 \cdots \int_{+\infty}^{u_{n-1}} \rd u_n \,.
\ee

The higher spin charges are defined by integrating the charge aspects against an arbitrary spin-$s$ function $\tau_s(z,\bar z)$ over the celestial sphere $S$,  namely
\be
Q_s(\tau_s):=\f{8}{\k}\int_S\tau_s(z,\bar z)q_s(z,\bar z)\,,
\ee
where we denote $\int_S:=\int_Sd^2z\sqrt{q}$, with $q$  the determinant of the leading order celestial sphere metric.
As celestial OPEs are usually derived conformally mapping  the celestial sphere to the celestial plane, with covariant spatial derivatives becoming simple partial derivatives $\p,\bar \p$, in the following we will adopt this setup as well to simplify some manipulations and  the comparison with previous literature. Where necessary, we will adopt also the convention $\p_1:=\p_{z_1}, \bar\p_1:=\bar\p_{z_1}$ and similarly for $z_2$.

The action of the quadratic charge aspects on the shear and News fields can be computed straightforwardly by means of the radiative phase space symplectic form, which yields   the fundamental Poisson bracket \cite{Ashtekar:1978zz, Ashtekar:1981sf, Ashtekar:2018lor}  
\be
 \label{NCbracket}
 \{   N(u_1,z_1), C(u_2, z_2)\}&= \f \k 2\delta(u_1-u_2) \delta^2(z_1,z_2)\,,
 \ee
 where $\kappa=\sqrt{32\pi G}$.
In a conformal primary basis, where one introduces the 
 {\it conformal primary gravitons} $N_{\Delta}^{{\pm}}$ of $\pm$-helicity corresponding to operators of dimension/helicity $(\Delta,\pm 2)$ defined as\footnote{We refer to \cite{Freidel:2022skz,Freidel:2023gue} for the precise details on $i\epsilon$-prescriptions and contour integrations.}
\be 
\label{conf-gr}
N_{\Delta}^-(z, \bar z) &:= -\frac{\Gamma(\Delta -1)}{2} \int_{-\infty}^{+\infty} \!\rd u\, u^{1-\Delta } {N}(u,z, \bar z), 
\cr
N_{\Delta}^+(z, \bar z) &:=  -\frac{\Gamma(\Delta -1)}{2} \int_{-\infty}^{+\infty} \!\rd u\, u^{1-\Delta } N^*(u,z, \bar z)\,,
\ee 
this action is given by \cite{Freidel:2021ytz}
 \be 
\label{qND}
 \{q^{2}_{s}(z_1, \bar z_1),N_{\Delta_2}^{{\pm}}(z_2, \bar z_2)\}=
\frac{ \kappa^2}{8 \,s!} \sum_{n = 0}^{s} (-1)^{ n -s}(\Delta_2 \pm 2)_{s-n} (s)_n (n+1)\p_{1}^{s - n}\delta^{(2)}(z_{12}, \bar z_{12}) \p_{2}^n N_{\Delta_2+1-s}^{{\pm}}(z_2, \bar z_2)\,,
\ee
where $z_{12}=z_1-z_2, ~ \bz_{12}=\bz_1-\bz_2$ and we have introduced the falling factorial  $(x)_n = x(x-1)\cdots(x-n+1)$. 
The fields $N_{\Delta}^{{\pm}}$ are proportional to conformal primary boost eigenstates  $O_{\Delta}^\pm$ of \cite{Pasterski:2016qvg, Pasterski:2017kqt}, explicitly
\be 
\label{propOG}
O_{\Delta}^\pm = i^{\Delta} \frac{8\pi}{i\kappa} N_{\Delta}^\pm.
\ee 
In terms of the Mellin transforms \eqref{CDelta}, we also have the relation
$N_\Delta^{\pm}=-\f{1}{4}(i^{-\Delta}\widehat C_\pm(\Delta)+i^{\Delta}{\widehat C}_\mp^*(\Delta))$.

The rescaling \eqref{propOG} is simply motivated by the fact that the residues of $N_\Delta^{-}$ at negative integer dimensions correspond exactly to the (sub)$^s$-leading soft graviton modes in \eqref{q1},
\be 
\label{csG-}
\mathrm{Res}_{\Delta= 1-s}\left( N_\Delta^-(z,\bz) \right) = \frac{(-1)^{s+1}}{2 {s!}}\int_{-\infty}^{+\infty} \!\rd u\, u^{s} {N}(u,z,\bz)
= N_s(z,\bz)\,,
\ee
where one uses the identity Res$_{\Delta=-n}\Gamma(\Delta)= \frac{(-1)^n}{n!}$; similarly, 
\be
\label{csG+}
\mathrm{Res}_{\Delta= 1-s}\left( N_\Delta^+(z,\bz) \right)
= \bar N_s(z,\bz)\,.
\ee

A main result of \cite{Freidel:2021ytz} was to show that the OPE of two conformal primary boost eigenstates in the antiholomorphic collinear limit
 \cite{Pate:2019lpp, Guevara:2021tvr}
\be 
\label{OPEs}
\begin{split}
O_{\Delta_1}^-(z_1, \bar z_1) O_{\Delta_2}^{\pm}(z_2, \bar z_2) \sim -\frac{\kappa}{2} \frac{1}{\bz_{12}} \sum_{n = 0}^{\infty} B(\Delta_1 - 1 + n,  \Delta_2 \pm 2 + 1) \frac{z_{12}^{n + 1}}{n!} \p^n O^{\pm}_{\Delta_1 + \Delta_2}(z_2, \bar z_2) + \mathcal{O}(\bz_{12}^0)\,, 
\end{split}
\ee
where $B(x,y)=\Gamma(x)\Gamma(y)/\Gamma(x+y)$ is the Euler beta function, implies the charge OPE/bracket correspondence 
\be
\la{key}
q^1_s(z_1, \bar z_1) N_{\Delta_2}^{\pm}(z_2, \bar z_2)
\leftrightarrow
\f 1i \{q^{2}_{s}(z_1, \bar z_1),N_{\Delta_2}^{{\pm}}(z_2, \bar z_2)\}
 \,.
\ee
This follows from taking the conformally soft limit of the first entry of \eqref{OPEs}.
This correspondence is the key observation that will allow us to compute celestial OPEs involving shadow operators starting from the hard charge bracket. As already mentioned above, a second important tool to achieve this goal is represented by celestial diamonds, which we will briefly review below. Before moving to the analog construction for YM theory, an important remark is in order.

As implied by the antiholomorphic collinear limit, the OPE \eqref{OPEs} is valid as long as $z, \bz$ are taken as independent real variables, corresponding to an analytic continuation to a $(2,2)$ signature in the bulk. When instead working with a Lorentzian signature, as in our case, $z$ and $\bz$ are complex conjugates, and the mixed-helicity OPE in \eqref{OPEs} receives extra contributions characterized by a pole  in $z_{12}$. This was derived in \cite{Fan:2019emx} for the case of Yang-Mills theory, starting from multi-gluon scattering amplitudes (see eq. (4.18) there). The case of gravity was discussed in \cite{Freidel:2021ytz}, where the authors propose the following expression  
\be 
\label{OPEsbis}
O_{\Delta_1}^-(z_1, \bar z_1) O_{\Delta_2}^{+}(z_2, \bar z_2) \sim &-\frac{\kappa}{2} \frac{1}{\bz_{12}} \sum_{n = 0}^{\infty} B(\Delta_1 - 1 + n,  \Delta_2 \pm 2 + 1) \frac{z_{12}^{n + 1}}{n!} \p^n O^{+}_{\Delta_1 + \Delta_2}(z_2, \bar z_2)
\cr
-&\frac{\kappa}{2} \frac{1}{z_{12}} \bigg(\bz_{12} B(\Delta_1 +3,  \Delta_2 \pm 2 - 1)   O^{-}_{\Delta_1 + \Delta_2}(z_2, \bar z_2) + \mathcal{O}(\bz_{12}^2)\bigg).
\ee
When taking the conformally soft limit on the first entry, these extra terms vanish for spins $s\leq3$, since the Euler beta function is regular in this range, reflecting the exact nature of the asymptotic evolution equations  \eqref{rec} (and therefore the expression \eqref{q2} of the hard charge aspects) for this range of spin. Whenever this fact, which applies also to  celestial OPEs in YM theory, will be relevant in our analysis, we will use extra care to point it out and deal with it.

\subsubsection{Yang-Mills theory}\la{subsec:YM}

We can apply the same analysis to Yang-Mills theory in 4-d Minkowski spacetime \cite{Freidel:2023gue}. We work with a non-abelian gauge theory with gauge group $G$ and related Lie algebra $\Ag$, whose generators $T_a$, with $a=1,\dots, \text{dim}(\Ag)$, satisfy
\be
[T_a,T_b]=i{f_{ab}}^cT_c,
\ee
where the structure constants are normalized such that
\be
\tR(T_aT_b)=\delta_{ab}.
\ee
The Yang-Mills equations take the form 
\be
d\star F, \quad F=dA+iA\wedge A,
\la{YM}
\ee
with the gauge field $A=A_\mu dx^\mu$ a 1-form valued in the adjoint representation of $\Ag$, i.e. $A_\mu=A_\mu^aT_a$.
We will denote the adjoint representation's action with $\text{Ad}(X)Y=[X,Y]_\Ag$.
We use retarded Bondi coordinates $(u,r,z, \bar z)$ in  which the Minkowski metric takes the form 
\be
ds^2=-du^2-2dudr+2r^2\gamma_{z\bar z}dzd\bar z
\ee
where $\gamma_{z\bar z}=2/(1+z\bar z)^2$ are the components of the round metric on the celestial sphere $S$.
As in the case of gravity, we will restrict for simplicity to the case when the celestial sphere is flattened to a plane, and the sphere metric becomes $\gamma_{z\bar z}=1$. Under this hypothesis, covariant derivatives on the sphere become partial derivatives.

In radial gauge $A_r=0$, the asymptotic expansion of the gauge field around $\mathscr{I}^+$ reads
\be
A_u=\sum_{n=0}^\infty \f{A^{(n)}_u}{r^n}, \quad A_z=\sum_{n=0}^\infty \f{A^{(n)}_z}{r^n}. 
\ee
Furthermore, we assume $A_u(z,\bz)|_{\mathscr{I}^+_-}=0$.
The asymptotic behaviour of the field strength $F$ is
\be
F_{ur}=\f{1}{r^2}\sum_{n=0}^\infty
\f{F_{ur}^{(n)}}{r^n}, \quad
F_{uz}=\sum_{n=0}^\infty
\f{F_{uz}^{(n)}}{r^n},\quad
F_{rz}=\f{1}{r^2}\sum_{n=0}^\infty
\f{F_{rz}^{(n)}}{r^n}\,,\quad
F_{z\bar z}=\sum_{n=0}^\infty\f{F_{z\bar z}^{(n)}}{r^n}\,. 
\ee

We define positive and negative energy modes of $A_z^{(0)}$ as
\be
\la{AD}
\widetilde A_+(\omega)=\int_{-\infty}^\infty du e^{i\omega u}A_z^{(0)}(u)\,, \quad \widetilde A_-(\omega)=\int_{-\infty}^\infty du e^{i\omega u}{A_z^{(0)*}(u)}\,, \quad \omega>0\,,
\ee
and the Mellin transforms
\be
\widehat A_\pm(\Delta)=\int_0^\infty d\omega \omega^{\Delta-1}\widetilde A_\pm(\omega).
\ee
One can decompose $A_z^{(0)}$ into positive and negative energy sectors as
\be
A_z^{(0)}(u)=A_+(u)+A^*_-(u)\,,
\ee
where 
\be
A_{\pm}(u)=\f{1}{2\pi}\int_{0}^\infty d\omega e^{-i\omega u}\widetilde A_{\pm}(\omega).
\ee
Analogously, one can decompose $F^{(0)}_{\bar z u}=-\partial_u A_{z}^{(0)*}$ into positive and negative energy modes, and define the related Mellin transform. This defines {\it conformal primary gluon} operators as
\be 
{F}^{-a}_\Delta(z,\bz)
&= - \Gamma(\Delta-1) \int_{-\infty}^{\infty} \rd u u^{1-\Delta} F^{(0)a}_{\bar z u}(u,z,\bz)\,,
\cr
{F}^{+a}_\Delta(z,\bz)
&= - \Gamma(\Delta-1) \int_{-\infty}^{\infty} \rd u u^{1-\Delta} F^{(0)a}_{ z u}(u,z,\bz)\,.
\la{Fpm}
\ee
These are related to the usual $O^{\pm a}_\Delta$ one finds in literature by a normalization factor (see \cite{Freidel:2021ytz} for details) 
\be
F_{\Delta}^{\pm a}=(-i)^{\Delta}\f{g_{YM}}{4\pi (-i)}O_{\Delta}^{\pm a}.
\la{norm}
\ee

By large-$r$ expanding the Yang-Mills equations \eqref{YM}, one can extract an infinite tower of $\Ag$-valued charge aspects with conformal dimension and spin $(\Delta, J)=(2,s)$ obeying the following recursive differential equations
\be
\la{Rdot}
\partial_u\mathcal{R}_s=\p\mathcal{R}_{s-1}+i[A^{(0)}_z,\mathcal{R}_{s-1}]_\Ag, \quad  s\ge0, \quad \mathcal{R}_{-1} =F^{(0)}_{\bar z u}.
\ee
In analogy with the case of gravity, \eqref{Rdot} are exact for $s=0,1$, while they represent a truncation of the full Yang--Mills equations for $s>1$; moreover, such charge aspects diverge in the limit $u\rightarrow-\infty$, so one introduces the renormalized charge aspects at $\mathscr{I}^+_-$
\be
r_s(z,\bz)=\lim_{u\rightarrow-\infty}\sum_{l=0}^s\f{(-u)^{s-l}}{(s-l)!}\p^{s-l}\mathcal{R}_l(u,z,\bz),
\ee
which are finite. These charge aspects have positive spin $s$, while their Hermitian conjugates $\bar r_s$ have negative spin $-s$.
They can be expanded in powers of the gauge field $A$ as
$r_s=r_s^1+r_s^2+\dots$, where we will focus on the first two contributions. Explicitly, the soft charge aspects read 
\be
r_s^1(z,\bz)=\p^{s+1}F_s(z,\bz), \quad \text{where} \quad F_s(z,\bz):=\frac{(-)^{s+1}}{s!}\int_{-\infty}^\infty duu^sF^{(0)}_{\bar z u}(z,\bz)
\la{r1}
\ee
are the (sub)$^s$-leading negative helicity {\it conformally soft gluons}. In fact, we have that
\be
\la{ResF}
\Res_{\Delta=1-s} F^{-a}_\Delta(z,\bz)=F^a_s(z,\bz)\,,
\quad
\Res_{\Delta=1-s} F^{+a}_\Delta(z,\bz)=\bar F^a_s(z,\bz)\,,
\ee
where $\bar F_s(z,\bz)$ is defined replacing $F^{(0)}_{\bar z u}$ with $F^{(0)*}_{\bar z u}$ in \eqref{r1}.
The quadratic charge aspects take the form
\be
  r^2_s(z,\bz)&= -i  \sum_{n=0}^{s}  \int_{-\infty}^\infty 
 \frac{(-u)^{s-n}}{(s-n)!} \p^{s-n} \left[A_z^{(0)}(u,z), (\pa_u^{-1} \pa)^{n} F_{\bz u}^{(0)}(u,z) \right]_{\Ag}\,.
 \la{r2}
\ee

The symplectic potential of the theory at $\mathscr{I}^+$ reads
\be
\Theta^{YM}=\f{1}{g_{YM}^2}\int_\mathscr{I^+}\tR[F^{(0)}_{\bar z u}(u,z,\bz)\delta A^{(0)}_z(u,z,\bz)+F^{(0)}_{ z u}(u,z,\bz)\delta A^{(0)}_{\bar z}(u,z,\bz)],
\la{Theta-YM}
\ee
from which we can read off the  Poisson bracket 
\be
\{F^{a(0)}_{\bar z u}(u,z,\bz), A^{b(0)}_z(u',z',\bz')\}=g_{YM}^2\delta^{ab}\delta(u-u')\delta^{2}(z,z')\,.
\la{FApoi}
\ee
Using this bracket, one can straightforwardly compute the action of the quadratic charge aspects \eqref{r2} on the conformal primary gluons and obtain for instance \cite{Freidel:2023gue}
\be 
\{ \bar r_{s}^{2a}(z_1,\bz_1),F^{\pm b}_{\Delta_2}(z_2,\bz_2) \}
= g_{YM}^2 f^{ab}{}_c 
\sum_{n=0}^s (-1)^{s-n}
\frac{(\Delta_2 \mp 1-1)_{s-n}}{(s-n)!}
\bar\pa_1^{s-n} \delta^{(2)}(z_{12}) \bar \pa_2^n F_{\Delta_2-s}^{\pm c}(z_2,\bz_2)\,.
\la{br2FD}
\ee

From this, one obtains the following same-helicity charge aspect bracket
\be 
\{ \bar r_{s_1}^{2a}(z_1,\bz_1),\bar r^{1b}_{s_2}(z_2,\bz_2) \}
= g_{YM}^2 f^{ab}{}_c 
\sum_{n=0}^{s_1} 
\frac{(s_1+s_2-n)_{s_1-n}}{(s_1-n)!}
\bar\pa_2^{s_2+1}\left(\bar\pa_1^{s_1-n} \delta^{(2)}(z_{12}) \bar \pa_2^n \bar F_{s_1+s_2}^{ c}(z_2,\bz_2)\right)\,.
\ee
On the other hand,  the tree level OPE for positive-helicity conformal primary gluons derived in \cite{Pate:2019lpp, Guevara:2021abz} (modulo the rescaling \eqref{norm}) is given by 
\be
F_{\Delta_1}^{a+}(z_1,\bz_1) 
F_{\Delta_2}^{b+}(z_2,\bz_2) \sim  
-i \frac{g_{YM}^2}{4\pi} \frac{f^{ab}{}_c }{z_{12}}
\sum_{n=0}^\infty 
\frac{\Gamma(\Delta_1 -1 +n ) \Gamma(\Delta_2-1)}{\Gamma(\Delta_1 + \Delta_2 + n - 2) } \frac{\bz_{12}^n\bar \pa_2^n }{n!}
F_{\Delta_1+\Delta_2-1}^{c+}(z_2,\bz_2)\,.
\la{eq:cope}
\ee 

By using the definition \eqref{r1}, \eqref{ResF} of the soft charge aspect, it is straightforward to verify \cite{Freidel:2023gue} that
\be
\la{keyYM}
\bar r_{s}^{1a}(z_1,\bz_1)F^{+ b}_{\Delta_2}(z_2,\bz_2)
\leftrightarrow 
{\f{1}{2i}}\{ \bar r_{s}^{2a}(z_1,\bz_1),F^{+ b}_{\Delta_2}(z_2,\bz_2) \}\,,
\ee
and similarly for negative-helicity gluons. The charge OPE/bracket correspondence \eqref{keyYM} is thus the equivalent of \eqref{key} for YM theory.

\subsection{Celestial diamonds}\la{sec:CD-gravity}

\begin{figure}[h!]
     \centering
      \begin{subfigure}[t]{0.61\textwidth}
      \centering
\includegraphics[width=\linewidth,height=50mm, keepaspectratio]{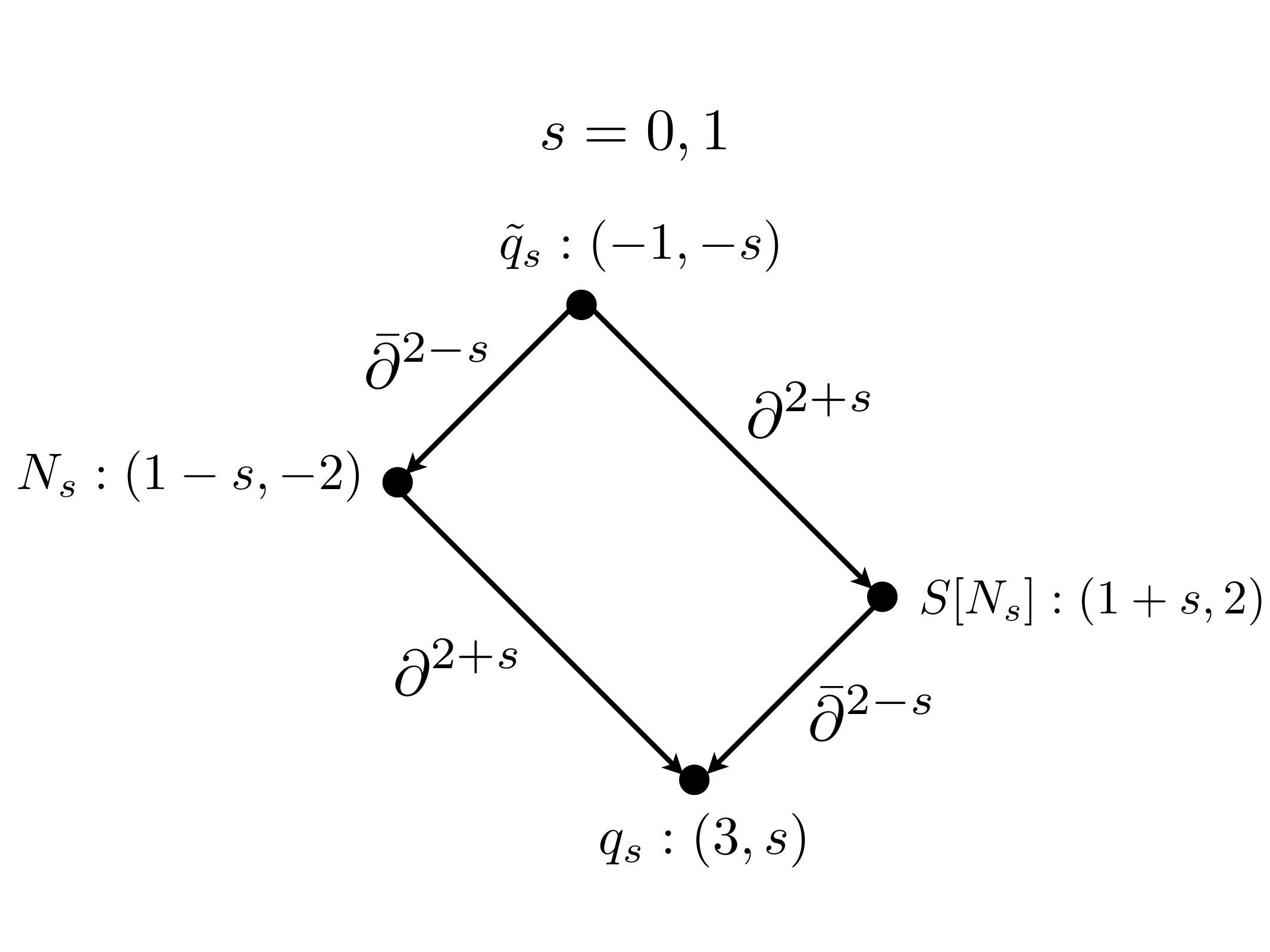}
 \caption{}
 \label{fig:CD-GR-2}
    \end{subfigure}
\quad
 \begin{subfigure}[t]{0.36\textwidth}
      \centering
\includegraphics[width=\linewidth,height=50mm, keepaspectratio]{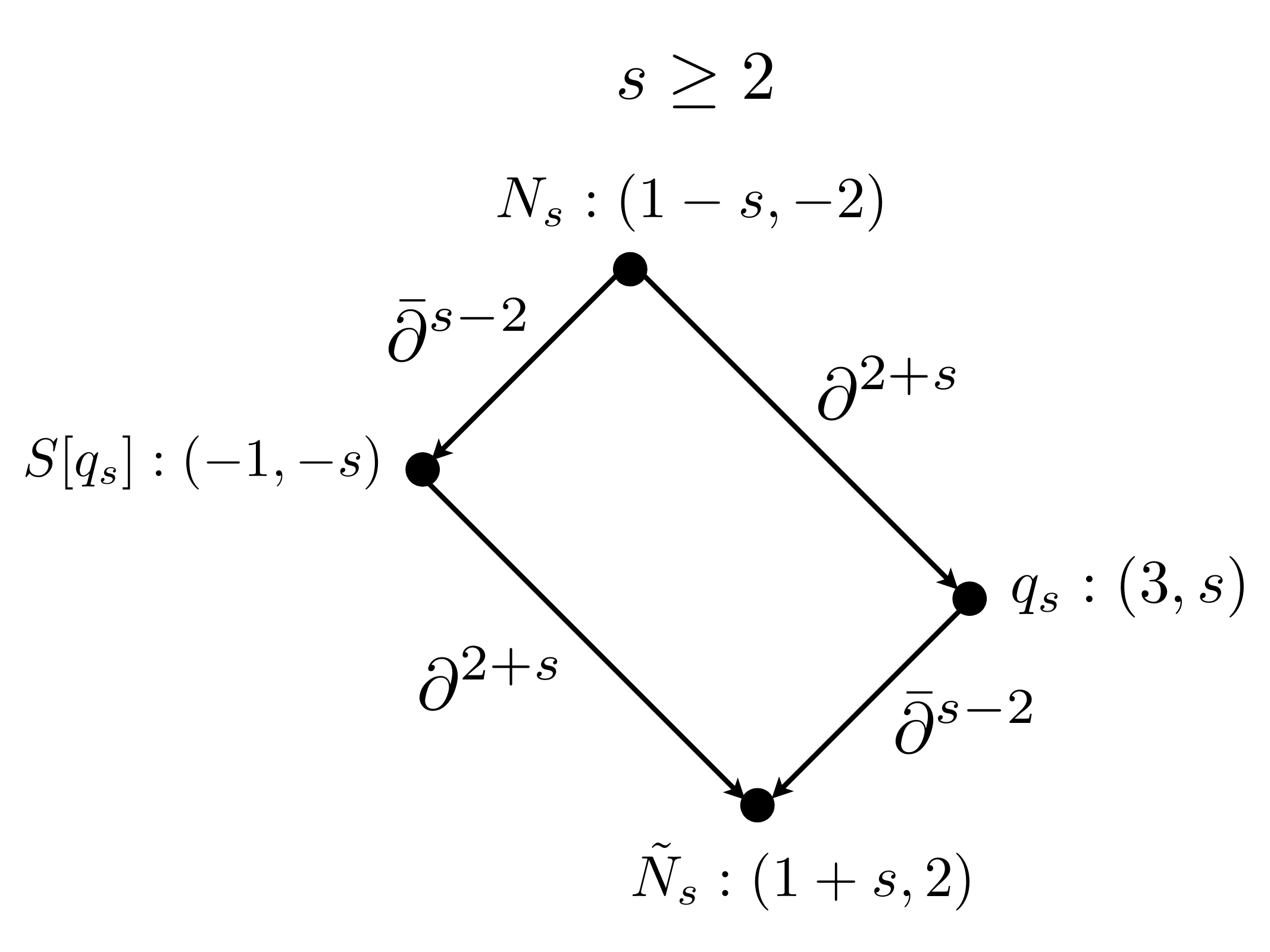}
 \caption{}
 \label{fig:CD-GR-1}
    \end{subfigure}
    \caption{ Celestial diamond associated with negative-helicity soft gravitons with (\subref{fig:CD-GR-2}) $s = 0, 1$ and 
    (\subref{fig:CD-GR-1}) $s\geq 2$.}
    \label{fig:CD-GR}
\end{figure}

\begin{figure}[h!]
     \centering
      \begin{subfigure}[t]{0.61\textwidth}
    \centering
\includegraphics[width=\linewidth,height=50mm, keepaspectratio]{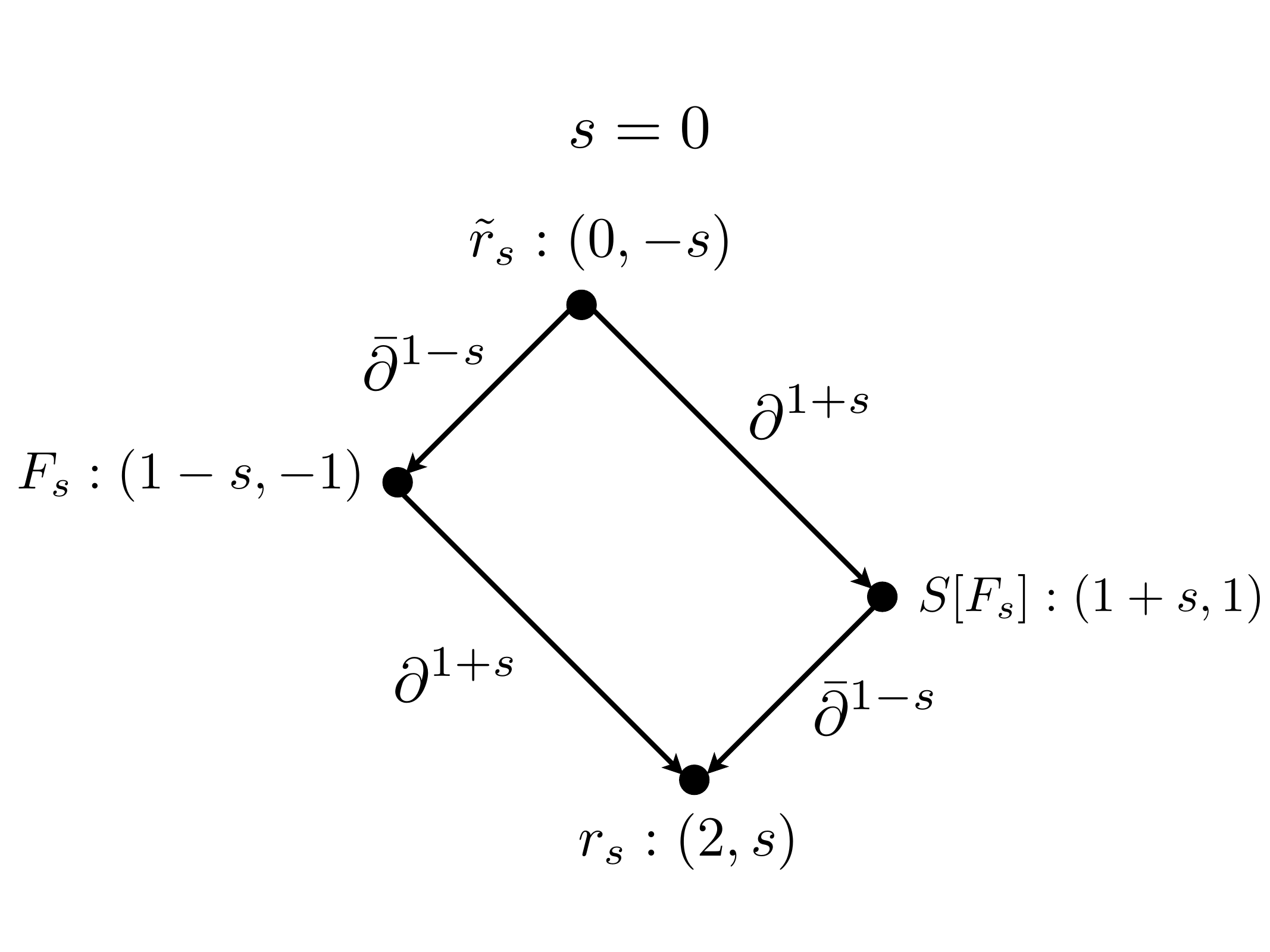}
 \caption{}  
 \label{fig:CD-YM-2}
    \end{subfigure}
\quad
 \begin{subfigure}[t]{0.36\textwidth}
      \centering
\includegraphics[width=\linewidth,height=50mm, keepaspectratio]{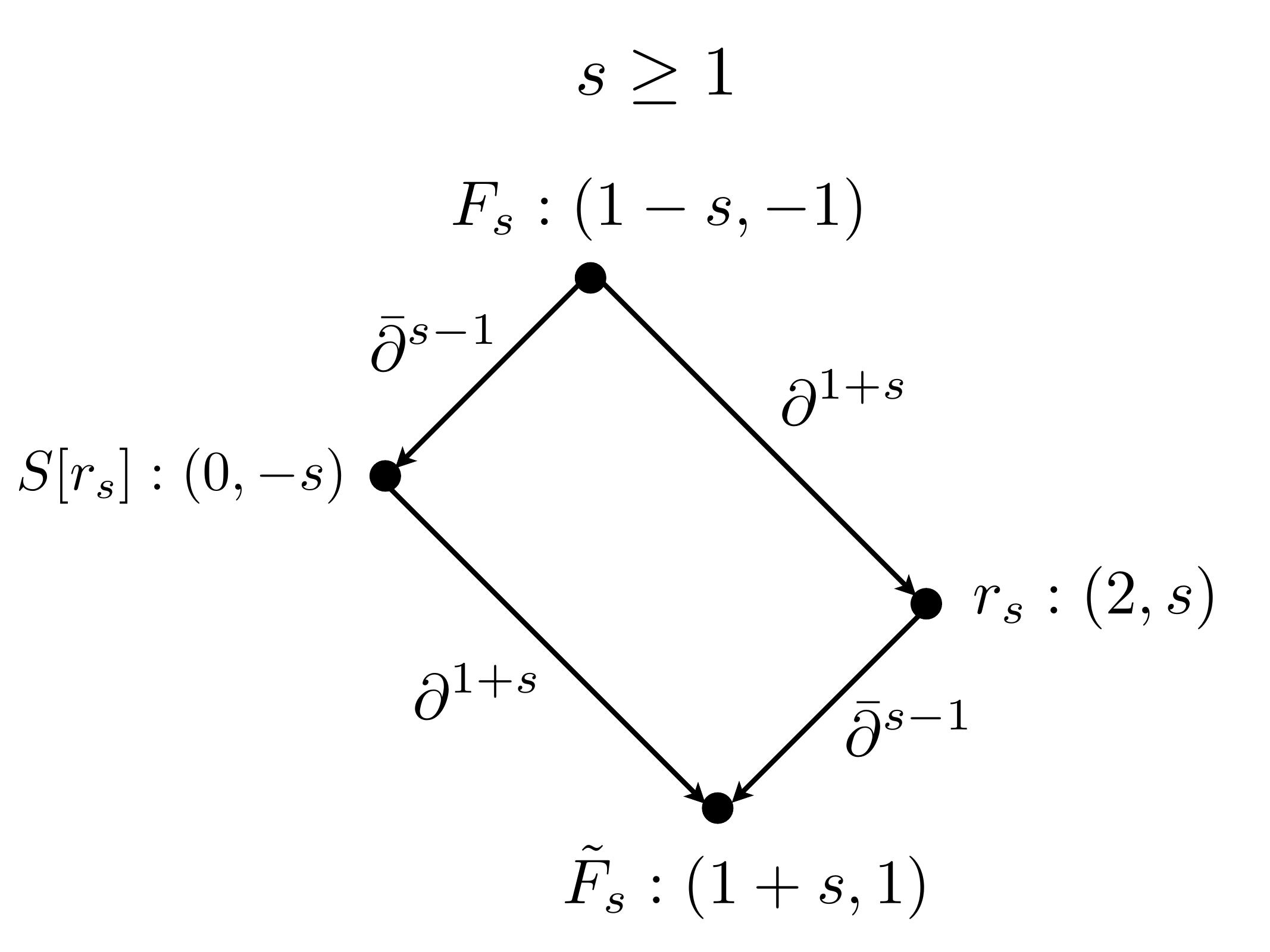}
 \caption{}
 \label{fig:CD-YM-1}
    \end{subfigure}
    \caption{ Celestial diamond associated with negative-helicity soft gluons with (\subref{fig:CD-YM-2}) $s = 0$ and (\subref{fig:CD-YM-1}) $s\geq1$.}
    
\label{fig:CD-YM}
\end{figure}

Celestial diamonds provide a graphical representation  in the $( \Delta,J)$ plane of conformal  primary descendants  operators belonging to global conformal multiplets in 2D celestial CFT (see \cite{Pasterski:2021fjn, Pasterski:2021dqe, Guevara:2021abz,Pano:2023slc,Freidel:2021ytz} for more details). In the case of gravity  and for the purposes of our analysis, there are two classes of memory diamonds according to whether $s\leq 1$ or $s\geq 2$.\footnote{The zero-area diamond of $s=2$ is a limiting case between the two.} In the first case, depicted in Fig. \ref{fig:CD-GR-2}, the left and right  corners are represented by the conformally soft graviton modes and their shadows respectively; they both descend from a generalized conformal primary \cite{Pasterski:2020pdk}, here denoted $\tilde q_s$ and  dubbed {\it dual  charge aspects},\footnote{These were referred to as  {\it conformally soft dressings} in \cite{Pasterski:2021fjn, Pasterski:2021dqe}.} with $(\Delta, J)=(-1,-s)$ and located at the top corner, and they both descend to the soft charge aspects at the bottom corner (also generalized conformal primaries). In the second case, the soft charge aspect operators are located on the right corner and their shadow on the left one; they both descend from the corresponding spin-$s$ soft graviton operators at the top corner and they both descend to the {\it dual soft graviton} operators, denoted $\tilde N_s$, with $(\Delta, J)=(1+s,2)$ and located at the bottom corner. This terminology follows \cite{Freidel:2021ytz}, where the notion of ``duality'' refers to the fact that both pairs of opposite corners  have weights related by $(\Delta, J) \leftrightarrow (2 - \Delta, -J)$ for arbitrary spin-$s$. In this regard, it is worth noting that only the left and right corner entries are indeed one the shadow transform of the other.
 The duality between the top and bottom corner entries has been given an interpretation by means  of SL$(2,\mathbb{C})$ representation theory  in \cite{Freidel:2021ytz}. In particular, representation spaces labeled by $(\Delta,J)$ and $(2-\Delta,-J)$ share a duality pairing through the following bilinear functional on  the complex plane
\be \la{bil_f}(\phi,\psi)= 
\int_{\mathbb{C}}d^2 z[\partial^{-\Delta-J-1}\bar\partial^{-\Delta+J+1}\phi]\psi,
 \ee
 where $\phi,\psi$ are elements of the representation space acting on $L^2(\mathbb{C})$, labelled by $(\Delta, J)$\footnote{Further details can be found in \cite{Gelfand-Book}.}.
 In the same work,   dual soft gravitons $\tN_s$ have been identified with the sub-leading component $\Psi_0^{(s-2)}$ of the $\Psi_0$ Weyl scalar   in the large-$r$ expansion around null infinity $\Psi_0=\sum_n^\infty \Psi_0^{(n)}/r^{5+n}$. {In the course of our analysis, we will devote a good amount of effort to comparing  the OPE structure of dual soft gravitons $\tN_s$ and gluons $\tF_s$ (see Fig. \ref{fig:CD-YM-1}) to the one of their shadow counterparts, providing evidence of their matching.}

In YM theory the two classes of celestial diamonds have the same pattern, with the limiting (zero-area) case corresponding now to $s=1$ and the number of spatial derivatives on each arrow varying as shown in Fig. \ref{fig:CD-YM}.

\section{Double-soft limit}\la{sec:double}

In the literature of celestial holography, an ambiguity concerning the double-soft gluon theorem has been discussed. In the context of $\cS$-matrix scattering amplitudes, when the two gluons becoming soft have opposite helicities, the soft theorem has two different forms depending on the order of the two soft limits \cite{He:2015zea}. The same ambiguity is encountered in the case of the OPE counterpart of the scattering problem \cite{Fan:2019emx}. We will focus on the OPE viewpoint, involving conformal primary gluons $F^{\pm a}_\Delta$.

In most works, when studying the behavior of two opposite helicity gluons labeled by indexes $1$ and $2$, the collinear limit defining the OPE is taken by considering the complex variables $(z,\bar z)$ as independent: this limit is called holomorphic when $z_{12}\rightarrow 0$, while keeping $\bar z_{12}$ fixed, and antiholomorphic in the opposite case \cite{Pate:2019lpp}.
The leading OPE of two opposite helicity conformal primary gluons in the holomorphic limit is

\be
O_{\Delta_1}^{+a}(z_1,\bz_1)O_{\Delta_2}^{-b}(z_2,\bz_2)\sim -ig_{YM}\f{f^{ab}{}_c}{ z_{12}}B(\Delta_1-1,\Delta_2+1)O_{\Delta_1+\Delta_2-1}^{-c}(z_2,\bz_2)\,.
\la{opeh}
\ee
When, otherwise, the variables $(z,\bar z)$ are considered as complex conjugates, the collinear limit takes both $z_{12},\bar z_{12}\rightarrow 0$. In this case, the leading OPE is 

\be
O_{\Delta_1}^{+a}(z_1,\bz_1)O_{\Delta_2}^{-b}(z_2,\bz_2)\sim& -ig_{YM}\f{f^{ab}{}_c}{ z_{12}}B(\Delta_1-1,\Delta_2+1)O_{\Delta_1+\Delta_2-1}^{-c}(z_2,\bz_2)
\cr
&-ig_{YM}\f{f^{ab}{}_c}{ \bar z_{12}}B(\Delta_1+1,\Delta_2-1)O_{\Delta_1+\Delta_2-1}^{+c}(z_2,\bz_2).
\la{ope}
\ee

We define the soft gluon currents as
\be
 j^{a}(z,\bz)&=\lim_{\Delta\rightarrow 1}(\Delta-1)O^{+a}_\Delta(z,\bz)\,,\\
\bar j^{a}(z,\bz)&=\lim_{\Delta\rightarrow 1}(\Delta-1)O^{-a}_\Delta(z,\bz)\,.
\ee
The OPE $\bar j^a(z_1,\bz_1) j^b(z_2,\bz_2)$, in principle, can be derived from \eqref{ope}, passing through a double conformally soft limit, but in practice it is not well-defined because the RHS depends on the order of the two limits. If we take the limit in $\Delta_1$ first, we get
\be
 j^{a}(z_1,\bz_1) \bar j^{b}(z_2,\bz_2)\sim& -ig_{YM}\f{f^{ab}{}_c}{ z_{12}} \bar j^{c}(z_2,\bz_2),
 \la{bj}
\ee
while, if we take the limit in $\Delta_2$ first, we get
\be
 j^{a}(z_1,\bz_1) \bar j^{b}(z_2,\bz_2)\sim& -ig_{YM}\f{f^{ab}{}_c}{\bar z_{12}} j^{c}(z_2,\bz_2).
 \la{j}
\ee

This is the ambiguity shown in \cite{He:2015zea} and \cite{Fan:2019emx}.
We want to exploit  the correspondence \eqref{keyYM} between the OPE and the charge bracket to provide a {\it prescription} to fix the ordering of the two limits. More precisely, when focusing on the mixed-helicity doubly-soft sector, the correspondence we want to implement, passing from Poisson brackets to commutators via the convention $\{\cdot,\cdot\}\to i[\cdot,\cdot]$, is 
\be
\bar r^{1a}_{s_1}(z_1,\bz_1) r_{s_2}^{1b}(z_2,\bz_2)\sim\f{1}{2}[\bar r^{2a}_{s_1}(z_1,\bz_1),r^{1b}_{s_2}(z_2,\bz_2)].
\la{id}
\ee
Therefore, first we  need to derive the RHS of \eqref{id}. As shown in Appendix \ref{App:YM}, this commutator is given by
\be
[\bar r^{2a}_{s_1}(z_1,\bz_1), r^{1b}_{s_2}(z_2,\bz_2)]
&={-}ig^2_{YM}f^{ab}{}_c\sum_{n=0}^{s_1}\frac{(s_1+s_2-n-2)_{s_1-n}}{(s_1-n)!} \p^{s_2+1}_2 (\bar\p^{s_1-n}_1\delta^2(z_1,z_2)\bar\p_2^{n}
F^c_{s_1+s_2}(z_2,\bz_2))\,.
\la{ymba}
\ee

With this tool at hand, let us start with the holomorphic collinear limit. From \cite{Himwich:2021dau} we know that the full expression of the OPE \eqref{opeh} including the descendants, after rescaling the operators with \eqref{norm}, is
\be
F_{\Delta_1}^{+a}(z_1,\bz_1)F_{\Delta_2}^{-b}(z_2,\bz_2)\sim& -i\f{g^2_{YM}}{4\pi}\f{f^{ab}{}_c}{ z_{12}}\sum_{n=0}^\infty \f{1}{n!} B(\Delta_1-1+n,\Delta_2+1)\bar z_{12}^n\bar \partial^n F_{\Delta_1+\Delta_2-1}^{-c}(z_2,\bz_2)\,.
\ee
We want to perform a double-soft limit and obtain the OPE  
\be
\bar F_{s_1}^{a}(z_1,\bz_1) F_{s_2}^{b}(z_2,\bz_2)=\lim_{\Delta_1\rightarrow 1-s_1,\Delta_2\rightarrow 1-s_2}(\Delta_1-1+s_1)(\Delta_2-1+s_2)F_{\Delta_1}^{+a}(z_1,\bz_1)F_{\Delta_2}^{-b}(z_2,\bz_2)\,.
\ee
If we take the limit in $\Delta_1$ first, we obtain
\be
\bar F_{s_1}^{a}(z_1,\bz_1) F_{s_2}^{b}(z_2,\bz_2)\sim& -i\f{g^2_{YM}}{4\pi}\f{f^{ab}{}_c}{ z_{12}}\sum_{n=0}^{s_1} \f{1}{n!} \frac{(s_1+s_2-n-2)_{s_1-n}}{(s_1-n)!}\bar z_{12}^n\bar \partial^n F_{s_1+s_2}^{c}(z_2,\bz_2),
\la{order1}
\ee
where we used that $\Gamma(z)$ has no singularities on the positive real axis, while $\Res_{z=-n}\Gamma(z)=\f{(-)^n}{n!}\,\forall n\in\mathbb{Z}_{+}$, and that $\Gamma(x+1)/\Gamma(x-n+1)=(x)_n$.

In the other order, when taking the limit on $\Delta_2$ first, we would get 
zero if $s_2=0,1$, as $\Res_{\Delta_2=1-s_2}\Gamma(\Delta_2+1)=0$ for $s_2<2$, which means that the two soft limits do not commute.  
In the case $s_2\ge2$, instead, we get the non-vanishing expression 
 \be
\bar F_{s_1}^{a}(z_1,\bz_1) F_{s_2}^{b}(z_2,\bz_2)\sim& -i\f{g^2_{YM}}{4\pi}\f{f^{ab}{}_c}{ z_{12}}\bigg[\sum_{n=0}^{s_1} \f{1}{n!} \frac{(s_1+s_2-n-2)_{s_1-n}}{(s_1-n)!}\bar z_{12}^n\bar \partial^n F_{s_1+s_2}^{c}(z_2,\bz_2)
\cr&
+\sum_{n=s_1+s_2-1}^{\infty} \f{1}{n!} \frac{(s_1+s_2-n-2)_{s_2-2}}{(s_2-2)!}\bar z_{12}^n\bar \partial^n F_{s_1+s_2}^{c}(z_2,\bz_2)\bigg],
 \la{order2}
 \ee
which is different from \eqref{order1} because of the second row.
Thus, for the holomorphic collinear limit, the conformally soft limits of the OPE  {\it do not} commute for any $s_2\geq 0$. 

Let us see what happens with the OPE of soft charges. By applying the differential operator $\partial_{2}^{s_2+1}\bar \partial_{1}^{s_1+1}$ to \eqref{order1}, we obtain

\be
 \bar r^{1a}_{s_1}(z_1,\bz_1) r_{s_2}^{1b}(z_2,\bz_2)\sim& -i\f{g^2_{YM}}{2}{f^{ab}{}_c}\sum_{n=0}^{s_1}  \frac{(s_1+s_2-n-2)_{s_1-n}}{(s_1-n)!} \partial^{s_2+1}_{2}(\bar \partial_{1}^{s_1-n}\delta^{(2)}(z_1,z_2) \bar \partial_{2}^n F_{s_1+s_2}^{c}(z_2,\bz_2)).
\la{order1r}
\ee

where we used 
\be
\f{1}{n!}\bar \partial_{1}^{s_1+1}\biggl(\f{\bar z_{12}^n}{ z_{12}}\biggr)=2\pi\bar\partial_{1}^{s_1-n}\delta^{(2)}(z_1,z_2).
\ee
The expression on the RHS of \eqref{order1r} is proportional to the bracket \eqref{ymba}, that is  the identification \eqref{id} is realized.
This holds for any $s_1,s_2$, only if the limit in $\Delta_1$ is {\it taken first}. If we take the limit in $\Delta_2$ first, instead, 
the above identification is not satisfied {(we have a vanishing OPE for $s_2=0,1$ and extra terms for $s_2\ge 2$)}.

Let us point out that a  similar non-commutativity
 of soft limits is found also for the same-helicity OPE, as soon as one includes the full tower of descendants in the OPE \eqref{eq:cope}. In fact,  if we take the soft limit of \eqref{eq:cope} in $\Delta_1$ first, we obtain 
\be
\bar F_{s_1}^{a}(z_1,\bz_1) 
\bar F_{s_2}^{b}(z_2,\bz_2) \sim  
-i \frac{g_{YM}^2}{4\pi} \frac{f^{ab}{}_c }{z_{12}}
\sum_{n=0}^{s_1} 
\f{(s_1+s_2-n)_{s_1-n}}{(s_1-n)!} \frac{\bz_{12}^n\bar \pa_2^n }{n!}
\bar F_{s_1+s_2}^{c}(z_2,\bz_2)\,,
\la{sh-order1}
\ee 
while if we take the limit in $\Delta_2$ first, we obtain
\be
\bar F_{s_1}^{a}(z_1,\bz_1) 
\bar F_{s_2}^{b}(z_2,\bz_2) \sim&  
-i \frac{g_{YM}^2}{4\pi} \frac{f^{ab}{}_c }{z_{12}}
\bigg[\sum_{n=0}^{s_1} 
\f{(s_1+s_2-n)_{s_1-n}}{(s_1-n)!} \frac{\bz_{12}^n\bar \pa_2^n }{n!}
\bar F_{s_1+s_2}^{c}(z_2,\bz_2)
\cr&
+\sum_{n=s_1+s_2+1}^{\infty} 
\f{(s_1+s_2-n)_{s_2}}{s_2!} \frac{\bz_{12}^n\bar \pa_2^n }{n!}
\bar F_{s_1+s_2}^{c}(z_2,\bz_2)\bigg]\,.
\la{sh-order2}
\ee
Here, terms in the second row represent an obstruction to the exact commutativity of the two soft limits, analogous to what happens for $s_2\ge2$ in the mixed-helicity case. {This non-commutativity issue was  noted also in \cite{Ball:2022bgg} for both same-helicity and mixed-helicity cases (see also \cite{Ball:2024oqa} for a review). In that work, it is argued that this ambiguity for the same-helicity sector does not affect the  commutator of the holomorphic currents constructed from a contour integral of the OPE, since the mode expansion of the obstruction terms does not contain any pole; this way,   the Jacobi identity of   the current algebra is preserved. However, we point out that greater care is required when applying this argument to the mixed-helicity case: here, for $s_2\le1$, the ambiguity manifests not through obstruction terms, but through the fact that only one ordering (first entry taken soft first) yields a non-vanishing OPE. We further emphasize that, unlike the holomorphic commutator, the ordering of the soft limits is crucial to recover the \textit{canonical} commutator of charge aspects, computed from the phase space Poisson bracket \eqref{FApoi}, from the OPE, even in the same-helicity case.} 
More precisely, the linear-order of the canonical  commutator of two YM charge aspects is given by the sum of two contributions, namely
\be
[\bar r^{a}_{s_1}(z_1,\bz_1), \bar r_{s_2}^{b}(z_2,\bz_2)]^1=[\bar r^{1a}_{s_1}(z_1,\bz_1), \bar r_{s_2}^{2b}(z_2,\bz_2)]
+[\bar r^{2a}_{s_1}(z_1,\bz_1), \bar r_{s_2}^{1b}(z_2,\bz_2)]\,.
\la{rrcom}
\ee
The bracket-OPE correspondence discussed so far, applies to \emph{each} of the two pieces on the RHS of \eqref{id-sh}. In particular, applying the derivative operators $\bar \p_1^{s_1+1}\bar\p_2^{s_2+1}$ on the first ordering \eqref{sh-order1} gives an exact match with the bracket-OPE correspondence 
\be
\bar r^{1a}_{s_1}(z_1,\bz_1) \bar r_{s_2}^{1b}(z_2,\bz_2)\sim\f{1}{2}[\bar r^{2a}_{s_1}(z_1,\bz_1),\bar r^{1b}_{s_2}(z_2,\bz_2)].
\la{id-sh}
\ee
This \emph{does not} happen for the second ordering \eqref{sh-order2}, which introduces an obstruction due to the extra terms in the second row.

One might then consider a looser notion of bracket-OPE correspondence for the full linear-order canonical commutator \eqref{rrcom}, consisting of the following combination of OPEs
\be
[\bar r^{a}_{s_1}(z_1,\bz_1), \bar r_{s_2}^{b}(z_2,\bz_2)]^1\sim
\f12\bar r^{1a}_{s_1}(z_1,\bz_1) \bar r^{1b}_{s_2}(z_2,\bz_2)- 
\f12 \bar r^{1b}_{s_2}(z_2,\bz_2)
\bar r^{1a}_{s_1}(z_1,\bz_1)\,.
\ee
One could then ask whether this correspondence is independent of the ordering of the two conformally-soft limits. The answer is negative: the correspondence is verified only when \emph{both} terms on the RHS are computed using the prescription {\it first entry goes soft first}. With the reverse ordering (or any mixed combination of orderings), the extra terms from \eqref{sh-order2} do not cancel, thus obstructing the correspondence. 
Let us also point out that we would have  commutativity of the two conformally soft limits only if the summation over the descendants in \eqref{eq:cope} was truncated at most  to $n=s_1{+s_2}$, i.e. if we restrict to the wedge, as already noticed in  \cite{Ball:2022bgg}. 
These results, considering the full tower of descendants, already provide a first indication that there is a preferred ordering for the double limit. 

So far we restricted to the holomorphic collinear limit. If one wants to go beyond it, thus considering $(z,\bar z)$ as complex conjugates, the expression of the same-helicity OPE \eqref{eq:cope} does not change, so the above observations still hold. In the case of mixed-helicity, instead
, the form of the OPE of the two conformal primary gluons becomes
\be
O_{\Delta_1}^{+a}(z_1,\bz_1)O_{\Delta_2}^{-b}(z_2,\bz_2)\sim& -ig_{YM}\f{f^{ab}{}_c}{ z_{12}}\sum_{n=0}^\infty \f{1}{n!} B(\Delta_1-1+n,\Delta_2+1)\bar z_{12}^n\bar \partial^n O_{\Delta_1+\Delta_2-1}^{-c}(z_2,\bz_2)
\cr
&-ig_{YM}\f{f^{ab}{}_c}{ \bar z_{12}} \left(B(\Delta_1+1,\Delta_2-1) O_{\Delta_1+\Delta_2-1}^{+c}(z_2,\bz_2)+\mathcal{O}(z_{12})\right).
\la{opef}
\ee

The first row here is the same as the holomorphic limit restriction. When one performs the conformally soft limit in $\Delta_1$ on the second row, this vanishes for $s_1< 2$. This means that beyond $s_1=1$, the conformally soft limit of the OPE receives corrections. If these come with corrections to the higher-spin charges, as expected since their evolution equations \eqref{Rdot} beyond $s_1=1$ are only truncations of the Yang-Mills equations, it might be possible to preserve the correspondence \eqref{id}. Such analysis is however beyond the scope of this manuscript. 

Nevertheless, if one takes the conformally soft limit of the second row of \eqref{opef} in $\Delta_2$ first, this is non-vanishing for all values of $s_2$, even spins $s_2=0,1$. 
Given the possible appearance of both helicity soft gluons in the OPE, when taking the double-soft limit in different orders,
 a scenario one could  contemplate is the following. 
In the mixed-helicity sector,
the linear order bracket between two charge aspects consists of the two contributions
\be
[\bar r^{a}_{s_1}(z_1,\bz_1), r^{b}_{s_2}(z_2,\bz_2)]^1=[\bar r^{2a}_{s_1}(z_1,\bz_1), r^{1b}_{s_2}(z_2,\bz_2)]+[\bar r^{1a}_{s_1}(z_1,\bz_1), r^{2b}_{s_2}(z_2,\bz_2)]\,.
\la{rrcomm}
\ee
A fair guess  could be  that the first contribution would correspond to the OPE $\bar r^{1a}_{s_1}(z_1,\bz_1)  r^{1b}_{s_2}(z_2,\bz_2)$, where the first entry goes soft first, while the second commutator in \eqref{rrcomm} would correspond to the same OPE but where the second entry goes soft first. What we show now is that in general this is {\it not} the case. 

We will only analyze the case $s_1=s_2=0$, which is interesting because in this regime the evolution equations of the higher-spin charges are exact, and also because this is the case in which the double-soft limit ambiguity is discussed in literature. 
Taking the double-soft limit of \eqref{opef} in the two orderings, we have
\be
\bar F^a_{0}(z_1,\bz_1)F^b_0(z_2,\bz_2)\sim \begin{cases}
  -i\f{g^2_{YM}}{4\pi}\f{f^{ab}{}_c}{ z_{12}} F_{0}^{c}(z_2,\bz_2) \quad \text{for} \quad \Delta_1\rightarrow 1 \quad \text{first,}\\
  -i\f{g^2_{YM}}{4\pi}\f{f^{ab}{}_c}{ \bar z_{12}} \bar F_{0}^{c}(z_2,\bz_2)\quad \text{for} \quad \Delta_2\rightarrow 1 \quad \text{first.}
\end{cases}
\la{FF}
\ee
We see that in the $\Delta_1\rightarrow 1$ first case the sum over descendants gets truncated to the primary only; even if the second row of \eqref{opef} already contains only the primary term, we assume that the same truncation for the $\Delta_2\rightarrow 1$ first case would remain valid even after inclusion of descendants for the positive-helicity gluon in \eqref{opef}.

These two expressions are respectively equivalent to \eqref{bj} and \eqref{j}, since 
\be
\bar F^a_0(z)=\f{g_{YM}}{4\pi}j^a(z).
\ee
If we apply $\partial_{2}\bar \partial_{1}$ to \eqref{FF},  we obtain

\be
\bar r^{1a}_{0}(z_1,\bz_1)r^{1b}_0(z_2,\bz_2)\sim \begin{cases}
  -i\f{g^2_{YM}}{2}f^{ab}{}_c\p_2(\delta^{(2)}(z_1,z_2) F_{0}^{c}(z_2,\bz_2)) \quad \text{for} \quad \Delta_1\rightarrow 1 \quad \text{first,}\\
  i\f{g^2_{YM}}{2}f^{ab}{}_c(\bar\p_{1}\delta^{(2)}(z_1,z_2) \bar F_{0}^{c}(z_2,\bz_2)+\f{1}{2\pi}\f{1}{\bar z^2_{12}}\p_2\bar F_{0}^{c}(z_2,\bz_2))\quad \text{for} \quad \Delta_2\rightarrow 1 \quad \text{first,}
  \la{casesYM}
\end{cases}
\ee
We observe that only for $\Delta_1\rightarrow 1$ taken first, the charge OPE/bracket correspondence 
\be
\bar r^{1a}_{0}(z_1,\bz_1) r_{0}^{1b}(z_2,\bz_2)\sim\f{1}{2}[\bar r^{2a}_{0}(z_1,\bz_1),r^{1b}_{0}(z_2,\bz_2)]
\la{id0}
\ee
is verified.
 In fact, if
we turn to the case $\Delta_2\rightarrow 1$ taken first, we can rewrite it as
\be
\bar r^{1a}_{0}(z_1,\bz_1) r_{0}^{1b}(z_2,\bz_2)&\sim
i\f{g^2_{YM}}{2}f^{ab}{}_c\bigg(\bar\p_{1}(\delta^{(2)}(z_1,z_2) \bar F_{0}^{c}(z_2,\bz_2))+\f{1}{2\pi}\f{1}{\bar z^2_{12}}\p_{2}\bar F_{0}^{c}(z_2,\bz_2)\bigg)\,.
\ee
This means
\be
\bar r^{1a}_{0}(z_1,\bz_1) r_{0}^{1b}(z_2,\bz_2)
&\sim
\f{1}{2}[\bar r^{1a}_{0}(z_1,\bz_1), r^{2b}_{0}(z_2,\bz_2)]+i\f{g^2_{YM}}{4\pi }\f{1}{\bar z^2_{12}}f^{ab}{}_c
\p_{2}\bar F_{0}^{c}(z_2,\bz_2)\,,
\la{br-ope}
\ee
from which we see that the  case $\Delta_2\rightarrow 1$ first does not realize a 
 charge OPE/bracket correspondence with the second term on the RHS of \eqref{rrcomm}. {We observe, however, that such scenario could be realized if the second line of \eqref{opef} were expanded around $(z_1, \bar z_1)$ instead of $(z_2, \bar z_2)$: at the level of primaries only, the two choices of expansion point in \eqref{opef} are equivalent. We would then start with
 \be
O_{\Delta_1}^{+a}(z_1,\bz_1)O_{\Delta_2}^{-b}(z_2,\bz_2)\sim& -ig_{YM}\f{f^{ab}{}_c}{ z_{12}}\sum_{n=0}^\infty \f{1}{n!} B(\Delta_1-1+n,\Delta_2+1)\bar z_{12}^n\bar \partial^n O_{\Delta_1+\Delta_2-1}^{-c}(z_2,\bz_2)
\cr
&-ig_{YM}\f{f^{ab}{}_c}{ \bar z_{12}} \left(B(\Delta_1+1,\Delta_2-1) O_{\Delta_1+\Delta_2-1}^{+c}(z_1,\bz_1)+\mathcal{O}(z_{12})\right),
\la{opef2}
\ee
instead of \eqref{opef}. Applying the double-soft limit to it, we would obtain
\be
\bar F^a_{0}(z_1,\bz_1)F^b_0(z_2,\bz_2)\sim \begin{cases}
  -i\f{g^2_{YM}}{4\pi}\f{f^{ab}{}_c}{ z_{12}} F_{0}^{c}(z_2,\bz_2) \quad \text{for} \quad \Delta_1\rightarrow 1 \quad \text{first,}\\
  -i\f{g^2_{YM}}{4\pi}\f{f^{ab}{}_c}{ \bar z_{12}} \bar F_{0}^{c}(z_1,\bz_1)\quad \text{for} \quad \Delta_2\rightarrow 1 \quad \text{first.}
\end{cases}
\la{FF2}
\ee
instead of \eqref{FF}\footnote{Alternatively, one could start with \eqref{opef}, and Taylor-expand the second line of \eqref{FF} around $(z_1,\bz_1)$, truncating to the primary contribution. This would lead to the same conclusion}. Then, acting with $\partial_{2}\bar \partial_{1}$ on the second line of \eqref{FF2} would precisely yield the correspondence
\be
\bar r^{1a}_{0}(z_1,\bz_1) r_{0}^{1b}(z_2,\bz_2)
&\sim
\f{1}{2}[\bar r^{1a}_{0}(z_1,\bz_1), r^{2b}_{0}(z_2,\bz_2)]\,,
\la{br-ope2}
\ee
without the obstruction term appearing in \eqref{br-ope}.} 

 To summarize, the results of this section  show that, both for the case of same-helicity and mixed-helicity gluons, the canonical approach can remove the ambiguities arising in the OPE approach when we include the full tower of descendants and both operators go soft, providing the prescription: {\it First entry goes soft first}. We will apply this rule in the following when studying OPEs involving shadow operators in the case where both entries are soft. Let us thus move to our second and main application of the charge OPE/bracket correspondence.

\section{Shadow OPE in gravity}\la{sec:garvity}

The shadow transform can be defined through a notion of   \textit{propagator} $G_n(z;z_1)$ such that 
\be 
D^n_{z_1}G_n(z;z_1)=\delta^2(z,z_1), 
\ee
and its complex conjugate $\bar G_n(z;z_1):=(G_n(z;z_1))^*$. On the celestial plane, for $n\ge1$, one has
\be
G_n(z_1;z)=\f{1}{2\pi(n-1)!}\f{(z-z_1)^{n-1}}{\bar z-\bar z_1}\,.
\ee
Shadow charge aspects and shadow soft gravitons are then given by
\be
S[q_s](z,\bz)&=
   \bar \p^{s-2}\int_{S_1}G_{s+2}(z_1;z)q_s(z_1,\bz_1), \quad &\text{for} \quad s\ge2\\
  S[N_s](z,\bz)&= \p^{s+2} \int_{S_1}\bar G_{2-s}(z_1;z)N_s(z_1,\bz_1)\quad &\text{for} \quad s\leq 1\,.
  \la{shaqN}
\ee
Complementary to these, we can express the dual  charge aspects $\tq_s$ and dual soft gravitons $\tN_s$ as
\be
\tq_s(z,\bz)&=
    \int_{ S_1}\int_{S_2}\bar G_{2-s}(z_1;z)G_{s+2}(z_2;z_1)q_s(z_2,\bz_2), \quad &\text{for} \quad s\leq 1\\
\tN_s(z,\bz)&= \p^{s+2} \bar \p^{s-2}N_s(z,\bz)\quad &\text{for} \quad s  \geq 2\,.
\la{dualqN}
\ee
Eq.s \eqref{shaqN},  \eqref{dualqN} express the web of relations depicted by the celestial diamonds in Fig. \ref{fig:CD-GR}. Shadow and dual soft charge aspects can then be expressed in terms of soft gravitons via \eqref{q1}. 

More generally, we  use the following definitions (see \eqref{Shadow}) of the shadow transform $S:(\Delta,J)\rightarrow(2-\Delta,-J)$ for conformal primary gravitons 
\be
S[N^-_{\Delta}](z_2,\bz_2)&=\f{K_{\Delta}}{2\pi}\int_{S_3}\f{1}{z_{23}^{4-\Delta}}\f{1}{\bar z_{23}^{-\Delta}}N^-_{\Delta}(z_3,\bz_3)\,, 
\la{scg}
\\
 N^-_{\Delta}(z_2,\bz_2)
&=\f{\tilde K_{2-\Delta}}{2\pi}\int_{S_3}\f{1}{z_{23}^{\Delta-2}}\f{1}{\bar z_{23}^{\Delta+2}}S[N^-_{\Delta}](z_3,\bz_3)\,,
 \la{scg2}
\ee
where the constants are
\be
K_{\Delta}&=(-)^{3-\Delta}\f{\Gamma(4-\Delta)}{\Gamma(1+\Delta)}, 
\la{KGR}
\\
\tilde K_{2-\Delta}&=(-)^{\Delta+1}\f{\Gamma(2+\Delta)}{\Gamma(3-\Delta)}\,.
\la{tKGR}
\ee 
{The corresponding expressions for the positive-helicity primaries $N^+_{\Delta}$ are identical to the above ones upon the exchange $z_{23}\leftrightarrow\bar z_{23}$.}
Let us point out that, taking the conformally soft limit of \eqref{scg}, we recover the   operators \eqref{shaqN} for $s\le1$;   for $s\ge2$, the two classes of operators given by the conformally soft limit of \eqref{scg} and \eqref{dualqN}
 only share the same conformal dimension and helicity $(\Delta,J)=(1+s,\,2)$. Nevertheless, one of the main focus of our analysis is to provide evidence that, when inserted in a correlation function, these two sets of operators exhibit the same OPE structure. 

Our strategy for deriving shadow celestial OPEs is to employ the charge OPE/bracket correspondence \eqref{key} together with \eqref{shaqN},  \eqref{dualqN}, \eqref{scg}. In order to test the validity of this strategy, we  apply it first to recover results in previous literature on OPEs involving the CCFT energy-momentum tensor. 

Before proceeding, we inform the reader that, in the rest of this manuscript, 
we will sometimes restrict to  conformal primary gravitons/gluons  such that the real part of their conformal dimension satisfies $\Re\Delta\le1$. This includes the complete basis of boost
eigenstates labeled  by the principal continuous series $\Delta=1+i\lambda, \lambda\in \R$ \cite{Pasterski:2017kqt}, as well as the set of memory observables in the discrete basis of \cite{Freidel:2022skz}; in particular,
 it allows us  to take the conformally soft limit  $\Delta\to1-s, s\in \N$ of both conformal primary gravitons and gluons within an OPE. At the same time, this restriction (which we will explicitly point out when applied) simplifies the expression of some  OPEs, whose analytic structure would otherwise require a more general treatment.

\subsection{Energy-momentum tensor }\la{sec:SET}

The CCFT energy-momentum tensor is defined as the shadow transform of the $s=1$ soft graviton operator
\be
T(z_1,\bz_1):=S [N_1](z_1,\bz_1)
&=-\f{3!}{2\pi}\int_{S}\f{1}{(z_1-z)^4}N_1(z,\bz)
\cr
&=\f{1}{2\pi}\int_{S}\f{1}{z_1-z}q_1(z,\bz)\,,
\la{1strow}
\ee
and it has been shown to be the generator of local conformal transformation on the celestial sphere \cite{Kapec:2016jld,Fotopoulos:2019tpe,Fotopoulos:2019vac}. Let us first verify this property by using our prescription to compute celestial OPEs involving shadow operators. 

Starting from the second line of \eqref{1strow}, we can use the bracket \eqref{qND} to write the OPE
\be
q^1_1(z_1,\bz_1)N^\pm_{\Delta_2}(z_2,\bz_2)
&\sim-\f{i\k}{8}\bigg[-(\Delta_2\pm2)\partial_1\delta^2(z_1,z_2)N^\pm_{\Delta_2}(z_2,\bz_2)+2\delta^2(z_1,z_2)\partial_2 N^\pm_{\Delta_2}(z_2,\bz_2)\bigg]\,,
\ee
from which
\be
T(z_1,\bz_1)N^\pm_{\Delta_2}(z_2,\bz_2)
&\sim-\f{i\k}{8}\f{1}{2\pi}\bigg[\f{(\Delta_2\pm2)}{(z_1-z_2)^2}N^\pm_{\Delta_2}(z_2,\bz_2)+\f{2}{z_1-z_2}\partial_2N^\pm_{\Delta_2}(z_2,\bz_2)\bigg]\,,
\la{TN}
\ee
which yields the expected conformal Ward identity for 
positive and negative helicity gravitons with conformal weights $2h_{2\pm} = \Delta_2 \pm 2$.

We can now proceed further and take a shadow transform on the second entry. We focus first on the same-helicity OPE.
For $\Re\Delta_2\le1$, we have (see Appendix \ref{App:TSN} for details)
\be
T(z_1,\bz_1) S[N^-_{\Delta_2}](z_2,\bz_2)
&\sim-\f{i\k}{16\pi}
\bigg(
-(4-\Delta_2)_3z_{12}^{\Delta_2-5}\bar z_{12}^{\Delta_2+1}\f{1}{2\pi}\int_{S_4}z_{41}^{1-\Delta_2}
z_{24}\bar z_{24}^{-1}\bar z_{41}^{-\Delta_2-1} S[N^-_{\Delta_2}](z_4,\bz_4)
\cr
&-(3-\Delta_2)_2z_{12}^{\Delta_2-4}\partial_{z_1}(\bar z_{12}^{\Delta_2+1})\f{1}{2\pi}\int_{S_4}z_{41}^{2-\Delta_2}
\bar z_{24}^{-1}\bar z_{41}^{-\Delta_2-1} S[N^-_{\Delta_2}](z_4,\bz_4)
\cr&+ z_{12}^{-2}(4-\Delta_2+2z_{12}\partial_{z_2})  S[N^-_{\Delta_2}](z_2,\bz_2)
\bigg)\,,
\la{tn1tn1b-2}
\ee
where the second row only contributes when $\Re\Delta_2\le-2$. {Upon the rescaling \eqref{propOG} of primary fields, the local part of this expression is in agreement with \cite{Himwich:2025bza} (see Eq. (3.16) there), up to a sign. In the aforementioned work, this sign mismatch is found via a different construction of the OPE of shadow operators, relying on an OPE block, and it is attributed to the alleged presence of additional singularity structures in correlators, beyond what is captured by the shadow transform of the OPE coefficients.}

In order to obtain the OPE between two copies of the energy-momentum tensor, we can take the conformally soft limit $\Delta_2\to 0)$ of \eqref{tn1tn1b-2} (see Appendix \ref{App:TT}) and we arrive at
\be
T(z_1,\bz_1)T(z_2,\bz_2)
\sim&
-\f{i\k}{8\pi}\bigg[
 \left(\f{2}{z_{12}^2}+\f{1}{z_{12}}\partial_{2}\right)T(z_2,\bz_2)
\cr
&
-\f{24 }{z_{12}^5}\left( \tilde q^1_1 (z_2,\bz_2)- \tilde q^1_1 (z_1,\bz_1)\right)
-\f{12}{z_{12}^4}(
\partial_{2} \tilde q^1_1 (z_2,\bz_2)
+\partial_{1} \tilde q^1_1 (z_1,\bz_1)
)
\bigg]\,.
\la{TT}
\ee
The first line on the RHS of \eqref{TT} reproduces the expected OPE of two copies of the energy-momentum tensor; the second line represents an extra  non-contact term\footnote{As a remark, if we compute \eqref{TT} with the opposite ordering of soft limits (i.e., taking the second entry soft first), the situation becomes more complicated: not only are the extra terms not removed, but additional undesired terms also survive.}. Such undesired contribution is not new of course; in fact, it can be rewritten as (see Appendix \ref{App:TT})
\be
\f{24 }{z_{12}^5}\left( \tilde q^1_1 (z_2,\bz_2)- \tilde q^1_1 (z_1,\bz_1)\right)
+\f{12}{z_{12}^4}(
\partial_{2} \tilde q^1_1 (z_2,\bz_2)
+\partial_{1} \tilde q^1_1 (z_1,\bz_1)
)
&=-\f1{\pi}\f{6}{z_{12}^2} \int_{S_3} \f{1}{z_{23}^2z_{13}^2}N_1(z_3,\bz_3)\,,
\la{TTextra}
\ee
which is exactly the obstruction term found in \cite{Fotopoulos:2019vac} (see Eq. (4.12) there). There the authors argued that, by means of the sub-leading soft theorem (due to the insertion of $N_1$) \cite{Kapec:2014opa} and conformal integrals of the form
\eqref{intid}, this undesired contribution vanishes due to  global conformal invariance. However, the validity of this conclusion has been questioned in \cite{Banerjee:2022wht} due to subtleties related to subleading divergent terms of a $\Gamma(0)$ function, appearing when performing the conformal integral. {Here, as in \cite{Banerjee:2022wht}, we include both operators $N_1$ and its shadow $T$ in the theory. In this setting, we provide further support for the conclusion that the obstruction term \eqref{TTextra} is non-vanishing, 
based on
 the structure of the relevant celestial diamond.
If we compute the OPE associated with the conservation law for the energy-momentum tensor, we obtain}
\be
T(z_1,\bz_1)\bar \p T(z_2,\bz_2)
\sim&
-\f{i\k}{4}\f{1}{2\pi}\bigg[-\f{24}{z_{12}^5}N_1(z_2,\bz_2)-\f{12}{z_{12}^4}\partial N_1(z_2,\bz_2)
\cr&+\f{2}{z_{12}^2}q^1_1(z_2,\bz_2)+\f{1}{z_{12}}\partial q^1_1(z_2,\bz_2)\bigg]\,;
\la{TbarpT}
\ee
we see that the second line on the RHS above gives rise to contact terms when inserted in a correlation function, due to the sub-leading graviton soft theorem, but the first line represents a true obstruction to a conservation law for tree level amplitudes, as its insertion does not vanish even for a point $(z_2,\bz_2)$ away from other operator insertions.
{As we show in Appendix \ref{App:TbarpT}, the result \eqref{TbarpT} is derived without passing through \eqref{TTextra}, but by making use of the celestial diamond relation \(\bar \p T=\p^3N_1\) (see Fig.~\ref{fig:CD-GR-2} for \(s=1\)). If we want to preserve both the right and left corners of the celestial diamond (namely, the sub-leading soft graviton $N_1$ and the stress-energy tensor $T$), the same expression should  follow also from simply acting with \(\bar\p_2\) on \eqref{TTextra}. This is indeed the case only if the obstruction term in the latter is non-vanishing, as otherwise  we would recover only the second line of \eqref{TbarpT} (up to contact terms).
}

In order to solve these issues,  the authors of \cite{Banerjee:2022wht} proposed a modified definition for the CCFT energy-momentum tensor, whose OPE with itself contains no obstruction terms. In such a construction, the modified energy-momentum tensor comes with a manifest Virasoro symmetry but, if $N_1$ is kept in the theory, the $\sl(2,\C)$ symmetry it generates is lost. Their reasoning is based on the introduction of an object $ {[\tilde q_1]}_S$ with the same conformal dimension and helicity as $ \tilde q^1_1 $, such that
\be
\bar\partial {[\tilde q_1]}_S= N_1 \quad \text{and} \quad \partial^3 {[\tilde q_1]}_S(z_1,\bz_1)N_1(z_2,\bz_2)\sim \text{Contact Terms}. \la{condP}
\ee
The definition they propose requires taking $z,\bz$ as independent variables and performing a contour integral from reference point\footnote{In \cite{Banerjee:2022wht} (see the discussion below eq. (3.9) there), the authors emphasize that the reference point $\bar z_0$ appearing in \eqref{qS} explicitly breaks the global conformal invariance, and they conveniently choose it to be $\bar z_0\rightarrow\infty$.} $\bz_0$ to $\bz$, explicitly 
\be
\la{qS}
 {[\tilde q_1]}_S(z,\bar z):=\int_{\bar z_0}^{\bar z}d\bar w N_1(z,\bar w).
\ee
Being able to perform a contour integral only on one of the celestial sphere coordinates is crucial in order to guarantee that the conservation law of the modified energy-momentum tensor $T_{mod}=T- \p^3  {[\tilde q_1]}_S$ is compatible with the vanishing of any non-contact term in the OPE $T_{mod}\bar\p T_{mod}$; in fact, the subtraction piece corresponds to a modification of the shadow integral kernel in the $(2,2)$ signature. 
The price to pay for rescuing the Virasoro symmetry associated to $T_{mod}$ this way though is that the $\sl(2,\C)$ symmetry associated to $N_1$ is no longer preserved, namely the left corner of the celestial diamond in Fig. \ref{fig:CD-GR-2} (for $s=1$) is removed as a local operator.  
This may represent a too drastic of a solution, as $N_1$ is associated to the soft component of the angular momentum charge, whose insertion in a $\cS$-matrix element generates the subleading soft graviton theorem \cite{Kapec:2014opa,Campiglia:2014yka,Freidel:2021dfs}; moreover, the operator $N_1$ defines the
subleading displacement memory near null infinity
\cite{Pasterski:2015tva,Nichols:2018qac}, an effect potentially measurable by future  detectors \cite{Grant:2022bla}.

Therefore, whether this analytic continuation and modification of 
the shadow transform kernel is the only viable prescription surely deserves  further investigation, but  that is beyond the scope of this work. 
We simply limit ourself to pointing out how the expression of the undesired non-contact contribution in the second line of \eqref{TT} in terms of $ \tilde q^1_1 $ and its spatial derivative might prove useful in finding a more general definition of modified energy-momentum tensor compatible with the expected OPEs. The possible usefulness of this rewriting of the obstruction term in \eqref{TT} is borne out by the following observation. If we use the standard definition of the Virasoro generators as the  coefficients of the mode expansion of the energy-momentum tensor and their commutator as a double contour integral of its OPE \cite{DiFrancesco:1997nk}, then one finds
\be
[L_n,L_m]=&\f{1}{(2\pi i)^2}\f{\k}{8\pi i}\oint_{0}dz_2z_2^{m+1}\oint_{z_2}dz_1z_1^{n+1}\bigg(\f{2T(z_2)}{z_{12}^2}+\f{\p T(z_2)}{z_{12}}
\cr&-\f{24}{z_{12}^5}(\tilde q_1^1(z_2)-\tilde q_1^1(z_1))-\f{12}{z_{12}^4}(\p_2\tilde q_1^1(z_2)+\p_1\tilde q_1^1(z_1)) \bigg)
\cr=&\f{1}{(2\pi i)^2}\f{\k}{8\pi i}\oint_{0}dz_2z_2^{m+1}\bigg(2(n+1)z_2^nT(z_2)+z_2^{n+1}\p T(z_2)
\cr&-2\p(z_2^{n+1})\p^3 \tilde q_1^1(z_2)-z_2^{n+1}\p^4 \tilde q_1^1(z_2) \bigg)
=0\,,
\ee
where we used $\p^3 \tilde q_1^1=T$ and the strong assumption of $\tilde q_1^1$ to be holomorphic. This shows that the second line of \eqref{TT} encodes the structure of a Virasoro algebra as well.

As a final check of our procedure, we compute the mixed-helicity $\bT T$ OPE. Following similar steps as above, {for $\Re\Delta_2\le1$}, we obtain first (see Appendix \ref{App:TbarSN})
\be
\bT(z_1,\bz_1) S[N^-_{\Delta_2}](z_2,\bz_2)
&\sim \f{i\k}{16\pi}\bigg[\Delta_2 \bar z_{12}^{-2} S[N^-_{\Delta_2}](z_2,\bz_2)-2 \bar z_{12}^{-1}\bar\partial S[N^-_{\Delta_2}](z_2,\bz_2)
\cr&-(\Delta_2+2)(\Delta_2+1)\Delta_2z_{12}^{\Delta_2-3}\bar z_{12}^{\Delta_2-1}\f{1}{2\pi}\int_{S_4}z_{24}^{-1}\bar z_{24}z_{41}^{3-\Delta_2}\bar z_{41}^{-\Delta_2-3} S[N^-_{\Delta_2}](z_4,\bz_4)
\cr
&+(\Delta_2+2)(\Delta_2+1)\bar\partial_{z_1}\left(z_{12}^{\Delta_2-3}\right)\bar z_{12}^{\Delta_2}\f{1}{2\pi}\int_{S_4}z_{24}^{-1}z_{41}^{3-\Delta_2}\bar z_{41}^{-\Delta_2-2} S[N^-_{\Delta_2}](z_4,\bz_4)
\bigg]\,.
\la{btn1tn1}
\ee
If we  take  now the second conformally soft limit $\Delta_2\to 0$, we arrive at (see Appendix \ref{App:TTbar})
\be
\bT(z_1,\bz_1) T(z_2,\bz_2)&\sim
-\f{i\k}{8\pi}\bigg[ \f1{\bar z_{12}}\bar\partial T(z_2,\bz_2)
\cr
&-2\pi\left(\partial_{2}^2\delta^2(z_1,z_2)\bar\partial \tilde q^1_1 (z_2,\bz_2)
-3\partial_{2}\delta^2(z_1,z_2)\bar\partial\partial \tilde q^1_1 (z_2,\bz_2)
-3\delta^2(z_1,z_2)\bar\partial\partial^2 \tilde q^1_1 (z_2,\bz_2)
\right)\bigg]\,.
\cr
\la{TbarT}
\ee
Again, this derivation reproduces  the result of \cite{Fotopoulos:2019vac}, with the RHS of  \eqref{TbarT}  giving only regular terms modulo delta-function contact
terms, including the first line on the RHS of \eqref{TbarT}---recall in fact that $\bar\partial T=q_1^1$, thus its insertion in  correlation functions also yields only delta-function contributions due to the sub-leading soft theorem---. We thus see that the mixed-helicity sector of the energy-momentum OPE does not pose any problem.

After this review of how OPEs involving the energy-momentum tensor, as obtained in previous literature, can be recovered using the algorithm laid out at the beginning of this section, we can now reliably apply it to compute OPEs involving a shadow soft operator  for the whole tower $s_1\geq 0$. 

Before moving to this more general case, let us point out another useful clarification of an ambiguity often times lurking behind the application of the shadow transform. This concerns the commutativity or not of performing the shadow transform before or after the conformally-soft limit. As shown in Appendix \ref{App:Commuting}, 
the same expressions for the OPEs \eqref{TT}, \eqref{TbarT} can be derived by performing the second conformally-soft limit first and then taking the shadow transform of the second entry. This provides good evidence that indeed the two operations commute with each other, and more will be provided shortly.

\subsection{Same-helicity sector}\la{sec:same-helicity-GR}

We want extend our procedure to compute shadow OPEs for a general $s_1\geq 0$, focusing first on the same-helicity sector. By `same-helicity' hereafter we refer to the starting point OPE between a soft charge and a conformal primary graviton, from which we can next derive OPEs involving a shadow transform on either one (in which case the OPE will acquire a mixed-helicity) or both entries (in which case we will still have the same helicity, although of opposite sign).  We start again from the charge aspect bracket \eqref{qND}, which yields the OPE
\be
q^1_{s_1}(z_1,\bz_1)N^-_{\Delta_2}(z_2,\bz_2)\sim&-\f{i\k}{8s_1!}\sum_{n=0}^{s_1}(-)^{n-s_1}(\Delta_2-2)_{s_1-n}(s_1)_n(n+1)\partial_{1}^{s_1-n}\delta^2(z_1,z_2)\partial_{2}^nN^-_{\Delta_2+1-s_1}(z_2,\bz_2)\,.
\la{qg}
\ee

Given the different structure of the two classes of celestial diamonds in Fig. \ref{fig:CD-GR}, let us treat the cases $s_1=0$ and $s_1\geq 2$ separately---the case $s_1=1$ has been studied already in Section \ref{sec:SET}---. For $s_1=0$, \eqref{shaqN} together with \eqref{qg} yield
\be
S[ N_0](z_1,\bz_1)N^-_{\Delta_2}(z_2,\bz_2)
&=-\f{i\k}{8}\f{1}{2\pi}\f{\bar z_1-\bar z_2}{z_1-z_2}N^-_{\Delta_2+1}(z_2,\bz_2)\,.
\la{tng}
\ee
If we take the conformally soft limit of the second entry, we get
\be
S[ N_0](z_1,\bz_1)N_{s_2}(z_2,\bz_2)
&\sim-\f{i\k}{8}\f{1}{2\pi}\f{\bar z_1-\bar z_2}{z_1-z_2}N_{s_2-1}(z_2,\bz_2)\,,
\ee
and, for ${s_2}\le1$,  taking the shadow transform of the second entry, 
\be
S[ N_0](z_1,\bz_1)S[N_{{s_2}}](z_2,\bz_2)
&\sim-\f{i\k}{16\pi}\f{1}{2\pi}(-)^{{s_2}}\f{(2+{s_2})!}{(1-{s_2})!}\int_{S_3}\f{(\bar z_2-\bar z_3)^{1-{s_2}}}{(z_2-z_3)^{3+{s_2}}}\f{\bar z_1-\bar z_3}{z_1-z_3}N_{{s_2}-1}(z_3,\bz_3)
\,.
\ee
This is non-trivial only for $s_2=1$, yielding
 (see Appendix \ref{App:SN01})
\be
S[ N_0](z_1,\bz_1)S[N_{1}](z_2,\bz_2)
&\sim 
\f{i\k}{8}\f{1}{2\pi}\f{3!}{2\pi}\int_{S}\f{1}{(z_2-z)^{4}}\f{\bar z_1-\bar z}{z_1-z}N_{0}(z,\bz)
\cr
&=-\f{i\k}{16\pi}\bigg[\f{1}{z_{12}}
\left(
\bar z_{12}\bar \partial\partial
+\partial 
+3\f{\bar z_{12}}{z_{12}}\bar \partial 
+3\f{1}{z_{12}}
\right)S[ N_{0}](z_2,\bz_2)
\cr
&+\f{6}{z_{12}^3}
\left(
\bar z_{12}\bar \partial\partial \tilde q^1_0 (z_2,\bz_2)
+\partial \tilde q^1_0 (z_2,\bz_2)
+\f{\bar z_{12}}{z_{12}}\bar \partial \tilde q^1_0 (z_2,\bz_2)
+\f{1}{z_{12}}( \tilde q^1_0 (z_2,\bz_2)- \tilde q^1_0 (z_1,\bz_1))
\right)
\bigg].
\cr
\la{SN01}
\ee

If instead we do not want to take the double-soft limit but compute the OPE with a shadow conformal primary graviton, we can take the shadow transform of the second entry in \eqref{tng}, which yields for a general $\Delta_2$ (see Appendix \ref{SN0SG})
\be
S[ N_0](z_1,\bz_1)S[N^-_{\Delta_2}](z_2,\bz_2)
&\sim-\f{i\k}{8}
\bigg[-(3-\Delta_2)_2\f{1}{2\pi}z_{12}^{\Delta_2-4}\bar z_{12}^{\Delta_2+2}\int_{S_3}z_{31}^{1-\Delta_2}
\bar z_{23}^{-2}\bar z_{31}^{-\Delta_2-1} S [N^-_{\Delta_2+1}](z_3,\bz_3)
\cr
&+z_{12}^{-2}\bar z_{12}^{\Delta_2+2}\left(3-\Delta_2+z_{12}\partial_{2}\right)\bar \partial_{2}\left(\bar z_{12}^{-\Delta_2-1}
S [N^-_{\Delta_2+1}](z_2,\bz_2)\right)\bigg].
\la{t0g}
\ee
Taking the conformally soft limit for $\Delta_2=0$, one recovers \eqref{SN01}, showing again the commutativity between conformally soft limit and shadow transform.

For $s_1\ge2$, the OPE \eqref{qg} together with \eqref{dualqN} yield
\be
\tilde N_{s_1}(z_1,\bz_1)N^-_{\Delta_2}(z_2,\bz_2)&\sim-\f{i\k}{8s_1!}\sum_{n=0}^{s_1}(-)^{n-s_1}(\Delta_2-2)_{s_1-n}(s_1)_n(n+1)\bar \partial_{1}^{s_1-2}\partial_{1}^{s_1-n}\delta^2(z_1,z_2)\partial_{2}^n N^-_{\Delta_2+1-s}(z_2,\bz_2).
\ee
Taking the shadow transform of the conformal primary graviton in the second entry, we immediately obtain
\be
\tilde N_{s_1}(z_1,\bz_1) S [N^-_{\Delta_2}](z_2,\bz_2)
\sim&
-\f{i\k}{8s_1!}\f{K_{\Delta_2}}{2\pi}\sum_{n=0}^{s_1}(-)^{n-s_1}
(\Delta_2-2)_{s_1-n}(s_1)_n(n+1)
\cr
&\times\bar \partial_{1}^{s_1-2}\partial_{1}^{s_1-n}\left(\f{\bar z_{12}^{\Delta_2}}{z_{12}^{4-\Delta_2}}\partial_{1}^n
N^-_{\Delta_2+1-s_1}(z_1,\bz_1)\right)\,.
\la{tNSG}
\ee
Eq. \eqref{tNSG} gives the OPE of a dual soft graviton with a shadow conformal primary. However,
let us stress again that, for $s_1\ge2$, the dual soft graviton and the shadow soft graviton operators are not the same object, $\tilde N_{s_1}\neq S[N_{s_1}]$. To derive the OPE of a   shadow soft graviton, rather than a dual soft graviton, one can start with the OPE involving a (sub)$^{s_1}$-leading soft graviton with a shadow conformal primary and then  eventually take the shadow on the first entry. Let us perform the first step and start again from \eqref{qg} to write\footnote{It is convenient to replace
\be
\partial_{1}^{s_1-n}\delta^2(z_1,z_2)
=\f1{2\pi(n+1)!}\partial_{1}^{s_1+2}\left(\f{(z_1-z_2)^{n+1}}{\bar z_1-\bar z_2}\right)
\ee
in \eqref{qg}.
}
\be
 N_{s_1}(z_1,\bz_1) N^-_{\Delta_2}(z_2,\bz_2)
 &\sim
 -\f{i\k}{16\pi s_1!}\sum_{n=0}^{s_1}\f{1}{n!}(-)^{n-s_1}(\Delta_2-2)_{s_1-n}(s_1)_n\f{(z_1-z_2)^{n+1}}{\bar z_1-\bar z_2}\partial_{2}^n N^-_{\Delta_2+1-s_1}(z_2,\bz_2).
 \la{NsND}
\ee
Taking the shadow of the second argument  for a general $\Delta_2$
we get (see Appendix \ref{App:NSG})
\be
  N_{s_1}(z_1,\bz_1) S[N^-_{\Delta_2}](z_2,\bz_2)
 \sim&
 -\f{i\k}{16\pi {s_1}!}\f{K_{\Delta_2}}{2\pi}\f{\tilde K_{1-\Delta_2+{s_1}}}{2\pi}\sum_{n=0}^{s_1}\f{1}{n!}(-)^{n-{s_1}}(\Delta_2-2)_{{s_1}-n}({s_1})_n
\bigg[(1+{s_1}-\Delta_2)_{n}
\cr
&\times
\int_{S_3}
\int_{S_4}
\f{1}{z_{13}^{-n-1}}\f{1}{z_{23}^{4-\Delta_2}}\f{1}{z_{34}^{\Delta_2-{s_1}-1+n}}\f{1}{\bar z_{13}}\f{1}{\bar z_{23}^{-\Delta_2}}\f{1}{\bar z_{34}^{\Delta_2+3-{s_1}}}
S[N^-_{\Delta_2+1-{s_1}}](z_4,\bz_4)
\cr
&+\sum_{k=1}^n(1+{s_1}-\Delta_2)_{n-k}\binom{n}{k}
\cr&\times\int_{S_3}\int_{S_4}\f{1}{z_{13}^{-n-1}}\f{1}{z_{23}^{4-\Delta_2}}\f{1}{z_{34}^{\Delta_2-{s_1}-1+n-k}}\f{1}{\bar z_{13}}\f{1}{\bar z_{23}^{-\Delta_2}}\partial_{z_3}^k\left(\f{1}{\bar z_{34}^{\Delta_2+3-{s_1}}}\right)S[N^-_{\Delta_2+1-{s_1}}](z_4,\bz_4)\bigg].
\cr
\la{NsSN}
\ee
The extra double integral in the last line on the RHS of the OPE above can contribute only in the case $\Re\Delta_2\geq s_1-2$. However, as shown in Appendix \ref{App:NsSNs}, when taking the conformally soft limit of the second entry as well, this contribution is never there for all $s_1,s_2\geq 0$. The resulting OPE reads
\be
  N_{s_1}(z_1,\bz_1) S[N_{s_2}](z_2,\bz_2)
 \sim&
 -\f{i\k}{16\pi }\f{(s_1+s_2)_{s_1}}{s_1!}\f{K_{1-{s_2}}}{2\pi}\f{\tilde K_{{s_2}+{s_1}}}{2\pi}\sum_{n=0}^{s_1}\f{({s_1})_n}{n!}
\cr
&\times
\int_{S_3}
\int_{S_4}
\f{1}{z_{13}^{-n-1}}\f{1}{z_{23}^{3+{s_2}}}\f{1}{z_{34}^{-{s_2}-{s_1}+n}}\f{1}{\bar z_{13}}\f{1}{\bar z_{23}^{s_2-1}}\f{1}{\bar z_{34}^{4-{s_2}-{s_1}}}
S[N_{{s_1}+{s_2}-1}](z_4,\bz_4)
.
\la{NsSNs}
 \ee
For $s_1\geq 1$ the conformal integral identity \eqref{intid} cannot be applied to the integral in square brackets above, as the condition \eqref{conh} cannot be satisfied. Therefore, we do not presently know how to simplify further the expression of the OPE  \eqref{NsSNs}. However, a local OPE expression can be obtained if, instead of taking the second soft limit in \eqref{NsSN}, we do a shadow transform of the first entry; in particular, we want to ask the question whether the OPE with a shadow conformal primary graviton of a dual soft graviton 
\eqref{tNSG} and of a shadow soft graviton have exactly the same form. The answer is in the affirmative, namely, 
as shown in Appendix \ref{App:SNsSND}, we have for $s_1\geq 2, \Delta_2\in \C$
\be
S[N_{s_1}](z_1,\bz_1)S[N^-_{\Delta_2}](z_2,\bz_2)
\sim&-\f{i\k}{8 {s_1}!}\f{K_{\Delta_2}}{2\pi}\sum_{n=0}^{s_1}(-)^{s_1-n}(\Delta_2-2)_{s_1-n}({s_1})_{n}(n+1)
\cr
&\times
\bar\partial_{1}^{s_1-2}\partial_{1}^{s_1-n}\left(\f{\bar z_{12}^{\Delta_2}}{z_{12}^{4-\Delta_2}}\partial_{1}^{n}N^-_{\Delta_2+1-{s_1}}(z_1,\bz_1)\right)\,.
\la{SNsSND}
\ee
This result is expected in the limiting case of the zero-area diamond $s_1=2$, where the definitions of the shadow soft graviton and of the dual soft graviton should coincide. The fact that the same correspondence of OPEs persists for higher spins is more surprising, and may hint at the possible decomposition of shadowed operators $S[N_s]$ with $s>2$ into a local part given by $\tilde N_s$, and a non-local part, which does not contribute in the OPE structure. We leave a more detailed investigation of this point to future work.

\subsection{Mixed-helicity sector}\la{sec:mixed-helicity-GR}

To study shadow OPEs in the mixed-helicity sector, the starting point is again the key bracket \eqref{qND}. In fact, from the complex conjugate of that bracket for a positive helicity graviton, we can obtain at the operatorial level the OPE
\be
\bar q^1_{s_1}(z_1,\bz_1)N^-_{\Delta_2}(z_2,\bz_2)\sim\f{\k}{8is_1!}\sum_{n=0}^{s_1}(-)^{n-s_1}(\Delta_2+2)_{s_1-n}(s_1)_n(n+1)\bar \partial_{1}^{s_1-n}\delta^2(z_1,z_2)\bar \partial_{2}^n N^-_{\Delta_2+1-s_1}(z_2,\bz_2)\,.
\la{bqN}
\ee
This implies
\be
\bar N_{s_1}(z_1,\bz_1)N^-_{\Delta_2}(z_2,\bz_2)\sim\f{\k}{8is_1!}\sum_{n=0}^{s_1}(-)^{n-s_1}(\Delta_2+2)_{s_1-n}(s_1)_n\f{1}{n!}\f{\bar z_{12}^{n+1}}{z_{12}}\bar \partial_{2}^n N^-_{\Delta_2+1-s_1}(z_2,\bz_2)\,.
\la{bng}
\ee
Before continuing, let us warn the reader that, for $s_1>3$, the OPE \eqref{bng} should receive a contribution including a pole in $\bz_{12}$, as already pointed out at the end of Section \ref{sec:gravity}. We will ignore these extra terms in the rest of this section, as we will be mostly interested in the case $s_1=0$. However, once the full quadratic charge aspect expression  is derived for a general spin, the complete structure of the mixed-helicity OPE \eqref{bng} can be obtained through its bracket action, and some of the OPE formulas below straightforwardly generalized. 

Taking the conformally soft limit on the second entry of \eqref{bng}, we obtain
\be
\bar N_{s_1}(z_1,\bz_1)N_{s_2}(z_2,\bz_2)
&\sim\f{\k}{8i}\sum_{n=0}^{s_1}\f{(s_1+s_2-n-4)_{s_1-n}}{(s_1-n)!}\f{1}{n!}\f{\bar z_{12}^{n+1}}{z_{12}}\bar \partial_{z_2}^nN_{s_1+s_2-1}(z_2,\bz_2)\,.
\la{sbng}
\ee
The OPE \eqref{sbng} can be used to derive 
(see Appendix \ref{App:bNtN})
\be
\bar N_{s_1}(z_1,\bz_1)\tilde N_{s_2}(z_2,\bz_2)
&\sim
\f{\k}{8i}\bigg[2\pi(s_1+s_2-2)_2\binom{s_1+s_2-4}{s_1}\sum_{n=0}^{s_2-4}
\f{1}{(s_1+n+2)_2}
\binom{s_2-4}{n}
\cr
&\times\partial_{2}^{s_2+2}\left(\bar\partial^{n}_{2}\delta^2(z_1,z_2)\bar \partial_{2}^{s_2-4-n}N_{s_1+s_2-1}(z_2,\bz_2)\right)
\cr
&
+\f{1}{s_1!}\partial_{z_2}^{s_2+2}\left(\f{\bar z_{12}^{s_1}}{z_{12}}(2-s_2-s_1+\bar z_{12}\bar\partial_{2})S[q^1_{s_1+s_2-1}](z_2,\bz_2)\right)
\bigg]\,,
\la{bNtN}
\ee
which is valid for all dual soft gravitons ($s_2\geq 2$). In order to compare the OPE structure of dual soft gravitons with the one where we insert a shadow soft graviton in the second entry instead, let us
restrict to the case $s_1=0$ in \eqref{bNtN}; after some manipulations (see Appendix \ref{App:bNtN}), this yields 
\be
\bar N_{0}(z_1,\bz_1)\tilde N_{s_2}(z_2,\bz_2)
\sim&\f{\k}{8i}\bigg[c.t.
-\f{1}{2\pi}(s_2+2)_2z_{12}^{-s_2-3}\int_{S_3}\left(\bar z_{12}+(s_2-2)\bar z_{23}\right)\f{z_{13}^{s_2}}{\bar z_{23}^2}\tilde N_{s_2-1}(z_3)
\cr&
+z_{12}^{-2}(s_2+2+z_{12}\partial_{z_2})(2-s_2+\bar z_{12}\bar\partial_{z_2})\tilde N_{s_2-1}(z_2,\bz_2)
\bigg]\,,
\la{nsdgen}
\ee
where $c.t.$ stands for the contact terms given by the sum contribution on the RHS of \eqref{bNtN}.

We thus study now the OPE involving a shadow soft graviton. Restricting again to the case $\Re\Delta_2\leq 1$, the OPE \eqref{bng} yields
(see Appendix \ref{App:bNSN} for the full expression for a general $\Delta_2\in \C$)
\be
\bar N_{s_1}(z_1,\bz_1)S[N^-_{\Delta_2}](z_2,\bz_2)
\sim&\f{\k}{8i{s_1}!}\f{K_{\Delta_2}}{2\pi}\f{\tilde K_{2-(\Delta_2+1-{s_1})}}{2\pi}\sum_{n=0}^{s_1}\f{1}{n!}(-)^{n-{s_1}}(\Delta_2+2)_{{s_1}-n}({s_1})_n({s_1}-\Delta_2-3)_n
\cr
&\times\int_{S_3}\int_{S_4}\f{1}{z_{13}}\f{1}{z_{23}^{4-\Delta_2}}\f{1}{z_{34}^{-{s_1}+\Delta_2-1}}\f{1}{\bar z_{13}^{-n-1}}\f{1}{\bar z_{23}^{-\Delta_2}}
\f{1}{\bar z_{34}^{3+\Delta_2-{s_1}+n}}S[N^-_{\Delta_2+1-{s_1}}](z_4,\bz_4).
\la{bNSN}
\ee
We immediately notice that, for $s_1=0$, the OPE \eqref{bng} is identical to \eqref{tng}, so when we take the shadow of the second argument, we get
\be
\bar N_0(z_1,\bz_1)S [N^-_{\Delta_2}](z_2,\bz_2)
&\sim
\f{\k}{8i}
\bigg[
-(3-\Delta_2)_2\f{1}{2\pi}z_{12}^{\Delta_2-4}\bar z_{12}^{2+\Delta_2}\int_{S_4}z_{41}^{1-\Delta_2}\bar z_{24}^{-2}\bar z_{41}^{-1-\Delta_2}S[N^-_{\Delta_2+1}](z_4,\bz_4)
\cr
&+z_{12}^{-2}\bar z_{12}^{2+\Delta_2}\left(3-\Delta_2+z_{12}\partial_{2}\right)\bar\partial_{2}\left(\bar z_{12}^{-1-\Delta_2}S[N^-_{\Delta_2+1}](z_2,\bz_2)\right)
\bigg]\,,
\la{bN0N}
\ee
which is identical to \eqref{t0g} and holds for a general $\Delta_2$. It is an interesting observation that the leading shadow negative-helicity soft graviton exhibits the same OPE than the leading positive-helicity soft graviton. This is possible as both operators $\bar N_{0}, S[N_0]$ have the same $(\Delta,J)=(1,0)$; clearly, this is no longer the case for $s_1\geq1$.

If we take the conformally soft limit on the second entry of \eqref{bNSN}. We get
\be
\bar N_{s_1}(z_1,\bz_1)S[N_{{s_2}}](z_2,\bz_2)\sim&\f{\k}{8i{s_1}!}\f{K_{1-{s_2}}}{2\pi}\f{\tilde K_{2-(2-{s_2}-{s_1})}}{2\pi}\sum_{n=0}^{s_1}\f{1}{n!}(-)^{n-{s_1}}(3-{s_2})_{{s_1}-n}(s_1)_n({s_1}+{s_2}-4)_n
\cr&\times\int_{S_3}\int_{S_4}\f{1}{z_{13}}\f{1}{z_{23}^{3+{s_2}}}\f{1}{z_{34}^{-{s_1}-{s_2}}}\f{1}{\bar z_{13}^{-n-1}}\f{1}{\bar z_{23}^{-\Delta_2}}
\f{1}{\bar z_{34}^{4-{s_2}-{s_1}+n}}S[N_{{s_1}+{s_2}-1}](z_4,\bz_4)\,,
\la{bNsSNs}
\ee
which holds for any $s_1,s_2\geq0$.
However, we can use the integral identity \eqref{intid} to compute the integral in $(z_3,\bar z_3)$ only for ${s_1}=0$ (otherwise we have $\bar h_1+\bar h_2+\bar h_4<2$)---this is why we explicitly worked out the case $s_1=0$ in \eqref{nsdgen}---. 
Therefore, restricting to this case, \eqref{bNsSNs} yields\footnote{After performing the integral over $(z_3,\bar z_3)$, we relabeled $z_4\to z_3$. }
\be
\bar N_0(z_1,\bz_1)S[N_{s_2}](z_2,\bz_2)
&\sim\f{\k}{8i}
\bigg[
-\f{1}{2\pi}(s_2+2)_2z_{12}^{-s_2-3}\bar z_{12}^{3-s_2}\int_{S_3}
\f{z_{13}^{s_2}}{\bz_{23}^2}\bz_{13}^{s_2-2}S[N_{s_2-1}](z_3,\bz_3)
\cr
&+z_{12}^{-2}(s_2+2+z_{12}\partial_{2})(2-s_2+\bar z_{12}\bar\partial_{2})S[N_{s_2-1}](z_2,\bz_2)
\bigg]\,.
\la{bN0SN}
\ee

Clearly, the second lines on the RHS of \eqref{nsdgen} and \eqref{bN0SN}    have exactly the same form upon the replacement $\tN_s\leftrightarrow S[N_s]$. In Appendix \ref{App:bNSN} we prove that in fact the first lines as well have the same form, namely one can show that
\be
\bar z_{12}^{3-s_2}\int_{S_3}
\f{z_{13}^{s_2}}{\bz_{23}^2}\bz_{13}^{s_2-2}S[N_{s_2-1}](z_3,\bz_3)=
\int_{S_3}(\bar z_{12}+(s_2-2)\bar z_{23}) \f{z_{31}^{s_2}}{\bar z_{23}^{2}}S[N_{s_2-1}](z_3,\bz_3)\,.
\ee


\section{Shadow OPE in Yang--Mills}\la{sec:YM}

We can apply our algorithm to derive OPE between shadow/soft operators also in the case of YM theory, using the celestial diamonds in Fig. \ref{fig:CD-YM}. In this case we have the relations 
\be
S[r_s](z,\bz)&=
   \bar \p^{s-1}\int_{S_1}G_{s+1}(z_1;z)r_s(z_1,\bz_1), \quad &\text{for} \quad s\ge1\\
  S[F_s](z,\bz)&= \p^{s+1} \int_{S_1}\bar G_{1-s}(z_1;z)F_s(z_1,\bz_1)\quad &\text{for} \quad s=0\,,
  \la{sharF}
\ee
as well as
\be
\tr_s(z,\bz)&=
    \int_{ S_1}\int_{S_2}\bar G_{1-s}(z_1;z)G_{s+1}(z_2;z_1)r_s(z_2,\bz_2), \quad &\text{for} \quad s=0\\
\tF_s(z,\bz)&= \p^{s+1} \bar \p^{s-1}F_s(z,\bz)\quad &\text{for} \quad s  \geq 1\,.
\la{dualrF}
\ee
In addition, we define the shadow conformal primary gluons as
\be
S[F^{-}_{\Delta}](z_2,\bz_2)=&\f{K_{\Delta}}{2\pi}\int_{S_3}\f{1}{z_{23}^{3-\Delta}}\f{1}{\bar z_{23}^{1-\Delta}}F^{-}_{\Delta}(z_3,\bz_3)\,,
\la{SFD}
\\
F^{-}_{\Delta}(z_2,\bz_2)=&\f{\tilde K_{2-\Delta}}{2\pi}\int_{S_3}\f{1}{z_{23}^{\Delta-1}}\f{1}{\bar z_{23}^{\Delta+1}}S[F^{-}_{\Delta}](z_3,\bz_3)\,,
\la{FD}
\ee
with the normalization constants given by
\be
K_\Delta=&(-)^{2-\Delta}\f{\Gamma(3-\Delta)}{\Gamma(\Delta)},
\la{KYM}
\\
\tilde K_{2-\Delta}=&(-)^{\Delta}\f{\Gamma(1+\Delta)}{\Gamma(2-\Delta)}\,.
\la{tKYM}
\ee
{The corresponding expressions for the positive-helicity primaries $F^+_{\Delta}$ are identical to the above ones upon the exchange $z_{23}\leftrightarrow \bar{z}_{23}$.}
Analogously to the case of gravity, in the conformally soft limit $\Delta\to 1-s$ \eqref{SFD} reduces to \eqref{sharF} only for $s=0$, while for $s\geq 1$ the two sets of operators $\tF_s, S[F_s]$ share the same weights $(\Delta,J)=(1+s,1)$ but otherwise they are  apriori different; nevertheless, we will provide evidence that they share the same OPE structure in this case as well.

\subsection{Same-helicity sector}\la{sec:same-helicity-YM}

We start again from the same-helicity sector, namely the sector that can be studied starting from the OPE
\be
F_{s_1}^{a}(z_1,\bz_1)F_{\Delta_2}^{-b}(z_2,\bz_2)
&\sim -i\f{g_{YM}^2}{4\pi s_1!}\f{f^{ab}{}_c}{\bar  z_{12}}\sum_{n=0}^{s_1}(-)^{s_1-n}  \f{(s_1)_n}{n!}(\Delta_2-2)_{s_1-n} z_{12}^n \partial^n F_{\Delta_2-s_1}^{-c}(z_2,\bz_2)\,,
\la{FsFD}
\ee
which follows from the bracket \eqref{br2FD} via the correspondence \eqref{keyYM}.
Assuming a general $\Delta_2\in \C$, if we take the shadow of the second entry, we get (see Appendix \ref{App:FsSFs})
\be
F_{{s_1}}^{a}(z_1,\bz_1)S[F_{\Delta_2}^{-b}](z_2,\bz_2)
\sim& -i\f{g_{YM}^2}{4\pi s_1!}f^{ab}{}_c\f{K_{\Delta_2}}{2\pi}\f{\tilde K_{2-\Delta_2+s_1}}{2\pi}\sum_{n=0}^{s_1}(-)^{{s_1}-n}  \binom{{s_1}}{n}(\Delta_2-2)_{{s_1}-n}\bigg[
\cr
&\times(1-\Delta_2+s_1)_n\int_{S_3}\int_{S_4}\f{1}{z_{13}^{-n}}\f{1}{z_{23}^{3-\Delta_2}}\f{1}{z_{34}^{-1+\Delta_2-s_1+n}}\f{1}{\bar  z_{13}} \f{1}{\bar z_{23}^{1-\Delta_2}} \f{1}{\bar z_{34}^{1+\Delta_2-s_1}}S[F^{-c}_{\Delta_2-s_1}](z_4,\bz_4)
\cr
&+\sum_{k=1}^{n}\binom{n}{k}(1-\Delta_2+s_1)_{n-k}\int_{S_3}\int_{S_4}\f{1}{z_{13}^{-n}}\f{1}{z_{23}^{3-\Delta_2}}\f{1}{z_{34}^{-1+\Delta_2-s_1+n-k}}\f{1}{\bar z_{23}^{1-\Delta_2}}
\cr
&\times\partial_{3}^k\left(\f{1}{\bar z_{34}^{1+\Delta_2-s_1}}\right)S[F^{-c}_{\Delta_2-s_1}](z_4,\bz_4)\bigg]\,.
\ee
Performing the conformally soft limit on the second entry, we further arrive at (see Appendix \ref{App:FsSFs}) 
\be
F_{{s_1}}^{a}(z_1\bz_1) S[F^{b}_{s_2}](z_2,\bz_2)
\sim& -i\f{g_{YM}^2}{4\pi}\f{({s_1}+{s_2})_{s_1}}{s_1!}f^{ab}{}_c\f{K_{1-s_2}}{2\pi}\f{\tilde K_{1+s_2+s_1}}{2\pi}\sum_{n=0}^{s_1}\f{({s_1})_n}{n!}
\cr
&\times \int_{S_3}\int_{S_4}\f{1}{z_{13}^{-n}}\f{1}{z_{23}^{2+s_2}}\f{1}{z_{34}^{-s_2-s_1+n}}\f{1}{\bar  z_{13}}\f{1}{\bar z_{23}^{s_2}}  \f{1}{\bar z_{34}^{2-s_2-s_1}}S[F^{c}_{s_1+s_2}](z_4,\bz_4)\,.
\la{FsSFs}
\ee
This is the YM equivalent of the OPE \eqref{NsSNs}.

Starting from the OPE \eqref{FsFD}, we can alternatively derive the OPE between a dual soft gluon and a negative-helicity conformal primary gluon. In light of different nature of the two $s=0$ and $s\geq1$ sectors, as pointed out above,  we proceed by studying the two separately. 

For $s_1=0$, taking the shadow of the first entry in \eqref{FsFD}, we have
\be
S[ F_{0}^{a}](z_1,\bz_1)F_{\Delta_2}^{-b}(z_2,\bz_2)
&\sim 
-i\f{g_{YM}^2}{4\pi}\f{f^{ab}{}_c}{z_{12}} F_{\Delta_2}^{-c}(z_2,\bz_2)\,.
\la{SF0FD}
\ee
 If we further take the shadow of the second argument for a general $\Delta_2\in \C$, we get (see Appendix \ref{App:SF0SF})
\be
S [F_{0}^{a}](z_1,\bz_1)S[F_{\Delta_2}^{-b}](z_2,\bz_2)
&\sim -i\f{g^2_{YM}}{4\pi}f^{ab}{}_c\bigg[
\f1{z_{12}}S[F_{\Delta_2}^{-c}](z_2,\bz_2)
\cr
&-\f{(2-\Delta_2)}{2\pi}z_{12}^{\Delta_2-3}\bar z_{12}^{\Delta_2}\int_{S_3}z_{31}^{1-\Delta_2}\bar z_{23}^{-1}\bar z_{31}^{-\Delta_2}S[F_{\Delta_2}^{-c}](z_3,\bz_3)\bigg]
\,.
\la{SF0SFD}
\ee
The OPE above can be used, taking the 
conformally soft limit $\Delta_2\to 1$, to derive
\be
S[ F_{0}^{a}](z_1,\bz_1)S[F_{0}^{b}](z_2,\bz_2)
\sim&-i\f{g^2_{YM}}{4\pi}f^{ab}{}_c \left[
\f{1}{z_{12}}S[F_0^c](z_2,\bz_2)
+\f{1}{ z_{12}^2}
\left(\tilde r_0^{1c}(z_2,\bz_2)-\tilde r_0^{1c}(z_1,\bz_1)
\right)
\right]\,.
\la{SF0SF0}
\ee
Eq. \eqref{SF0SF0} is reminiscent of  the structure of the OPE \eqref{TT} of the energy-momentum tensor with itself in gravity, with the contribution in $z_{12}^{-2}$ on the RHS above, which we can write as
\be
\f{1}{ z_{12}^2}
\left(\tilde r_0^{1c}(z_2,\bz_2)-\tilde r_0^{1c}(z_1,\bz_1)
\right)=\f1{2\pi}\f{1}{ z_{12}} \int_{S_3} \f1{z_{23}z_{13}}F_0(z_3,\bz_3)\,,
\ee
and representing an obstruction to the realization of a holomorphic Kac--Moody symmetry.
 Moreover, to continue the analogy with gravity \eqref{TbarpT}, we see from the OPE
\be
S[ F_{0}^{a}](z_1,\bz_1)\bar\partial S[F_{0}^{b}](z_2,\bz_2)
\sim&-i\f{g^2_{YM}}{4\pi}f^{ab}{}_c \bigg[
\f{1}{ z_{12}^2}
F^{c}_0(z_2,\bz_2)
+\f{1}{z_{12}}r_0^{1c}(z_2,\bz_2)
\bigg]
\ee
that the first term in the square brackets above represents an obstruction to the holomorphicity condition $\bar \partial S[F_0]=0$.

{Of course, these concerns are unmotivated in the case of YM theory. In fact,  as we know very well from \cite{He:2015zea},  a holomorphic Kac-Moody current insertion is a positive helicity soft gluon insertion, not a shadow transform of a negative soft gluon   insertion; this follows immediately from the structure of the OPE \eqref{bFsFD}, which
 corresponds to the insertion of a holomorphic Kac--Moody current. 
Taking the double-soft limit $\Delta_1,\Delta_2\to1$ of \eqref{eq:cope} then yields 
the Ward identity
of a holomorphic Kac--Moody symmetry for the gauge group  $G$ \cite{He:2015zea}, with no obstruction to the  holomorphicity condition. The reason why  leading soft gluon operators are the natural candidates to play the role of (anti-)holomorphic  Kac--Moody currents is clear from the conformal dimension of the operator appearing on the RHS of the OPE of two primary gluons. In fact, we see for instance in \eqref{eq:cope} that the primary gluon on the RHS has conformal dimension $\Delta_1+\Delta_2-1$. This means that the expected OPE of a Kac--Moody current requires a first entry insertion such that $\Delta_1=1$, which happens to be exactly the conformal dimension of a Kac--Moody current. In this case then, the positive-helicity leading soft gluons $\bF_0$, defined as $\Res_{\Delta=1} F^{+}_\Delta$, is the natural object to consider as a holomorphic Kac--Moody current;\footnote{The operator $\bF_0$ has also the correct spin $J=1$ of  a holomorphic Kac--Moody current.}  there is    no need to take any shadow transform, despite the fact that  the operators $S[F_0]$ and $\bF_0$ have the same OPE structure when appearing as the first entry of the OPE {(we will clarify further this point in the next section)}. 

The case of gravity is different as, on the RHS of \eqref{OPEs}, the primary graviton operator has conformal dimension $\Delta_1+\Delta_2$, requiring a first entry insertion on the LHS of an operator with $\Delta_1=0$ in order to be compatible with an energy-momentum tensor action; the natural candidate would then be the positive helicity sub-leading soft graviton $\bar N_1=\Res_{\Delta=0} N^{+}_\Delta$. However, in this case $\Delta_1=0$ does not match the conformal dimension of an energy-momentum tensor, which is $\Delta=2$. Hence, in gravity, the shadow transform of a negative-helicity sub-leading soft graviton comes to the rescue and it allows us to construct an operator with the right $(\Delta,J)=(2,2)$ and able to reproduce the OPE of a  positive-helicity energy-momentum tensor. At the same time, this need to involve the shadow transform is also at the origin of the issue with the $TT$-OPE revisited in Section \ref{sec:SET}.
}

 For ${s_1}\ge1$,  we can use  \eqref{dualrF} to write first
\be
\tilde F_{{s_1}}^{a}(z_1,\bz_1)F_{\Delta_2}^{-b}(z_2,\bz_2)\sim& -i\f{g_{YM}^2}{2 s_1!}f^{ab}{}_c\sum_{n=0}^{s_1}(-)^{{s_1}-n}  ({s_1})_n(\Delta_2-2)_{{s_1}-n} \bar \partial^{{s_1}-1}_{1}\partial_{1}^{{s_1}-n}\delta^2(z_1,z_2) \partial^n F_{\Delta_2-{s_1}}^{-c}(z_2,\bz_2)\,.
\ee
Taking the shadow transform of the second entry then, it is straightforward to  derive the OPE 

\be
\tilde F_{{s_1}}^{a}(z_1,\bz_1)S[F_{\Delta_2}^{-b}](z_3,\bz_3)\sim& -i\f{g_{YM}^2}{4\pi s_1!} K_{\Delta_2}f^{ab}{}_c\sum_{n=0}^{s_1}(-)^{{s_1}-n}  ({s_1})_n(\Delta_2-2)_{{s_1}-n}
\cr
&\times\bar \partial^{{s_1}-1}_{1}\partial_{1}^{{s_1}-n}\left(\f{\bar z_{12}^{\Delta_2-1}}{z_{12}^{3-\Delta_2}} \partial^n F_{\Delta_2-s}^{-c}(z_1,\bz_1)\right)\,.
\la{tFsSFD}
\ee

As in the case of gravity (Section \ref{sec:same-helicity-GR}), we can now ask whether replacing the dual soft gluon in the first entry above with a shadow soft gluon yields the same OPE. Not surprisingly, as shown in Appendix \ref{App:SFsSFD}, the answer is again in the affirmative, namely
\be
S[F_{{s_1}}^{a}](z_1,\bz_1)S[F_{\Delta_2}^{-b}](z_2,\bz_2)\sim& -i\f{g_{YM}^2}{4\pi s_1!}K_{\Delta_2}f^{ab}{}_c\sum_{n=0}^{s_1}(-)^{{s_1}-n}  ({s_1})_n(\Delta_2-2)_{{s_1}-n}
\cr
&\times\bar \partial^{{s_1}-1}_{1}\partial_{1}^{{s_1}-n}\left(\f{\bar z_{12}^{\Delta_2-1}}{z_{12}^{3-\Delta_2}} \partial^n F_{\Delta_2-s}^{-c}(z_1,\bz_1)\right)\,.
\la{SFsSFD}
\ee

\subsection{Mixed-helicity sector}\la{sec:mixed-helicity-YM}

In this case, the starting point OPE is
\be
\bF_{s_1}^{a}(z_1,\bz_1)F_{\Delta_2}^{-b}(z_2,\bz_2)
&\sim -i\f{g_{YM}^2}{4\pi s_1!}\f{f^{ab}{}_c}{  z_{12}}\sum_{n=0}^{s_1}(-)^{s_1-n}  \f{(s_1)_n}{n!}(\Delta_2)_{s_1-n} \bz_{12}^n \bar \partial^n F_{\Delta_2-s_1}^{-c}(z_2,\bz_2)\,,
\la{bFsFD}
\ee
that follows immediately from the bracket action \eqref{br2FD}.
 For $s_1=0$, this gives  the OPE
\be
\bar F_{0}^{a}(z_1,\bz_1)F_{\Delta_2}^{-b}(z_2,\bz_2)\sim& -i\f{g^2_{YM}}{4\pi}f^{ab}{}_{c}\f{1}{ z_{12}}F_{\Delta_2}^{-c}(z_2,\bz_2)\,.
\la{bf0g}
\ee
 The RHS of \eqref{bf0g} is identical to \eqref{SF0FD}; consequently, if we take the shadow of the second argument, we get,
\be
\bar F_{0}^{a}(z_1,\bz_1)  S[F_{\Delta_2}^{-b}](z_2,\bz_2)
&\sim
 -i\f{g^2_{YM}}{4\pi}f^{ab}{}_{c}\bigg[
 \f1{z_{12}}S[F_{\Delta_2}^{-c}](z_2,\bz_2)
 -\f{1}{2\pi}(2-\Delta_2)z_{12}^{\Delta_2-3}\bar z_{12}^{\Delta_2}\int_{S_3}
 \f{z_{31}^{1-\Delta_2}}{\bar z_{23}\bar z_{31}^{\Delta_2}}S[F_{\Delta_2}^{-c}](z_3,\bz_3)\bigg]\,.
 \la{bF0SFD}
\ee
As in the case of gravity, the two operators $S[F_0], \bF_0$, sharing the same $(\Delta, J)=(1,1)$, exhibit the same OPEs.
{As we did in the case of gravity, it is interesting to compare these observations with the findings of \cite{Himwich:2025bza}. There the correspondence of OPEs \eqref{bF0SFD} and \eqref{SF0SFD} finds confirmation (see, respectively, the conformally soft limits $\Delta_1\rightarrow 1$ of Eq.'s (3.21) and (3.22) in \cite{Himwich:2025bza}). However, the expression they find corresponds only to the local part of the RHS of \eqref{bF0SFD}, up to a sign difference. Such a mismatch is exactly analogous to what we observed for the case of the energy-momentum tensor in gravity.}

If we take the conformally soft limit of the second entry, we get
\be
\bar F_{0}^{a}(z_1,\bz_1)  S[F_{s_2}^{b}](z_2,\bz_2)
&\sim
 -i\f{g^2_{YM}}{4\pi}f^{ab}{}_{c}\bigg[
 \f1{z_{12}}S[F^c_{s_2}](z_2,\bz_2)
 +\f{(s_2+1)}{2\pi}\f{\bar z_{12}^{1-s_2}}{z_{12}^{s_2+2}}\int_{S_3}
 \f{z_{13}^{s_2}\bar z_{13}^{s_2-1}}{\bar z_{23}}S[F^c_{s_2}](z_3,\bz_3)\bigg]\,.
 \la{bF0SFs}
\ee
{Let us now come back to an observation made above. If we take the double leading soft limit of \eqref{eq:cope}, we obtain
\be
\bar F_{0}^{a}(z_1,\bz_1)  \bar F_{0}^{b}(z_2,\bz_2)
&\sim
 -i\f{g^2_{YM}}{4\pi}f^{ab}{}_{c}
 \f1{z_{12}}\bar F^c_{0}(z_2,\bz_2)\,,
 \la{bF0bF0}
\ee
which is the expected Ward identity of a holomorphic Kac--Moody symmetry. 
If the correspondence in the OPE structures of $\bar F_0$ and $S[F_0]$ were completely general, one might expect that replacing $\bar F_0\rightarrow S[F_0]$ in the second entry of \eqref{bF0bF0}, would preserve the same structure on the RHS. However, what we actually obtain specializing \eqref{bF0SFs} to the case $s_2=0$ is
\be
\bar F_{0}^{a}(z_1,\bz_1)  S[F_{0}^{b}](z_2,\bz_2)
&\sim
 -i\f{g^2_{YM}}{4\pi}f^{ab}{}_{c}\bigg[
 \f1{z_{12}}S[F^c_{0}](z_2,\bz_2)
 +\f{1}{2\pi}\f{\bar z_{12}}{z_{12}^{2}}\int_{S_3}
 \f{1}{\bar z_{13}\bar z_{23}}S[F^c_{0}](z_3,\bz_3)\bigg]\,,
 \la{bF0SF0}
\ee
which is manifestly different.  
This shows that the correspondence $S[F_0]\sim \bar F_0$ in OPEs only applies to the first entry.}

For $s_2\ge1$, as we show in Appendix \ref{App:bF0SFs}, the OPE \eqref{bF0SFs} can be recast in the form
\be
\bar F_{0}^{a}(z_1,\bz_1)  S[F_{s_2}^{b}](z_2,\bz_2)
&\sim
 -i\f{g^2_{YM}}{4\pi}f^{ab}{}_{c}\left[
 \f1{z_{12}}S[F^c_{s_2}](z_2,\bz_2)
 +\f{(s_2+1)}{2\pi}\f{1}{z_{12}^{s_2+2}}
 \int_{S_3}
 \f{z_{13}^{s_2}}{\bar z_{23}}
 S[F^c_{s_2}](z_3,\bz_3)
\right]\,.
 \la{bF0SFs-2}
\ee

We now would like  to compare this OPE with the one where in the second entry we replace the shadow soft gluon with the dual soft gluon $\tF_{s_2}$, with $s_2\geq1$. To this purpose, we can work in the more general case of an arbitrary $s_1\geq 0$ and derive first (see Appendix \ref{App:bFstrs})
\be
\bar F_{s_1}^{a}(z_1,\bz_1) S[r_{s_2}^{1b}](z_2,\bz_2)
&\sim
 -i\f{g^2_{YM}}{4\pi}f^{ab}{}_{c} \bigg[
 \f{1}{s_1!}\f{\bar z_{12}^{s_1}}{z_{12}} S[r_{s_1+s_2}^{c}](z_2,\bz_2)
 \cr
 &-2\pi (s_1+s_2-1)
\left(\begin{matrix}
s_1+s_2-2\\
s_1
\end{matrix}\right)
\sum_{n=0}^{s_2-2}\f{(s_2-2)_n}{(s_1+n+1)n!}
\bar\partial_{2}^{n}\delta^2(z_1,z_2)\bar \partial^{s_2-2-n}_2 F_{s_1+s_2}^{c}(z_2,\bz_2)
\bigg]\,,
\cr
\la{ftilder}
\ee
where the last sum contribution vanishes for $s_2=1$, consistently with the fact that $S[r_1]=F_1$. Next, if we apply $\partial_{2}^{s_2+1}$ to \eqref{ftilder}, we obtain
\be
\bar F_{s_1}^a(z_1,\bz_1) \tilde F_{s_2}^b(z_2,\bz_2)
&\sim -i\f{g^2_{YM}}{4\pi}f^{ab}{}_{c} \bigg[
\sum_{k=0}^{s_2+1}\binom{s_2+1}{k}\f{k!}{s_1!}\f{\bar z_{12}^{s_1}}{z_{12}^{k+1}} \partial_{2}^{s_2+1-k}S[r_{s_1+s_2}^{1c}](z_2,\bz_2)
\cr
&-2\pi (s_2+s_1-1)
\left(\begin{matrix}
s_2+s_1-2\\
s_1
\end{matrix}\right)
\sum_{n=0}^{s_2-2}\f{(s_2-2)n}{(s_1+n+1)n!}
\partial_{2}^{s_2+1}\left(\bar\partial_{2}^{n}\delta^2(z_1,z_2)\bar \partial_2^{s_2-2-n} F_{s_1+s_2}^{c}(z_2,\bz_2)\right)
\bigg]\,.
\la{fbftgen}
\cr
\ee
For the case $s_1=0$, \eqref{fbftgen} yields
\be
\bar F_0^a(z_1,\bz_1) \tilde F_{s_2}^b(z_2,\bz_2)
&\sim  -i\f{g^2_{YM}}{4\pi}f^{ab}{}_{c} \bigg[
\f{1}{z_{12}} \tilde F_{s_2}^{c}(z_2,\bz_2)
+\sum_{k=1}^{s_2+1}
\f{(s_2+1)_k}{z_{12}^{k+1}} \partial_{2}^{s_2+1-k}S[r_{s_2}^{c}](z_2,\bz_2)
\cr
&
-2\pi (s_2-1)
\sum_{n=0}^{s_2-2}\f1{(n+1)}
\left(\begin{matrix}
s_2-2\\
n
\end{matrix}\right)\partial_{2}^{s_2+1}\left(\bar\partial_{2}^{n}\delta^2(z_1,z_2)\bar \partial_2^{s_2-2-n} F_{s_2}^{c}(z_2,\bz_2)\right)
\bigg]\,.
\la{bF0tFs}
\ee
If we ignore for the moment the contact terms appearing in the second line on the RHS of \eqref{bF0tFs}, we see that the  terms {$\propto z_{12}^{-1}$} in both \eqref{bF0SFs-2} and \eqref{bF0tFs} have the same form, upon the replacement $S[F_s]\leftrightarrow \tF_s$. A natural question is wether, upon the same replacement, also the sum term in the first line of \eqref{bF0tFs} can be matched with the integral term on the RHS of \eqref{bF0SFs-2}. To answer the question, we use the celestial diamond in Fig. \ref{fig:CD-YM-1} to rewrite
\be
\sum_{k=1}^{s_2+1}
\f{(s_2+1)_k}{z_{12}^{k+1}} \partial_{2}^{s_2+1-k}S[r_{s_2}^{c}](z_2,\bz_2)
&=
\f1{2\pi s_2!}\sum_{k=1}^{s_2+1}
\f{(s_2+1)_k}{z_{12}^{k+1}} 
\int_{S_3}\partial_{2}^{s_2+1-k}\left( \f{z_{23}^{s_2}}{\bz_{23}}\right)
\tF_{s_2}^{c}(z_3,\bz_3)
\cr
&=
\f{(s_2+1)}{2\pi }
\f1{z_{12}^{s_2+2}}
\int_{S_3} 
\f{z_{13}^{s_2}}{\bz_{23}}
\tF_{s_2}^{c}(z_3,\bz_3)\,.
\ee
We thus see that, up to contact terms in the second line on the RHS of \eqref{bF0tFs}, the OPE \eqref{bF0SFs-2} and \eqref{bF0tFs} have the same form, upon the replacement $S[F_s]\leftrightarrow \tF_s$, in complete analogy to the gravity case (Section \ref{sec:mixed-helicity-GR}).

For completeness, if we take the shadow transform of the first entry in \eqref{bF0SFD}, we obtain 
\be
S[\bar F_{0}^{a}](z_1,\bz_1)  S[F_{\Delta_2}^{-b}](z_2,\bz_2)
&\sim
 -i\f{g^2_{YM}}{4\pi}f^{ab}{}_{c}\bigg[
 \f1{z_{12}}S[F_{\Delta_2}^{-c}](z_2,\bz_2)
 +\f{1}{2\pi}\Delta_2 z_{12}^{\Delta_2-2}\bar z_{12}^{\Delta_2-1}\int_{S_3}
 \f{z_{31}^{2-\Delta_2}}{ z_{23}\bar z_{31}^{\Delta_2+1}}S[F_{\Delta_2}^{-c}](z_3,\bz_3)\bigg]\,.
 \la{SbF0SFD}
\ee

Finally, let us point out that, as in the case of the energy-momentum tensor in gravity, the holomorphicity condition $\bar \partial S[F_0]=0$ does not receive any obstruction from the mixed-helicity OPE
\be
S[\bar F_{0}^{a}](z_1,\bz_1)  \bar \p S[F_{0}^{b}](z_2,\bz_2)
&\sim
 -i\f{g^2_{YM}}{4\pi}f^{ab}{}_c\partial_{2}\left(\f{1}{\bar z_{12}}F_{0}^{c}(z_2,\bz_2)\right)\,,
\ee
as the RHS  generates only contact terms when inserted in a correlation function where all of the other operators are hard.

\section{Conclusions}\la{sec:conc}

The charge OPE/bracket correspondence \eqref{OPE-bra} represents a useful and insightful tool to translate between  CCFT techniques and phase space formalism, allowing to   connect results in scattering amplitudes to implications of asymptotic charge evolution equations and the Poisson structure of radiative modes. 

We have first exploited this correspondence to fix a prescription to perform  the double-soft limit of  celestial OPEs. The prescription corresponds to a sequential limit where the first entry of the OPE goes soft first. It has been pointed out in several places in the literature that a sequential limit would violate the symmetry structure of one of the two helicity sectors. The fate of asymptotic symmetries  in the mixed-helicity sector will be the main topic of our forthcoming companion paper \cite{MH-bracket}, where an interplay between mixed-helicity bracket and shadow charges will be revealed. We hope that such connection described through techniques of the covariant phase space framework can be applied to investigate the associativity of mixed-helicity OPEs via the charge OPE/bracket correspondence (see \cite{Mago:2021wje,Ren:2022sws,Ball:2022bgg,Ball:2023sdz} for the issue of associativity of the
celestial OPE).

As a second step of this work, we have applied the charge OPE/bracket correspondence, together with the celestial diamond organization of CCFT primaries and their descendants, to derive OPEs involving shadow operators. 
The agnostic nature of the charge bracket towards collinear limits in momentum space allowed us to deal with the non-local nature of the shadow transform and reveal a non-local analytic structure of the OPE. 

Moreover, we have verified that, up to possible contact terms, dual soft gravitons and gluons exhibit the same OPE structure than shadow soft gravitons and gluons; more precisely, starting from an OPE involving $\tN_s$ (or $\tF_s$) one can obtain the equivalent OPE involving $S[N_s]$ (or $S[F_s]$) simply replacing $\tN_s\to S[N_s]$ everywhere. This is not an obvious property of  CCFT as dual soft gravitons/gluons, defined as the bottom corner of celestial diamonds for $s\geq2, s\geq 1$ respectively, do share the same weights $(\Delta,J)$ than shadow soft gravitons/gluons, but they are apriori different operators. In fact, the duality pairing and the shadow transform correspond to different maps between  $SL(2,\C)$ representations acting on $L^2(\C)$ \cite{Freidel:2021ytz}; nevertheless, their OPE similarity may prove to be a useful technical tool in computing shadow OPEs, given the local expressions of the dual operators (bottom corners) in terms of their soft modes partners (top corners), as compared to the non-local nature of their corresponding shadow operators. 
It would be desirable to develop a more precise group theoretical understanding of this celestial OPE property, as well as a more physical intuition of its origin. 

The shadow transform plays a key role in the study of soft symmetries, providing a definition of energy-momentum tensor for the dual CFT \cite{Kapec:2016jld,Fotopoulos:2019tpe,Fotopoulos:2019vac}.
Moreover, the shadow transform of  conformal primary operators has appeared in the definition of boundary observables constructed from  an Euclidean path integral in the asymptotic boundary limit of an asymptotically flat space-time 
\cite{Jain:2023fxc}, \cite{Banerjee:2024yir}, in analogy with the
AdS-CFT correspondence.   
Advantages of using the shadow conformal
primary basis over the standard one, when computing  celestial four-point amplitudes, have been pointed out in \cite{Fan:2021isc,Chang:2022jut,Surubaru:2025qhs,Bhattacharyya:2025nfp}.
At the same time, shadow-transformed operators pose the puzzle of an over-complete basis for conformal primary states \cite{Pasterski:2017kqt}.
These are all indications that a deeper characterization of the sought after CCFT cannot occur without a clearer understanding of the role of shadow operators and their relation to non-shadowed basis elements of opposite  (w.r.t. the operator being shadowed) helicity.
We expect that the OPEs derived in this work can prove useful to such investigation.


\section*{Acknowledgments}

We are grateful to Ana-Maria Raclariu for her comments on a preliminary version of this manuscript. 
We would also like to thank Laurent Freidel, Elizabeth Himwich, Prahar Mitra, Monica Pate, Andrea Puhm,  Emilio Trevisani for helpful discussions and insights.

\appendix

\section{mixed-helicity charge bracket in Yang--Mills}\la{App:YM}

If we require the vector potential to belong to the Schwartz space $\mathcal{S}$, a discrete basis can be used to describe the phase space of the Yang-Mills theory \cite{Freidel:2023gue}. In complete analogy with the case of gravity, we can introduce memory observables $\mathscr{F}_\pm(n,z,\bz)$, and Goldstone bosons $\mathscr{A}_\pm(n,z,\bz), n\in\mathbb{Z}_+$. This can be done by first decomposing the fields as
 \be
 \label{eq:gluon-dec}
 A_z^{(0)}(u,z,\bz)&
{  = A_+(u,z,\bz)+A_-^*(u,z,\bz)}\,,\\
 F_{\bz u}^{(0)}(u,z,\bz)&
 { = F_-(u,z,\bz)+ F_+^*(u,z,\bz)}\,,
 \ee
and then\footnote{The rigorous definition of the memory observables $\mathscr{F}_\pm(n,z,\bz) $ requires a contour prescription; the precise details can be found in \cite{Freidel:2023gue}.}
\be
 A_{\pm} (u,z,\bz)
=\f{1}{2i\pi }  \sum_{\ell=0}^\infty \frac{(-iu)^\ell}{\ell!}   \mathscr{A}_\pm(\ell,z,\bz)\,,
\qquad
 \mathscr{F}_\pm(n,z,\bz) 
 := \frac{i^n}{n!}  \int_{-\infty}^{+\infty} du\, u^n  F_{\pm}  (u,z,\bz)\,.
 \la{AnFn}
 \ee

In this discrete basis,  the symplectic potential  \eqref{Theta-YM}
can be written    as $\Theta^{YM}=\Theta^{YM}_++\Theta^{YM}_-$, where
\be
\Theta^{YM}_\pm=\f{1}{2\pi ig^2_{YM}}\sum_{n=0}^\infty\int_S\tR[\mathscr{F}_\pm(n,z,\bz)\delta\mathscr{A}^*_\pm(n,z,\bz)],
\ee
from which we can read off the only non-trivial commutator of the quantum theory
\be
[\mathscr{F}^a_\pm(n,z,\bz),\mathscr{A}^{b\dagger}_\pm(m,z',\bz')]=2\pi g^2_{YM}\delta^{ab}\delta_{nm}\delta^{2}(z,z').
\la{ymcomm}
\ee

 Similarly, the charge aspects can be decomposed in terms of positive and negative energy parts  
\be
r^k_s(z,\bz)=  r^{k}_{s+}(z,\bz)+ r^{k}_{s-}(z,\bz)\,,
\label{eq:ymc}
\ee
where  the linear and quadratic charge aspect operators are expressed as
\be
r^1_{s\pm}(z,\bz)&=-i^{-s}\p^{s+1}\mathscr{F}^\dagger_{\pm}(s,z,\bz)\,,
\la{r1-app}
\\
r^2_{s\pm}(z,\bz)&=-\f{i^{-s}}{{2\pi}} \sum_{\ell=0}^\infty \sum_{n=0}^{s} 
\f{(\ell+n)_\ell}{\ell!}
 \p^{n} \left[\mathscr{A}_\pm(\ell,z,\bz), \p^{s-n}  \mathscr{F}^\dagger_\pm(s+\ell,z,\bz) \right]_\Ag\,
 \cr
  &=-\f{i^{s}}{2\pi}\sum_{\ell=0}^\infty\sum_{n=0}^{s} 
(-)^{n}
\f{(s +\ell-1)_{s-n}}{(s-n)!} \p^{s-n} \left[   \mathscr{F}_\pm(s+\ell,z,\bz),  \p^{n} \mathscr{A}^\dagger_\pm(\ell,z,\bz) \right]_\Ag\,.
\la{r2-app}
\ee
The alternative expression of the quadratic charge aspect in the second line of \eqref{r2-app} was not derived in \cite{Freidel:2023gue}, but it can be obtained using the same identities applied in the case of gravity in \cite{Freidel:2022skz} (see Appendix G.1 there); we omit the explicit derivation here. 
 
 Given this expression of the quadratic charge aspect, we can use the commutator \eqref{ymcomm} to compute the action
 \be
 [  r^{2a}_{s_1+}(z_1,\bz_1),\mathscr{F}^b_+(s_2,z_2,\bz_2)]
 &=-i
 g^2_{YM}i^{s_1}f^{ab}{}_c\sum_{n=0}^{s_1} 
(-)^{n}
\f{(s_1 +s_2-1)_{s_1-n}}{(s_1-n)!} \p_1^{s_1-n}   \left(\mathscr{F}^c_+(s_1+s_2,z_1,\bz_1)  \p_1^{n} \d^2(z_1,z_2)\right)
\cr
&=-i g^2_{YM}i^{s_1}f^{ab}{}_c
\sum_{\ell=0}^{s_1}
\left(\sum_{n=\ell}^{s_1}
(-)^{s_1+n}
\f{(s_1 +s_2-1)_{n}}{n!} 
\f{(n)_{\ell}}{\ell!} 
\right)
\p_1^{\ell}\mathscr{F}^c_+(s_1+s_2,z_1,\bz_1) \p_1^{s_1-\ell} \d^2(z_1,z_2)
\cr
&=-i g^2_{YM} i^{s_1}f^{ab}{}_c
\sum_{\ell=0}^{s_1}
\f1{(s_1+s_2-\ell-1)}\f{(s_1 +s_2-1)!}{\ell!(s_1-\ell)!(s_2-2)!} 
\p_1^{\ell}\mathscr{F}^c_+(s_1+s_2,z_1,\bz_1) \p_1^{s_1-\ell} \d^2(z_1,z_2)
\cr
&=-i g^2_{YM}i^{-s_1}f^{ab}{}_c
\sum_{p=0}^{s_1}
\left(\sum_{\ell=0}^{s_1-p}
\f{(-)^{\ell+p}}{(s_1+s_2-\ell-1)}\f{(s_1 +s_2-1)!}{\ell!p!(s_1-\ell-p)!(s_2-2)!} 
\right)
\cr
&\times
\p_2^{s_1-p}\mathscr{F}^c_+(s_1+s_2,z_2,\bz_2) \p_1^{p} \d^2(z_1,z_2)
\cr
&=- g^2_{YM} i^{s_1+1}f^{ab}{}_c
\sum_{p=0}^{s_1}
\f{(s_1+s_2-p-2)_{s_2-2}}{(s_2-2)!}
\p_2^{p}\mathscr{F}^c_+(s_1+s_2,z_2,\bz_2) \p_1^{s_1-p} \d^2(z_1,z_2)\,,
\la{r2F}
 \ee
 where we used
 \be
 \sum_{n=\ell}^{s_1}
(-)^{s_1+n}
\f{(s_1 +s_2-1)_{n}}{n!} 
\f{(n)_{\ell}}{\ell!} &=
\f1{(s_1+s_2-\ell-1)}\f{(s_1 +s_2-1)!}{\ell!(s_1-\ell)!(s_2-2)!}\,,
 \\
 \sum_{\ell=0}^{s_1-p}
\f{(-)^{\ell+p}}{(s_1+s_2-\ell-1)}\f{(s_1 +s_2-1)!}{\ell!p!(s_1-\ell-p)!(s_2-2)!} 
&=(-)^{s_1}\f{(s_2+p-2)_p}{p! }\,.
 \ee

 The action \eqref{r2F}, can be used to compute in a straightforward manner the commutator
 \be
[r^{2a}_{s_1+}(z_1,\bz_1),\bar r^{1b}_{s_2+}(z_2,\bz_2)]
=&i^{1+s_1+s_2}g^2_{YM}f^{ab}{}_{c}\sum_{n=0}^s
\f{(s_1+s_2-n-2)_{s_2-2}}{(s_2-2)!}
\bar \p^{s_2+1}_{2}(\p^{s_1-n}_1\delta^2(z_1,z_2)\p_{2}^{n}\mathscr{F}^c_+(s_1+s_2,z_2,\bz_2)).
\la{posrcomm}
\ee
A similar result for the negative-energy counterpart of \eqref{posrcomm} can be obtained by performing the  substitutions $r_{s+}\rightarrow\bar r_{s-}$, $\p\rightarrow\bar\p$, $\mathscr{F}_+(s)\rightarrow\mathscr{F}_-(s)$.
Putting the two sectors together, one obtains
\be
[r^{2a}_{s_1}(z_1,\bz_1),\bar r^{1b}_{s_2}(z_2,\bz_2)]
=&{-}ig^2_{YM}f^{ab}{}_{c}\sum_{n=0}^s
\f{(s_1+s_2-n-2)_{s_2-2}}{(s_2-2)!}
\bar \p^{s_2+1}_{2}(\p^{s_1-n}_1\delta^2(z_1,z_2)\p_{2}^{n} \bar F^c_{s'+s}(z_2,\bz_2)),
\la{ymba-app}
\ee
where one uses that $F_s(z,\bz)=
-i^{-s}\mathscr{F}^\dagger_+(s,z,\bz)-i^s\mathscr{F}_-(s,z,\bz)$. The Hermitian conjugate of this commutator yields \eqref{ymba}.

\section{Shadow OPE in Gravity}\la{App:GR-OPE}

In this Appendix, as well as in Appendix \ref{App:YM-OPE}, we will  repeatedly use the integral identity \cite{Dolan:2011dv}
\be
\f{1}{2\pi}\int_{S_3}\f{1}{z_{31}^{h_1}}\f{1}{z_{32}^{h_2}}\f{1}{z_{34}^{h_4}}\f{1}{\bar z_{31}^{\bar h_1}}\f{1}{\bar z_{32}^{\bar h_2}}\f{1}{\bar z_{34}^{\bar h_4}}=C_{124}z_{12}^{h_4-1}z_{24}^{h_1-1}z_{41}^{h_2-1}\bar z_{12}^{\bar h_4-1}\bar z_{24}^{\bar h_1-1}\bar z_{41}^{\bar h_2-1}
\la{intid}
\ee
with
\be
C_{124}=\f{\Gamma(1-h_1)\Gamma(1-h_2)\Gamma(1-h_4)}{\Gamma(\bar h_1)\Gamma(\bar h_2)\Gamma(\bar h_4)}=\f{\Gamma(1-\bar h_1)\Gamma(1-\bar h_2)\Gamma(1-\bar h_4)}{\Gamma( h_1)\Gamma( h_2)\Gamma( h_4)}\,,
\la{C124}
\ee
which holds for
\be
h_1+h_2+h_4=\bar h_1+\bar h_2+\bar h_4=2 \quad \text{and} \quad h_i-\bar h_i\in \mathbb Z,\, i=1,2,4.
\la{conh}
\ee
Also, in order to avoid extra cluttering, we adopt the notation $(z):=(z,\bz)$.

\subsection{$T S[N^-_{\Delta}]$}\la{App:TSN}

We start from \eqref{TN} and \eqref{scg}, \eqref{scg2} and we want to derive the OPEs \eqref{tn1tn1b-2}. 
Assuming $\Re\Delta_2\le1$, we have
\be
T(z_1) S[N^-_{\Delta_2}](z_2)\sim&-\f{i\k}{8}\f{1}{2\pi}\f{K_{\Delta_2}}{2\pi}\int_{S_3}\f{\bar z_{23}^{\Delta_2}}{z_{23}^{4-\Delta_2}}\bigg[(\Delta_2-2)\f{1}{z_{13}^2}N^-_{\Delta_2}(z_3)+2\f{1}{z_{13}}\partial_{z_3}N^-_{\Delta_2}(z_3)\bigg]
\cr
\sim&-\f{i\k}{8}\f{1}{2\pi}\f{K_{\Delta_2}}{2\pi}\f{\tilde K_{2-\Delta_2}}{2\pi}\int_{S_3}\f{\bar z_{23}^{\Delta_2}}{z_{23}^{4-\Delta_2}}\bigg[(\Delta_2-2)
\f{1}{z_{13}^2}+2\f{1}{z_{13}}\partial_{z_3}\bigg]
\cr&\times\int_{S_4}\f{1}{z_{34}^{\Delta_2-2}}\f{1}{\bar z_{34}^{\Delta_2+2}} S[N^-_{\Delta_2}](z_4)
\cr
\sim&-\f{i\k}{8}\f{1}{2\pi}\f{K_{\Delta_2}}{2\pi}\f{\tilde K_{2-\Delta_2}}{2\pi}(\Delta_2-2)\int_{S_4}\int_{S_3}\f{\bar z_{32}^{\Delta_2}}{z_{32}^{4-\Delta_2}}\bigg[
\f{1}{z_{31}}+2\f{1}{z_{34}}\bigg]
\cr&\times\f{1}{z_{31}}\f{1}{z_{34}^{\Delta_2-2}}\f{1}{\bar z_{34}^{\Delta_2+2}} S[N^-_{\Delta_2}](z_4)
\cr
\sim&-\f{i\k}{8}\f{1}{2\pi}\f{K_{\Delta_2}}{2\pi}\f{\tilde K_{2-\Delta_2}}{2\pi}\int_{S_4}[
(\Delta_2-2)\partial_{z_1}+2\partial_{z_4}]\bigg(\int_{S_3}\f{1}{z_{31}}\f{1}{z_{32}^{4-\Delta_2}}
\cr&\times\f{1}{z_{34}^{\Delta_2-2}}\f{1}{\bar z_{32}^{-\Delta_2}}\f{1}{\bar z_{34}^{\Delta_2+2}}\bigg) S[N^-_{\Delta_2}](z_4)
\cr
\sim&-\f{i\k}{8}\f{1}{2\pi}\f{K_{\Delta_2}}{2\pi}\f{\tilde K_{2-\Delta_2}}{2\pi}\f{1}{(\Delta_2-3)}\int_{S_4}[
(\Delta_2-2)\partial_{z_1}+2\partial_{z_4}]\partial_{z_4}\bigg(\int_{S_3}\f{1}{z_{31}}\f{1}{z_{32}^{4-\Delta_2}}
\cr&\times\f{1}{z_{34}^{\Delta_2-3}}\f{1}{\bar z_{32}^{-\Delta_2}}\f{1}{\bar z_{34}^{\Delta_2+2}}\bigg) S[N^-_{\Delta_2}](z_4)
\cr\sim&-\f{i\k}{8}\f{1}{2\pi}\f{K_{\Delta_2}}{2\pi}\tilde K_{2-\Delta_2}C_{124}\f{1}{(\Delta_2-3)}\int_{S_4}[
(\Delta_2-2)\partial_{z_1}+2\partial_{z_4}]
\cr&\times\partial_{z_4}\bigg(z_{12}^{\Delta_2-4}z_{41}^{3-\Delta_2}\bar z_{12}^{\Delta_2+1}\bar z_{24}^{-1}\bar z_{41}^{-\Delta_2-1}\bigg) S[N^-_{\Delta_2}](z_4)
\cr\sim&-\f{i\k}{8}\f{1}{2\pi}\f{K_{\Delta_2}}{2\pi}\tilde K_{2-\Delta_2}C_{124}\f{1}{(\Delta_2-3)}\int_{S_4}[
(\Delta_2-2)\partial_{z_1}+2\partial_{z_4}]
\bigg((3-\Delta_2)z_{12}^{\Delta_2-4}z_{41}^{2-\Delta_2}
\cr&\times\bar z_{12}^{\Delta_2+1}\bar z_{24}^{-1}\bar z_{41}^{-\Delta_2-1}
-2\pi z_{12}^{\Delta_2-4}z_{41}^{3-\Delta_2}\bar z_{12}^{\Delta_2+1}\delta^2(z_2,z_4)\bar z_{41}^{-\Delta_2-1}\bigg) S[N^-_{\Delta_2}](z_4).
\ee
This becomes
\be
T(z_1) S[N^-_{\Delta_2}](z_2)
\sim&-\f{i\k}{16\pi}
\bigg(
-(4-\Delta_2)_3z_{12}^{\Delta_2-5}\bar z_{12}^{\Delta_2+1}\f{1}{2\pi}\int_{S_4}z_{41}^{1-\Delta_2}
z_{24}\bar z_{24}^{-1}\bar z_{41}^{-\Delta_2-1} S[N^-_{\Delta_2}](z_4)
\cr&-(3-\Delta_2)_2z_{12}^{\Delta_2-4}\partial_{z_1}(\bar z_{12}^{\Delta_2+1})\f{1}{2\pi}\int_{S_4}z_{41}^{2-\Delta_2}
\bar z_{24}^{-1}\bar z_{41}^{-\Delta_2-1} S[N^-_{\Delta_2}](z_4)
\cr&+ z_{12}^{-2}(4-\Delta_2+2z_{12}\partial_{z_2})  S[N^-_{\Delta_2}](z_2)
\bigg)\,,
\la{tn1tn1b-2-app}
\ee
where the second row on the RHS above only contributes when $\Re\Delta_2\le-2$.

\subsection{$\bT S[N^-_{\Delta}]$}\la{App:TbarSN}

To compute the OPE \eqref{btn1tn1}, we start with
\be
\bar q^1_1(z_1)N^-_{\Delta_2}(z_2)\sim&-\f{i\k}{8}\sum_{n=0}^1(-)^{n-1}(\Delta_2+2)_{1-n}(1)_n(n+1)\bar \partial_{z_1}^{1-n}\delta^2(z_1,z_2)\bar\partial_{z_2}^nN^-_{\Delta_2}(z_2)
\cr
\sim&-\f{i\k}{8}\bigg[-(\Delta_2+2)\bar\partial_{z_1}\delta^2(z_1,z_2)N^-_{\Delta_2}(z_2)+2\delta^2(z_1,z_2)\bar\partial_{z_2}N^-_{\Delta_2}(z_2)\bigg]\,,
\ee
from which
\be
\bT(z_1)N^-_{\Delta_2}(z_2)
\sim&-\f{i\k}{16\pi}\bigg[-(\Delta_2+2)\int_S\f{1}{\bar z_1-\bar z}\bar\partial_{z_1}\delta^2(z_1,z_2)N^-_{\Delta_2}(z_2)+2\int_S\f{1}{\bar z_1-\bar z}\delta^2(z_1,z_2)\bar\partial_{z_2}N^-_{\Delta_2}(z_2)\bigg]
\cr
\sim&-\f{i\k}{16\pi}\bigg[(\Delta_2+2)\f{1}{(\bar z_1-\bar z')^2}N^-_{\Delta_2}(z_2)+2\f{1}{\bar z_1-\bar z'}\bar\partial_{z_2}N^-_{\Delta_2}(z_2)\bigg]\,.
\ee

If we take the shadow of the second argument with 
$\Re(\Delta_2)\le 1$, we have
\be
\bT(z_1) S[N^-_{\Delta_2}](z_2)
\sim&-\f{i\k}{16\pi}\f{K_{\Delta_2}}{2\pi}\int_{S_3}\f{\bar z_{23}^{\Delta_2}}{z_{23}^{4-\Delta_2}}\bigg[(\Delta_2+2)\f{1}{\bar z_{13}^2}N^-_{\Delta_2}(z_3)+2\f{1}{\bar z_{13}}\bar\partial_{z_3}N^-_{\Delta_2}(z_3)\bigg]
\cr\sim&-\f{i\k}{16\pi}\f{K_{\Delta_2}}{2\pi}\f{\tilde K_{2-\Delta_2}}{2\pi}\int_{S_4}\int_{S_3}\f{1}{\bar z_{23}^{-\Delta_2}}\f{1}{z_{23}^{4-\Delta_2}}\bigg[(\Delta_2+2)\f{1}{\bar z_{13}^2}+2\f{1}{\bar z_{13}}\bar\partial_{z_3}\bigg]
\cr&\times\bigg[\f{1}{z_{34}^{\Delta_2-2}}\f{1}{\bar z_{34}^{\Delta_2+2}} S[N^-_{\Delta_2}](z_4)\bigg]
\cr\sim&-\f{i\k}{16\pi}\f{K_{\Delta_2}}{2\pi}\f{\tilde K_{2-\Delta_2}}{2\pi}(\Delta_2+2)\int_{S_4}\int_{S_3}\f{1}{z_{32}^{4-\Delta_2}}\f{1}{z_{34}^{\Delta_2-2}}\bigg[\f{1}{\bar z_{31}^2}\f{1}{\bar z_{32}^{-\Delta_2}}\f{1}{\bar z_{34}^{\Delta_2+2}}+2\f{1}{\bar z_{31}}\f{1}{\bar z_{32}^{-\Delta_2}}\f{1}{\bar z_{34}^{\Delta_2+3}}\bigg]
\cr&\times S[N^-_{\Delta_2}](z_4)
\cr\sim&-\f{i\k}{16\pi}\f{K_{\Delta_2}}{2\pi}\f{\tilde K_{2-\Delta_2}}{2\pi}\f{1}{(\Delta_2+1)}\int_{S_4}((\Delta_2+2)\bar\partial_{z_1}\bar\partial_{z_4}+2\bar\partial_{z_4}^2)\int_{S_3}\f{1}{z_{32}^{4-\Delta_2}}\f{1}{z_{34}^{\Delta_2-2}}\bigg[\f{1}{\bar z_{31}}\f{1}{\bar z_{32}^{-\Delta_2}}\f{1}{\bar z_{34}^{\Delta_2+1}}\bigg]
\cr&\times S[N^-_{\Delta_2}](z_4)
\cr\sim&-\f{i\k}{16\pi}K_{\Delta_2}\f{\tilde K_{2-\Delta_2}}{2\pi}C_{124}\f{1}{(\Delta_2+1)}\int_{S_4}((\Delta_2+2)\bar\partial_{z_1}\bar\partial_{z_4}+2\bar\partial_{z_4}^2)\bigg[z_{12}^{\Delta_2-3}z_{24}^{-1}z_{41}^{3-\Delta_2}\bar z_{12}^{\Delta_2}\bar z_{41}^{-\Delta_2-1}\bigg]
\cr&\times S[N^-_{\Delta_2}](z_4)
\cr\sim&-\f{i\k}{16\pi}K_{\Delta_2}\f{\tilde K_{2-\Delta_2}}{2\pi}C_{124}\f{1}{(\Delta_2+1)}\bigg[(\Delta_2+2)\bar\partial_{z_1}\bigg[-2\pi\int_{S_4} z_{12}^{\Delta_2-3}\delta^2(z_2,z_4)z_{41}^{3-\Delta_2}\bar z_{12}^{\Delta_2}\bar z_{41}^{-\Delta_2-1}
\cr&-(\Delta_2+1)\int_{S_4}z_{12}^{\Delta_2-3}z_{24}^{-1}z_{41}^{3-\Delta_2}\bar z_{12}^{\Delta_2}\bar z_{41}^{-\Delta_2-2}\bigg]
 S[N^-_{\Delta_2}](z_4)
\cr&-2\int_{S_4}\bigg[-2\pi z_{12}^{\Delta_2-3}\delta^2(z_2,z_4)z_{41}^{3-\Delta_2}\bar z_{12}^{\Delta_2}\bar z_{41}^{-\Delta_2-1}
\cr&-(\Delta_2+1)z_{12}^{\Delta_2-3}z_{24}^{-1}z_{41}^{3-\Delta_2}\bar z_{12}^{\Delta_2}\bar z_{41}^{-\Delta_2-2}\bigg]
\bar\partial_{z_4} S[N^-_{\Delta_2}](z_4)\bigg]
\cr\sim&-\f{i\k}{16\pi}\bigg[-\Delta_2 \bar z_{12}^{-2} S[N^-_{\Delta_2}](z_2)+2 \bar z_{12}^{-1}\bar\partial S[N^-_{\Delta_2}](z_2)
\cr&+(\Delta_2+2)(\Delta_2+1)\Delta_2z_{12}^{\Delta_2-3}\bar z_{12}^{\Delta_2-1}\f{1}{2\pi}\int_{S_4}z_{24}^{-1}\bar z_{24}z_{41}^{3-\Delta_2}\bar z_{41}^{-\Delta_2-3} S[N^-_{\Delta_2}](z_4)
\cr&-(\Delta_2+2)(\Delta_2+1)\f{1}{2\pi}\bar\partial_{z_1}\left(z_{12}^{\Delta_2-3}\right)\bar z_{12}^{\Delta_2}\int_{S_4}z_{24}^{-1}z_{41}^{3-\Delta_2}\bar z_{41}^{-\Delta_2-2} S[N^-_{\Delta_2}](z_4)
\bigg]\,.
\la{btn1tn1-2-app}
\ee

\subsection{$TT$}\la{App:TT}

If we take the subleading conformally soft limit (residue in $\Delta_2=0)$ of \eqref{tn1tn1b-2-app}, we get
\be
T(z_1)T(z_2)
\sim&-\f{i\k}{16\pi}\bigg[ -24z_{12}^{-5}\bar z_{12}\f{1}{2\pi}\int_{S_4}\f{ z_{14}}{\bar z_{14}}\f{ z_{42}}{ \bar z_{42}}T(z_4)
+ z_{12}^{-2}(4+2z_{12}\partial_{z_2})T(z_2)\bigg]
\cr\sim&-\f{i\k}{16\pi}\bigg[ -24z_{12}^{-5}\f{1}{2\pi}\int_{S_4}\left(\f{ z_{42}- z_{14}}{\bar z_{14}}+\f{ z_{42}- z_{14}}{ \bar z_{42}}\right)\partial_{z_4}^2 \tilde q^1_1 (z_4)
+ z_{12}^{-2}(4+2z_{12}\partial_{z_2})T(z_2)\bigg]
\cr\sim&-\f{i\k}{16\pi}\bigg[ 24z_{12}^{-5}\f{1}{2\pi}\int_{S_4}\left(\f{2}{\bar z_{14}}-2\pi z_{12}\delta^2(z_1,z_4)+\f{2}{\bar z_{14}}-2\pi z_{12}\delta^2(z_4,z_2)\right)\partial_{z_4} \tilde q^1_1 (z_4)
\cr&+ z_{12}^{-2}(4+2z_{12}\partial_{z_2})T(z_2)\bigg]
\cr
\sim&-\f{i\k}{8\pi}\bigg[
z_{12}^{-2}(2+z_{12}\partial_{z_2})T(z_2)
\cr
&
-24 z_{12}^{-5}\left(- \tilde q^1_1 (z_1)+ \tilde q^1_1 (z_2)\right)
-12z_{12}^{-4}(\partial_{z_1} \tilde q^1_1 (z_1)+\partial_{z_2} \tilde q^1_1 (z_2))
 \bigg]\,.
\ee
Let us write the terms in the last line above as
\be
\f{24}{z_{12}^5} \left(  \tilde q^1_1 (z_2)
-  \tilde q^1_1 (z_1)
\right)
&=\f1{2\pi}\f{24}{z_{12}^5} \int_{S}\left( \f1{z_2-z}N_1(z)
-\f1{z_1-z}N_1(z)
\right)
\cr
&=\f1{2\pi}\f{24}{z_{12}^4} \int_{S} \f1{(z_2-z)(z_1-z)}N_1(z)\,,
\ee
and analogously
\be
\f{12}{z_{12}^4} \left( \bar\partial_{z_2} \tilde q^1_1 (z_2)+
 \bar\partial_{z_1} \tilde q^1_1 (z_1)
\right)
&=-\f1{2\pi}\f{12}{z_{12}^4} \int_{S}\left( \f1{(z_2-z)^2}N_1(z)
+\f1{(z_1-z)^2}N_1(z)
\right)
\cr
&=-\f1{2\pi}\f{12}{z_{12}^4} \int_{S} \f{z_1^2-2z_1z+z_2^2-2z_2z+2z^2}{(z_2-z)^2(z_1-z)^2}N_1(z).
\ee
We thus have
\be
\f{24}{z_{12}^5} \left(  \tilde q^1_1 (z_2)
-  \tilde q^1_1 (z_1)
\right)+\f{12}{z_{12}^4} \left( \bar\partial_{z_2} \tilde q^1_1 (z_2)+
 \bar\partial_{z_1} \tilde q^1_1 (z_1)
\right)
&=-\f1{\pi}\f{6}{z_{12}^2} \int_{S} \f{1}{(z_2-z)^2(z_1-z)^2}N_1(z)\,.
\ee

\subsection{$T \bar \p T$}\la{App:TbarpT}

The OPE \eqref{TbarpT} can be computed starting from the conformally soft limit $\Delta_2\rightarrow0$ of \eqref{TN}, i.e.
\be
T_1(z_1)N_1(z_2) \sim&-\f{i\k}{4}\f{1}{2\pi}\int_S\f{1}{z_1-z}\bigg[\partial_{z}\delta^2(z,z_2)N_1(z_2)+\delta^2(z,z_2)\partial_{z_2}N_1(z_2)\bigg]
\cr
\sim&-\f{i\k}{8\pi}\int_S\bigg[-\f{1}{(z_1-z)^2}\delta^2(z,z_2)+\f{1}{z_1-z}\delta^2(z,z_2)\partial_{z_2}\bigg]N_1(z_2)
\cr\sim&-\f{i\k}{4}\f{1}{2\pi}\bigg[-\f{1}{(z_1-z_2)^2}+\f{1}{z_1-z_2}\partial_{z_2}\bigg]N_1(z_2), \la{tnn}
\ee
and applying\footnote{From the celestial diamond in Fig. \ref{fig:CD-GR-2}, we see that $\bar \p T=\p^3 N_1$} $\partial_{z_2}^3$
\be
T_1(z_1)\p^3 N_1(z_2)
\sim&-\f{i\k}{4}\f{1}{2\pi}\bigg[-\partial_{z_2}^3\left(\f{1}{(z_1-z_2)^2}\right)N_1(z_2)-3\partial_{z_2}^2\left(\f{1}{(z_1-z_2)^2}\right)\partial N_1(z_2)
\cr&-3\partial_{z_2}\left(\f{1}{(z_1-z_2)^2}\right)\partial^2 N_1(z_2)
-\left(\f{1}{(z_1-z_2)^2}\right)\partial^3 N_1(z_2)+\partial_{z_2}^3\left(\f{1}{z_1-z_2}\right)\partial N_1(z_2)
\cr&+3\partial_{z_2}^2\left(\f{1}{z_1-z_2}\right)\partial^2N_1(z_2)+3\partial_{z_2}\left(\f{1}{z_1-z_2}\right)\partial^3N_1(z_2)+\f{1}{z_1-z_2}\partial^4N_1(z_2)\bigg]
\cr\sim&-\f{i\k}{4}\f{1}{2\pi}\bigg[-24\f{1}{(z_1-z_2)^5}N_1(z_2)-12\f{1}{(z_1-z_2)^4}\partial N_1(z_2)
\cr&+2\f{1}{(z_1-z_2)^2}\partial^3N_1(z_2)+\f{1}{z_1-z_2}\partial^4N_1(z_2)\bigg]\,.
\ee


\subsection{$\bT T$}\la{App:TTbar}

If we take the residue in $\Delta_2=0$ of \eqref{btn1tn1-2-app}, we get
\be
\bT(z_1) T(z_2)\sim&-\f{i\k}{16\pi}\bigg[2 \bar z_{12}^{-1}\bar\partial T(z_2)-\partial_{z_1}^2\delta^2(z_1,z_2)\int_{S_4}z_{24}^{-1}z_{41}^{3}\bar z_{41}^{-2} T(z_4)
\cr\sim&-\f{i\k}{16\pi}\bigg[2 \bar z_{12}^{-1}\bar\partial T(z_2)-\partial_{z_1}^2\left(\delta^2(z_1,z_2)\int_{S_4}z_{24}^{-1}z_{41}^{3}\bar z_{41}^{-2} T(z_4)\right)
\cr&-6\partial_{z_1}\left(\delta^2(z_1,z_2)\int_{S_4}z_{24}^{-1}z_{41}^{2}\bar z_{41}^{-2} T(z_4)\right)
\cr&-6\delta^2(z_1,z_2)\int_{S_4}z_{24}^{-1}z_{41}\bar z_{41}^{-2} T(z_4)
\bigg]
\cr\sim&-\f{i\k}{16\pi}\bigg[2 \bar z_{12}^{-1}\bar\partial T(z_2)+\partial_{z_1}^2\delta^2(z_1,z_2)\bar\partial_{z_2}\int_{S_4}z_{42}^{2}\bar z_{42}^{-1} T(z_4)
\cr&+6\partial_{z_1}\delta^2(z_1,z_2)\bar\partial_{z_2}\int_{S_4}z_{42}\bar z_{42}^{-1} T(z_4)
\cr&+6\delta^2(z_1,z_2)\bar\partial_{z_2}\int_{S_4}\bar z_{42}^{-1} T(z_4)
\bigg]
\cr\sim&-\f{i\k}{8\pi}\bigg[ \bar z_{12}^{-1}\bar\partial T(z_2)-2\pi\partial_{z_2}^2\delta^2(z_1,z_2)\bar\partial \tilde q^1_1 (z_2)
\cr&-6\pi\partial_{z_2}\delta^2(z_1,z_2)\bar\partial\partial \tilde q^1_1 (z_2)
-6\pi\delta^2(z_1,z_2)\bar\partial\partial^2 \tilde q^1_1 (z_2)
\bigg]\,.
\ee

\subsection{Commuting shadow transform and soft limit}\la{App:Commuting}

The same expressions for the OPEs \eqref{TT}, \eqref{TbarT} can be derived by performing the second conformally-soft limit first and then taking the shadow transform of the second entry. We start by showing the commutativity of these two operations in the same-helicity sector. 
In this case, the relevant charge aspect bracket is 
\be
\{q^2_1(z),q^1_1(z')\}&=\f\k8\sum_{n=0}^1
(2-n)\binom{1+n}{1}\p_{z'}^3(\p^{1-n}_{z'}N_1(z')\p^n_z\delta^2(z,z'))
\cr&=\f\k4 \p_{z'}^3\left(
\p_{z'}N_1(z')\delta^2(z,z')+N_1(z')\p_z\delta^2(z,z')\right).
\ee
Recalling the relation \eqref{1strow}, we can compute the $TT$-OPE from the double shadow transform
\be
T(z_1)T(z_2)
&=-i\f{1}{4\pi^2}\int_S\int_{S'}\f{1}{z_1-z}\f{1}{z_2-z'}\{q^2_1(z),q^1_1(z')\}
\cr
&=\f{i\kappa^2}{16\pi^2}\int_S\partial_{z}^3\left(\f{1}{z_2-z}\right)\left(
\f{1}{z_1-z}\partial_{z}\bar\partial_{z} \tilde q^1_1 (z)-\partial_z\left(\f{1}{z_1-z}\right) \bar\partial_{z} \tilde q^1_1 (z)\right)
\cr
&=-\f{i\kappa^2}{16\pi^2}\int_S\left(\bar\partial_{z}\left[\partial_{z}^3\left(\f{1}{z_2-z}\right)
\f{1}{z_1-z}\right]\partial_{z} \tilde q^1_1 (z)-\bar\partial_{z}\left[\partial_{z}^3\left(\f{1}{z_2-z}\right)\partial_z\left(\f{1}{z_1-z}\right)\right]  \tilde q^1_1 (z)\right)
\cr
&=\f{i\kappa^2}{8\pi}\int_S\left(\left[\partial_{z}^3\left(\delta^2(z_2,z)\right)
\f{1}{z_1-z}\right]\partial_{z} \tilde q^1_1 (z)-\left[\partial_{z}^3\left(\delta^2(z_2,z)\right)\partial_z\left(\f{1}{z_1-z}\right)\right]  \tilde q^1_1 (z)\right)
\cr
&+\f{i\kappa^2}{8\pi}\int_S\left(\left[\partial_{z}^3\left(\f{1}{z_2-z}\right)
\delta^2(z_1,z)\right]\partial_{z} \tilde q^1_1 (z)-\left[\partial_{z}^3\left(\f{1}{z_2-z}\right)\partial_z\left(\delta^2(z_1,z)\right)\right]  \tilde q^1_1 (z)\right)
\cr
&=-\f{i\kappa^2}{8\pi}\bigg[
\partial_{z_2}^3\left(\f{1}{z_1-z_2}\partial_{z_2} \tilde q^1_1 (z_2)\right)-\partial_{z_2}^3\left(\partial_{z_2}\left(\f{1}{z_1-z_2}\right)  \tilde q^1_1 (z_2)\right)
\cr
&-\partial_{z_1}^3\left(\f{1}{z_2-z_1}\right)
\partial_{z_1} \tilde q^1_1 (z_1)-\partial_{z_1}\Bigg(\partial_{z_1}^3\left(\f{1}{z_2-z_1}\right)  \tilde q^1_1 (z_1)\Bigg )\bigg]
\cr
&=-\f{i\kappa^2}{8\pi}\bigg[
\f{1}{z_{12}}\partial_{z_2}T(z_2)+\f{2}{z_{12}^2}T(z_2)
-\f{12}{z_{12}^4}\left(\partial_{z_2} \tilde q^1_1 (z_2)+\partial_{z_1} \tilde q^1_1 (z_1)
\right)
-\f{24}{z_{12}^5} \left(  \tilde q^1_1 (z_2)
-  \tilde q^1_1 (z_1)
\right)
\bigg]\,,
\la{set}
\ee
where in the first passage we used  the celestial diamond relation $N_1=\bar \partial \tq_1$.
We thus recover exactly \eqref{TT}. 

In the mixed-helicity sector, the relevant charge aspect bracket is
\be
 \{ \bar q_1^2(z)  , q_1^1(z')\}=-\f\k {4}\partial_{z'}^3    \big(N_1(z) \partial_{\bar z} 
 \delta^2(z,z')\big)\,,
\ee
and the $\bT T$-OPE can be obtained 
from 
\be
\bT(z_1) T(z_2) 
&=i\f\k {4}\f{1}{4\pi^2}\int_S\int_{S'}\f{1}{\bar z_1-\bar z}\f{1}{z_2-z'}\partial_{z'}^3    \big(N_1(z) \partial_{\bar z} 
 \delta^2(z,z')\big)
\cr
&=i\f\k {4}\f{1}{4\pi^2}\int_S\partial_{z}^3\left(\f{1}{z_2-z}\right)    \partial_{\bar z}\left(\f{1}{\bar z_1-\bar z} N_1(z)\right)
\cr
&=-i\f\k {4}\f{1}{2\pi}\partial_{z_2}^3\left(\f{1}{\bar z_{12}} N_1(z_2)\right)
\cr
&=-i\f\k {4}\f{1}{2\pi}\bigg(\f{1}{\bar z_{12}} \bar \partial \tilde N_1(z_2)-6\pi\delta^2(z_1,z_2)\partial^2\bar \partial  \tilde q^1_1 (z_2)-6\pi\partial_{z_2}\delta^2(z_1,z_2)\partial\bar \partial  \tilde q^1_1 (z_2)
-2\pi\partial_{z_2}^2\delta^2(z_1,z_2)\bar \partial  \tilde q^1_1 (z_2)\bigg)\,,
\cr
\la{mset}
\ee
where in the last passage we used again the celestial diamond relation $N_1=\bar \partial \tq_1$; hence again, we 
exactly reproduce the expression \eqref{TbarT}. Notice that the second to the last row coincides with the (complex conjugate) expression given in Eq. (4.23) of \cite{Fotopoulos:2019vac}.

\subsection{$S[N_0] S[N_1]$}\la{App:SN01}

The OPE \eqref{SN01} is obtained as follows
\be
S[N_0](z_1)S[N_{1}](z_2)\sim&\f{i\k}{8}\f{1}{2\pi}\f{3!}{2\pi}\int_{S'}\f{1}{(z_2-z')^{4}}\f{\bar z_1-\bar z'}{z_1-z'}N_{0}(z')
\cr\sim&\f{i\k}{8}\f{1}{2\pi}\f{3!}{2\pi}\int_{S'}\bar \partial^2_{z'}\left(\f{1}{(z_2-z')^{4}}\f{\bar z_1-\bar z'}{z_1-z'}\right) \tilde q^1_0 (z')
\cr\sim&\f{i\k}{8}\f{1}{2\pi}\f{3!}{2\pi}\int_{S'}\bigg(-2\pi\f{1}{3!}\bar \partial_{z'}\partial_{z'}^3\delta^2(z_2,z')\f{\bar z_1-\bar z'}{z_1-z'}-4\pi\f{1}{3!}\partial_{z'}^3\delta^2(z_2,z')\f{1}{z'-z_1}
\cr&+2\pi\f{1}{(z_2-z')^{4}}\delta^2(z_1,z')\bigg) \tilde q^1_0 (z')
\cr\sim&\f{i\k}{8}\f{1}{2\pi}\f{3!}{2\pi}\int_{S'}\bigg(-2\pi\f{1}{3!}\delta^2(z_2,z')\bar \partial_{z'}\partial_{z'}^3\left(\f{\bar z_1-\bar z'}{z_1-z'} \tilde q^1_0 (z')\right)-4\pi\f{1}{3!}\delta^2(z_2,z')\partial_{z'}^3\left(\f{1}{z_1-z'} \tilde q^1_0 (z')\right)
\cr&+2\pi\f{1}{(z'-z_2)^{4}}\delta^2(z_1,z') \tilde q^1_0 (z')\bigg)
\cr\sim&\f{i\k}{8}\f{3!}{2\pi}\bigg[-\f{1}{3!}\partial_{z_2}^3\left(\f{\bar z_{12}}{z_{12}}\bar \partial_{z_2} \tilde q^1_0 (z_2)\right)-\f{1}{3!}\partial_{z_2}^3\left(\f{1}{z_{12}} \tilde q^1_0 (z_2)\right)
+\f{1}{z_{12}^{4}} \tilde q^1_0 (z_1)\bigg]
\cr\sim&-\f{i\k}{16\pi}\partial_{z_2}^3\bigg[\f{\bar z_{12}}{z_{12}}\bar \partial_{z_2} \tilde q^1_0 (z_2)+\f{1}{z_{12}}( \tilde q^1_0 (z_2)- \tilde q^1_0 (z_1))
\bigg]
\cr\sim&\f{i\k}{8}\f{3!}{2\pi}\bigg[-\f{\bar z_{12}}{z_{12}^4}\bar \partial_{z_2} \tilde q^1_0 (z_2)-\f{\bar z_{12}}{z_{12}^3}\bar \partial_{z_2}\partial_{z_2} \tilde q^1_0 (z_2)-\f{1}{2}\f{\bar z_{12}}{z_{12}^2}\bar \partial_{z_2}\partial_{z_2}^2 \tilde q^1_0 (z_2)-\f{1}{6}\f{\bar z_{12}}{z_{12}}\bar \partial_{z_2}\partial_{z_2}^3 \tilde q^1_0 (z_2)
\cr&-\f{1}{z_{12}^4} \tilde q^1_0 (z_2)-\f{1}{z_{12}^3}\partial_{z_2} \tilde q^1_0 (z_2)-\f{1}{2}\f{1}{z_{12}^2}\partial_{z_2}^2 \tilde q^1_0 (z_2)-\f{1}{6}\f{1}{z_{12}}\partial_{z_2}^3 \tilde q^1_0 (z_2)
+\f{1}{z_{12}^{4}} \tilde q^1_0 (z_1)\bigg]
\cr\sim&-\f{i\k}{16\pi}\bigg[\f{\bar z_{12}}{z_{12}}\bar \partial\partial S[N_{0}](z_2)+\f{1}{z_{12}}\partial S[N_{0}](z_2)+3\f{\bar z_{12}}{z_{12}^2}\bar \partial S[N_{0}](z_2)+3\f{1}{z_{12}^2} S[N_{0}](z_2)
\cr&+6\f{\bar z_{12}}{z_{12}^3}\bar \partial\partial \tilde q^1_0 (z_2)+6\f{1}{z_{12}^3}\partial \tilde q^1_0 (z_2)+6\f{\bar z_{12}}{z_{12}^4}\bar \partial \tilde q^1_0 (z_2)
+6\f{1}{z_{12}^4}( \tilde q^1_0 (z_2)- \tilde q^1_0 (z_1))
\bigg].
\ee


\subsection{$S[N_0] S[N^-_{\Delta_2}]$}\la{SN0SG}

In order to derive the OPE \eqref{t0g}, we take the shadow transform of the second entry in \eqref{tng}; restricting to $\Re \Delta_2\leq 1$, we have
\be
S[N_0](z_1)S[N^-_{\Delta_2}](z_2)\sim&\f{\k}{8i}\f{K_{\Delta_2}}{2\pi}\f{\tilde K_{2-(\Delta_2+1)}}{2\pi}\int_{S_4}
\cr
&
\bigg[
\int_{S_3}\f{1}{z_{13}}\f{1}{z_{23}^{4-\Delta_2}}\f{1}{z_{34}^{\Delta_2-1}}\f{1}{\bar z_{13}^{-1}}\f{1}{\bar z_{23}^{-\Delta_2}}
\f{1}{\bar z_{34}^{3+\Delta_2}}\bigg]S[N^-_{\Delta_2+1}](z_4)
\cr\sim&\f{\k}{8i}\f{K_{\Delta_2}}{2\pi}\f{\tilde K_{2-(\Delta_2+1)}}{2\pi}\int_{S_4}
\cr
&
\bigg[
\int_{S_3}\f{1}{z_{13}}\f{1}{z_{23}^{4-\Delta_2}}\f{1}{(3-\Delta_2)_2}\partial_{z_4}^{2}\f{1}{z_{34}^{\Delta_2-3}}\f{1}{\bar z_{13}^{-1}}\f{1}{\bar z_{23}^{-\Delta_2}}
\f{1}{\bar z_{34}^{3+\Delta_2}}\bigg]S[N^-_{\Delta_2+1}](z_4)
\cr\sim&\f{\k}{8i}\f{1}{(3-\Delta_2)_2}\f{K_{\Delta_2}}{2\pi}\f{\tilde K_{2-(\Delta_2+1)}}{2\pi}\int_{S_4}
\cr&\partial_{z_4}^{2}\bigg[\int_{S_3}\f{1}{z_{31}}\f{1}{z_{32}^{4-\Delta_2}}\f{1}{z_{34}^{\Delta_2-3}}\f{1}{\bar z_{31}^{-1}}\f{1}{\bar z_{32}^{-\Delta_2}}
\f{1}{\bar z_{34}^{3+\Delta_2}}\bigg]S[N^-_{\Delta_2+1}](z_4)
\cr\sim&\f{\k}{8i}\f{1}{(3-\Delta_2)_2}\f{K_{\Delta_2}}{2\pi}\tilde K_{2-(\Delta_2+1)}C_{124}z_{12}^{\Delta_2-4}\bar z_{12}^{2+\Delta_2}\int_{S_4}
\cr&\partial_{z_4}^{2}\bigg[z_{41}^{3-\Delta_2}\bar z_{24}^{-2}\bar z_{41}^{-1-\Delta_2}\bigg]S[N^-_{\Delta_2+1}](z_4)
\cr\sim&\f{\k}{8i}\f{1}{(3-\Delta_2)_2}\f{K_{\Delta_2}}{2\pi}\tilde K_{2-(\Delta_2+1)}C_{124}z_{12}^{\Delta_2-4}\bar z_{12}^{2+\Delta_2}\int_{S_4}
\cr
&
\bigg[
\partial_{z_4}^{2}z_{41}^{3-\Delta_2}\bar z_{24}^{-2}\bar z_{41}^{-1-\Delta_2}+2\partial_{z_4}z_{41}^{3-\Delta_2}\partial_{z_4}\left(\bar z_{24}^{-2}\bar z_{41}^{-1-\Delta_2}\right)+z_{41}^{3-\Delta_2}\partial_{z_4}^2\left(\bar z_{24}^{-2}\bar z_{41}^{-1-\Delta_2}\right)
\bigg]S[N^-_{\Delta_2+1}](z_4)
\cr\sim&\f{\k}{8i}\f{1}{(3-\Delta_2)_2}\f{K_{\Delta_2}}{2\pi}\tilde K_{2-(\Delta_2+1)}C_{124}z_{12}^{\Delta_2-4}\bar z_{12}^{2+\Delta_2}\int_{S_4}
\bigg[(3-\Delta_2)_2z_{41}^{1-\Delta_2}\bar z_{24}^{-2}\bar z_{41}^{-1-\Delta_2}
\cr
&
-4\pi(3-\Delta_2)z_{41}^{2-\Delta_2}\bar\partial_{z_4}\delta^2(z_2,z_4)\bar z_{41}^{-1-\Delta_2}-2\pi z_{41}^{3-\Delta_2}\bar\partial_{z_4}\partial_{z_4}\delta^2(z_2,z_4)\bar z_{41}^{-1-\Delta_2}
\bigg]S[N^-_{\Delta_2+1}](z_4)
\cr\sim&\f{\k}{8i}\f{1}{(3-\Delta_2)_2}\f{K_{\Delta_2}}{2\pi}\tilde K_{2-(\Delta_2+1)}C_{124}z_{12}^{\Delta_2-4}\bar z_{12}^{2+\Delta_2}
\bigg[
\cr
&
(3-\Delta_2)_2\int_{S_4}z_{41}^{1-\Delta_2}\bar z_{24}^{-2}\bar z_{41}^{-1-\Delta_2}S[N^-_{\Delta_2+1}](z_4)
-4\pi(3-\Delta_2)\bar\partial_{z_2}\left(z_{12}^{2-\Delta_2}\bar z_{12}^{-1-\Delta_2}S[N^-_{\Delta_2+1}](z_2)\right)
\cr
&-2\pi\bar\partial_{z_2}\partial_{z_2}\left(z_{12}^{3-\Delta_2}\bar z_{12}^{-1-\Delta_2}S[N^-_{\Delta_2+1}](z_2)\right)
\bigg]\,;
\ee
if we use the explicit form of the constants \eqref{tKGR}, \eqref{KGR}, \eqref{C124}, we arrive at 
\be
S[N_0](z_1)S[N^-_{\Delta_2}](z_2)
\sim&\f{\k}{8i}
\bigg[
-(3-\Delta_2)_2\f{1}{2\pi}z_{12}^{\Delta_2-4}\bar z_{12}^{2+\Delta_2}\int_{S_4}z_{41}^{1-\Delta_2}\bar z_{24}^{-2}\bar z_{41}^{-1-\Delta_2}S[N^-_{\Delta_2+1}](z_4)
\cr
&+z_{12}^{-2}\bar z_{12}^{2+\Delta_2}\bigg(3-\Delta_2+z_{12}\partial_{z_2}\bigg)\bar\partial_{z_2}\left(\bar z_{12}^{-1-\Delta_2}S[N^-_{\Delta_2+1}](z_2)\right)
\bigg]\,.
\la{ge1}
\ee

Otherwise, if we take the case $\Re\Delta_2>1$, we obtain
\be
S[N_0](z_1)S[N^-_{\Delta_2}](z_2)\sim&-\f{i\k}{8}\f{1}{2\pi}\f{K_{\Delta_2}}{2\pi}\int_{S_3}\f{\bar z_{23}^{\Delta_2}}{z_{23}^{4-\Delta_2}}\f{\bar z_{13}}{z_{13}}N^-_{\Delta_2+1}(z_3)
\cr\sim&-\f{i\k}{8}\f{1}{2\pi}\f{K_{\Delta_2}}{2\pi}\f{\tilde K_{2-\Delta_2-1}}{2\pi}\int_{S_4}\bigg[\int_{S_3}\f{1}{z_{31}}\f{1}{z_{32}^{4-\Delta_2}}\f{1}{z_{34}^{\Delta_2-1}}
\f{1}{\bar z_{31}^{-1}}\f{1}{\bar z_{32}^{-\Delta_2}}\f{1}{\bar z_{34}^{\Delta_2+3}}\bigg]
S[N^-_{\Delta_2+1}](z_4)
\cr\sim&-\f{i\k}{8}\f{1}{2\pi}\f{K_{\Delta_2}}{2\pi}\f{\tilde K_{2-\Delta_2-1}}{2\pi}\f{1}{(3-\Delta_2)(2-\Delta_2)}\partial^2_{z_2}\int_{S_4}\bigg[\int_{S_3}\f{1}{z_{31}}\f{1}{z_{32}^{2-\Delta_2}}\f{1}{z_{34}^{\Delta_2-1}}
\cr&\times\f{1}{\bar z_{31}^{-1}}\f{1}{\bar z_{32}^{-\Delta_2}}\f{1}{\bar z_{34}^{\Delta_2+3}}\bigg]
S[N^-_{\Delta_2+1}](z_4)
\cr\sim&-\f{i\k}{8}\f{K_{\Delta_2}}{2\pi}\tilde K_{2-\Delta_2-1}C_{124}\f{1}{(3-\Delta_2)(2-\Delta_2)}\partial^2_{z_2}\int_{S_4}\bigg[z_{12}^{\Delta_2-2}z_{41}^{1-\Delta_2}
\bar z_{12}^{\Delta_2+2}\bar z_{24}^{-2}\bar z_{41}^{-\Delta_2-1}
\cr&\bigg]
S[N^-_{\Delta_2+1}](z_4)
\cr\sim&-\f{i\k}{8}\f{K_{\Delta_2}}{2\pi}\tilde K_{2-\Delta_2-1}C_{124}\f{1}{(3-\Delta_2)(2-\Delta_2)}\int_{S_4}\bigg[\partial^2_{z_2}z_{12}^{\Delta_2-2}z_{41}^{1-\Delta_2}
\bar z_{12}^{\Delta_2+2}\bar z_{24}^{-2}\bar z_{41}^{-\Delta_2-1}
\cr&+2\partial_{z_2}z_{12}^{\Delta_2-2}z_{41}^{1-\Delta_2}
\bar z_{12}^{\Delta_2+2}\partial_{z_2}\bar z_{24}^{-2}\bar z_{41}^{-\Delta_2-1}
+z_{12}^{\Delta_2-2}z_{41}^{1-\Delta_2}
\bar z_{12}^{\Delta_2+2}\partial_{z_2}^2\bar z_{24}^{-2}\bar z_{41}^{-\Delta_2-1}\bigg]
S[N^-_{\Delta_2+1}](z_4)
\cr\sim&-\f{i\k}{8}\f{K_{\Delta_2}}{2\pi}\tilde K_{2-\Delta_2-1}C_{124}\f{1}{(3-\Delta_2)_2}\int_{S_4}\bigg[(3-\Delta_2)_2z_{12}^{\Delta_2-4}z_{41}^{1-\Delta_2}
\bar z_{12}^{\Delta_2+2}\bar z_{24}^{-2}\bar z_{41}^{-\Delta_2-1}
\cr&+4\pi(2-\Delta_2)z_{12}^{\Delta_2-3}z_{41}^{1-\Delta_2}
\bar z_{12}^{\Delta_2+2}\bar \partial_{z_4}\delta^2( z_2, z_4)\bar z_{41}^{-\Delta_2-1}
+2\pi z_{12}^{\Delta_2-2}z_{41}^{1-\Delta_2}
\cr&\times\bar z_{12}^{\Delta_2+2}\bar \partial_{z_4}\partial_{z_2}\delta^2( z_2, z_4)\bar z_{41}^{-\Delta_2-1}\bigg]
S[N^-_{\Delta_2+1}](z_4)
\cr\sim&-\f{i\k}{8}\f{K_{\Delta_2}}{2\pi}\tilde K_{2-\Delta_2-1}C_{124}\f{1}{(3-\Delta_2)_2}\bigg[
\cr&(3-\Delta_2)_2 z_{12}^{\Delta_2-4}\bar z_{12}^{\Delta_2+2}\int_{S_4}z_{41}^{1-\Delta_2}
\bar z_{24}^{-2}\bar z_{41}^{-\Delta_2-1}S[N^-_{\Delta_2+1}](z_4)
\cr&-4\pi(2-\Delta_2)z_{12}^{\Delta_2-3}
\bar z_{12}^{\Delta_2+2}\bar \partial_{z_2}\left(z_{21}^{1-\Delta_2}\bar z_{21}^{-\Delta_2-1}S[N^-_{\Delta_2+1}](z_2)\right)
\cr&-2\pi 
 z_{12}^{\Delta_2-2}\bar z_{12}^{\Delta_2+2}\partial_{z_2}\bar \partial_{z_2}\left(z_{21}^{1-\Delta_2}\bar z_{21}^{-\Delta_2-1}S[N^-_{\Delta_2+1}](z_2)\right)\bigg]
\cr\sim&\f{i\k}{8}\bigg[(3-\Delta_2)_2\f{1}{2\pi}z_{12}^{\Delta_2-4}\bar z_{12}^{\Delta_2+2}\int_{S_4}z_{41}^{1-\Delta_2}
\bar z_{24}^{-2}\bar z_{41}^{-\Delta_2-1}S[N^-_{\Delta_2+1}](z_4)
\cr&-z_{12}^{\Delta_2-3}
\bar z_{12}^{\Delta_2+2}(4-2\Delta_2+z_{12}\partial_{z_2})\bar \partial_{z_2}\left(z_{21}^{1-\Delta_2}\bar z_{21}^{-\Delta_2-1}S[N^-_{\Delta_2+1}](z_2)\right)
\bigg]
.
\ee

With further manipulations, this expression reduces to \eqref{ge1}.

\subsection{$N_{s_1} S[N^-_{\Delta_2}]$}\la{App:NSG}

The OPE \eqref{NsSN} is obtained from \eqref{NsND} computing
\be
  N_{s_1}(z_1,\bz_1) S[N^-_{\Delta_2}](z_2,\bz_2)
  \sim&-\f{i\k}{16\pi {s_1}!}\f{K_{\Delta_2}}{2\pi}\sum_{n=0}^{s_1}\f{1}{n!}(-)^{n-{s_1}}(\Delta_2-2)_{{s_1}-n}({s_1})_n\int_{S_3}\f{\bar z_{23}^{\Delta_2}}{z_{23}^{4-\Delta_2}}\f{z_{13}^{n+1}}{\bar z_{13}}\partial_{z_3}^n N^-_{\Delta_2+1-{s_1}}(z_3)
\cr
\sim&-\f{i\k}{16\pi {s_1}!}\f{K_{\Delta_2}}{2\pi}\f{\tilde K_{2-\Delta_2-1+{s_1}}}{2\pi}\sum_{n=0}^{s_1}\f{1}{n!}(-)^{n-{s_1}}(\Delta_2-2)_{{s_1}-n}({s_1})_n
\cr
&\times\int_{S_3}\f{\bar z_{23}^{\Delta_2}}{z_{23}^{4-\Delta_2}}\f{z_{13}^{n+1}}{\bar z_{13}}\partial_{z_3}^n\int_{S_4}\f{1}{z_{34}^{\Delta_2-{s_1}-1}}\f{1}{\bar z_{34}^{\Delta_2+3-{s_1}}}S[N^-_{\Delta_2+1-{s_1}}](z_4)
  \cr
 \sim&
 -\f{i\k}{16\pi {s_1}!}\f{K_{\Delta_2}}{2\pi}\f{\tilde K_{1-\Delta_2+{s_1}}}{2\pi}\sum_{n=0}^{s_1}\f{1}{n!}(-)^{n-{s_1}}(\Delta_2-2)_{{s_1}-n}({s_1})_n
\bigg[(1+{s_1}-\Delta_2)_{n}
\cr
&\times\int_{S_3}
\int_{S_4}\f{1}{z_{13}^{-n-1}}\f{1}{z_{23}^{4-\Delta_2}}\f{1}{z_{34}^{\Delta_2-{s_1}-1+n}}\f{1}{\bar z_{13}}\f{1}{\bar z_{23}^{-\Delta_2}}\f{1}{\bar z_{34}^{\Delta_2+3-{s_1}}}
S[N^-_{\Delta_2+1-{s_1}}](z_4)
\cr
&+\sum_{k=1}^n(1+{s_1}-\Delta_2)_{n-k}\binom{n}{k}
\cr&\times\int_{S_3}\int_{S_4}\f{1}{z_{13}^{-n-1}}\f{1}{z_{23}^{4-\Delta_2}}\f{1}{z_{34}^{\Delta_2-{s_1}-1+n-k}}\f{1}{\bar z_{13}}\f{1}{\bar z_{23}^{-\Delta_2}}\partial_{z_3}^k\left(\f{1}{\bar z_{34}^{\Delta_2+3-{s_1}}}\right)S[N^-_{\Delta_2+1-{s_1}}](z_4)\bigg].
\cr
\la{NsSN-app}
\ee


\subsection{$N_{s_1} S[N_{s_2}]$}\la{App:NsSNs}

If we take the conformally soft limit on the second entry in the OPE \eqref{NsSN-app}, we get
\be
N_{s_1}(z_1)S[N_{{s_2}}] (z_2)\sim&-\f{i\k}{16\pi {s_1}!}\f{K_{1-{s_2}}}{2\pi}\f{\tilde K_{{s_2}+{s_1}}}{2\pi}\sum_{n=0}^{s_1}\f{1}{n!}(-)^{n-{s_1}}(-{s_2}-1)_{{s_1}-n}({s_1})_n
\bigg[
({s_1}+{s_2})_{n}
\cr&\times\int_{S_3}\int_{S_4}\f{1}{z_{13}^{-n-1}}\f{1}{z_{23}^{3+{s_2}}}\f{1}{z_{34}^{-{s_2}-{s_1}+n}}\f{1}{\bar z_{13}}\f{1}{\bar z_{23}^{-1+{s_2}}}\f{1}{\bar z_{34}^{4-{s_2}-{s_1}}}S[N_{{s_1}+{s_2}-1}](z_4)
\cr&+\sum_{k=1}^n({s_1}+{s_2})_{n-k}\binom{n}{k}
\cr&\times\int_{S_3}\int_{S_4}\f{1}{z_{13}^{-n-1}}\f{1}{z_{23}^{3+{s_2}}}\f{1}{z_{34}^{-{s_2}-{s_1}+n-k}}\f{1}{\bar z_{13}}\f{1}{\bar z_{23}^{-1+{s_2}}}\partial_{z_3}^k\left(\f{1}{\bar z_{34}^{4-{s_2}-{s_1}}}\right)S[N_{{s_1}+{s_2}-1}](z_4)
\bigg]
\cr\sim&-\f{i\k}{16\pi {s_1}!}\f{K_{1-{s_2}}}{2\pi}\f{\tilde K_{{s_2}+{s_1}}}{2\pi}\sum_{n=0}^{s_1}\f{1}{n!}(-)^{n-{s_1}}(-{s_2}-1)_{{s_1}-n}({s_1})_n
\bigg[
({s_1}+{s_2})_{n}
\cr
&\times\int_{S_3}\int_{S_4}\f{1}{z_{13}^{-n-1}}\f{1}{z_{23}^{3+{s_2}}}\f{1}{z_{34}^{-{s_2}-{s_1}+n}}\f{1}{\bar z_{13}}\f{1}{\bar z_{23}^{-1+{s_2}}}\f{1}{\bar z_{34}^{4-{s_2}-{s_1}}}S[N_{{s_1}+{s_2}-1}](z_4)
\cr
&+\theta(3-s_1-s_2)\sum_{k=1}^n({s_1}+{s_2})_{n-k}\binom{n}{k}\cr&\times
\int_{S_3}\int_{S_4}\f{1}{z_{13}^{-n-1}}\f{1}{z_{23}^{3+{s_2}}}\f{1}{z_{34}^{-{s_2}-{s_1}+n-k}}\f{1}{\bar z_{13}}\f{1}{\bar z_{23}^{-1+{s_2}}}
\f{2\pi}{(3-{s_1}-{s_2})!}\cr&\times(-)^{k-1}\partial_{z_4}^{k-1}\bar \partial_{z_4}^{3-{s_1}-{s_2}}\left(\delta^2(z_3,z_4)\right)S[N_{{s_1}+{s_2}-1}](z_4)
\bigg]
\cr
\sim&-\f{i\k}{16\pi {s_1}!}\f{K_{1-{s_2}}}{2\pi}\f{\tilde K_{{s_2}+{s_1}}}{2\pi}\sum_{n=0}^{s_1}\f{1}{n!}(-)^{n-{s_1}}(-{s_2}-1)_{{s_1}-n}({s_1})_n
\cr
&\times
\bigg[({s_1}+{s_2})_{n}
\int_{S_3}\int_{S_4}\f{1}{z_{13}^{-n-1}}\f{1}{z_{23}^{3+{s_2}}}\f{1}{z_{34}^{-{s_2}-{s_1}+n}}\f{1}{\bar z_{13}}\f{1}{\bar z_{23}^{-1+{s_2}}}\f{1}{\bar z_{34}^{4-{s_2}-{s_1}}}S[N_{{s_1}+{s_2}-1}](z_4)
\cr
&+\theta(3-s_1-s_2)\sum_{k=1}^n(-)^{k-1}({s_1}+{s_2})_{n-k}\binom{n}{k}\sum_{p=0}^{k-1}\binom{k-1}{p}(-)^{3-{s_1}-{s_2}}
\cr&\times\sum_{q=0}^{3-{s_1}-{s_2}}\binom{3-{s_1}-{s_2}}{q}\int_{S_3}\int_{S_4}
\f{1}{z_{13}^{-n-1}}\f{1}{z_{23}^{3+{s_2}}}\bar\partial_{z_4}^{3-{s_1}-{s_2}-q}\partial_{z_4}^{k-1-p}\left(z_{34}^{{s_2}+{s_1}-n+k}\right)
\cr
&\times\f{1}{\bar z_{13}}\f{1}{\bar z_{23}^{-1+{s_2}}}
\f{2\pi}{(3-{s_1}-{s_2})!}\delta^2(z_3,z_4)\bar\partial_{z_4}^q\partial_{z_4}^pS[N_{{s_1}+{s_2}-1}](z_4)\bigg]\,,
\ee
where $\theta(x)$ is the Heaviside function; 
the sum over $q$ in the second contribution in the square brackets above is different from zero only for $q=3-s_1-s_2$, and we thus arrive at
\be
N_{s_1}(z_1)S[N_{{s_2}}] (z_2)
\sim&-\f{i\k}{16\pi {s_1}!}\f{K_{1-{s_2}}}{2\pi}\f{\tilde K_{{s_2}+{s_1}}}{2\pi}\sum_{n=0}^{s_1}\f{1}{n!}(-)^{n-{s_1}}(-{s_2}-1)_{{s_1}-n}({s_1})_n
\cr
&\times
\bigg[({s_1}+{s_2})_{n} \int_{S_3}
\int_{S_4}\f{1}{z_{13}^{-n-1}}\f{1}{z_{23}^{3+{s_2}}}\f{1}{z_{34}^{-{s_2}-{s_1}+n}}\f{1}{\bar z_{13}}\f{1}{\bar z_{23}^{-1+{s_2}}}\f{1}{\bar z_{34}^{4-{s_2}-{s_1}}}S[N_{{s_1}+{s_2}-1}](z_4)
\cr&+\theta(3-s_1-s_2,0)\sum_{k=1}^n(-)^{k-1}({s_1}+{s_2})_{n-k}\binom{n}{k}\sum_{p=0}^{k-1}\binom{k-1}{p}(-)^{3-{s_1}-{s_2}}
\cr
&\times\int_{S_3}\int_{S_4}
\f{1}{z_{13}^{-n-1}}\f{1}{z_{23}^{3+{s_2}}}(-)^{k-1-p}({s_2}+{s_1}-n+k)_{k-1-p}z_{34}^{{s_2}+{s_1}-n+1+p}
\cr
&\times\f{1}{\bar z_{13}}\f{1}{\bar z_{23}^{-1+{s_2}}}
\f{2\pi}{(3-{s_1}-{s_2})!}\delta^2(z_3,z_4)\bar\partial_{z_4}^{3-{s_1}-{s_2}}\partial_{z_4}^pS[N_{{s_1}+{s_2}-1}](z_4)\bigg]
\cr\sim&-\f{i\k}{16\pi {s_1}!}\f{K_{1-{s_2}}}{2\pi}\f{\tilde K_{{s_2}+{s_1}}}{2\pi}\sum_{n=0}^{s_1}\f{1}{n!}(-)^{n-{s_1}}(-{s_2}-1)_{{s_1}-n}({s_1})_n
({s_1}+{s_2})_{n}
\cr&\times\int_{S_4}
\int_{S_3}\f{1}{z_{13}^{-n-1}}\f{1}{z_{23}^{3+{s_2}}}\f{1}{z_{34}^{-{s_2}-{s_1}+n}}\f{1}{\bar z_{13}}\f{1}{\bar z_{23}^{-1+{s_2}}}\f{1}{\bar z_{34}^{4-{s_2}-{s_1}}}S[N_{{s_1}+{s_2}-1}](z_4)
\,.
\ee 
Here, in the last step, we use that the power of $z_{34}$ is such that ${s_2}+{s_1}-n+1+p\ge {s_2}+1\ge 1$, so integrating it against the Dirac delta always yields zero. This derivation proves that the potential additional terms one has when $\Delta_2=1-s_2\ge {s_1}-2$ do never contribute for  ${s_1},{s_2}\ge0$.


\subsection{$S[N_{s_1}] S[N_{\Delta_2}]$}\la{App:SNsSND}

The OPE \eqref{SNsSND} can be derived starting from the first line of \eqref{NsSN-app} and computing
\be
S[N_{s_1}](z_1)S[N^-_{\Delta_2}](z_2)
\sim&-\f{i\k}{16\pi {s_1}!}\lim_{\Delta_1\rightarrow 1-s_1}\f{K_{\Delta_2}}{2\pi}\f{K_{\Delta_1}}{2\pi}\sum_{n=0}^{s_1}\f{1}{(s_1-n)!}(-)^{n}(\Delta_2-2)_{n}({s_1})_{s_1-n}
\cr
&\times\int_{S_4}\int_{S_3}\f{1}{z_{14}^{4-\Delta_1}}\f{1}{z_{43}^{\Delta_1+n-2}}\f{1}{\bar z_{14}^{-\Delta_1}}\f{1}{\bar z_{43}}\f{\bar z_{23}^{\Delta_2}}{z_{23}^{4-\Delta_2}}\partial_{z_3}^{s_1-n}N^-_{\Delta_2+1-{s_1}}(z_3)
\cr
\sim&-\f{i\k}{16\pi {s_1}!}\lim_{\Delta_1\rightarrow 1-s_1}\f{K_{\Delta_2}}{2\pi}\f{K_{\Delta_1}}{2\pi}\sum_{n=0}^{s_1}\f{1}{(s_1-n)!}(-)^{n}(\Delta_2-2)_{n}({s_1})_{s_1-n}
\cr
&\times\int_{S_4}\int_{S_3}\f{1}{z_{14}^{4-\Delta_1}}\f{1}{(\Delta_1+n-1)_n}\partial_{z_3}^n\f{1}{z_{43}^{\Delta_1-2}}\f{1}{\bar z_{14}^{-\Delta_1}}\f{1}{\bar z_{43}}\f{\bar z_{23}^{\Delta_2}}{z_{23}^{4-\Delta_2}}\partial_{z_3}^{s_1-n}N^-_{\Delta_2+1-{s_1}}(z_3)
\cr
\sim&-\f{i\k}{16\pi {s_1}!}\lim_{\Delta_1\rightarrow 1-s_1}\f{K_{\Delta_2}}{2\pi}\f{K_{\Delta_1}}{2\pi}\sum_{n=0}^{s_1}\f{1}{(s_1-n)!}(-)^{n}(\Delta_2-2)_{n}({s_1})_{s_1-n}\f{1}{(2-\Delta_1)_n}
\cr
&\times\int_{S_4}\int_{S_3}\f{1}{z_{14}^{4-\Delta_1}}\f{1}{z_{43}^{\Delta_1-2}}\f{1}{\bar z_{14}^{-\Delta_1}}\f{1}{\bar z_{43}^{2+\Delta_1-2+s_1}}\partial_{z_3}^n\left(\f{\bar z_{23}^{\Delta_2}}{z_{23}^{4-\Delta_2}}\partial_{z_3}^{s_1-n}N^-_{\Delta_2+1-{s_1}}(z_3)\right)
\cr
\sim&-\f{i\k}{16\pi {s_1}!}\lim_{\Delta_1\rightarrow 1-s_1}\f{K_{\Delta_2}}{2\pi}\f{K_{\Delta_1}}{2\pi}\sum_{n=0}^{s_1}\f{1}{(s_1-n)!}(-)^{n}(\Delta_2-2)_{n}({s_1})_{s_1-n}
\cr
&\times\f{1}{(2-\Delta_1)_n}\f{1}{(\Delta_1+s_1-1)_{s_1-2}}\int_{S_4}\int_{S_3}\f{1}{z_{14}^{4-\Delta_1}}\f{1}{z_{43}^{\Delta_1-2}}\f{1}{\bar z_{14}^{-\Delta_1}}\bar\partial_{z_3}^{s_1-2}\f{1}{\bar z_{43}^{2+\Delta_1}}
\cr
&\times\partial_{z_3}^n\left(\f{\bar z_{23}^{\Delta_2}}{z_{23}^{4-\Delta_2}}\partial_{z_3}^{s_1-n}N^-_{\Delta_2+1-{s_1}}(z_3)\right)
\cr
\sim&-\f{i\k}{16\pi {s_1}!}\lim_{\Delta_1\rightarrow 1-s_1}\f{K_{\Delta_2}}{2\pi}\f{K_{\Delta_1}}{2\pi}\sum_{n=0}^{s_1}\f{1}{(s_1-n)!}(-)^{n}(\Delta_2-2)_{n}({s_1})_{s_1-n}
\cr
&\times\f{1}{(2-\Delta_1)_n}\f{1}{(-\Delta_1-2)_{s_1-2}}\int_{S_4}\int_{S_3}\f{1}{z_{14}^{4-\Delta_1}}\f{1}{z_{43}^{\Delta_1-2}}\f{1}{\bar z_{14}^{-\Delta_1}}\f{1}{\bar z_{43}^{2+\Delta_1}}
\cr
&\times \bar\partial_{z_3}^{s_1-2}\partial_{z_3}^n\left(\f{\bar z_{23}^{\Delta_2}}{z_{23}^{4-\Delta_2}}\partial_{z_3}^{s_1-n}N^-_{\Delta_2+1-{s_1}}(z_3)\right).
\la{SNsSND-app}
\ee
{Here, in taking the shadow of $N_{s_1}$, we defined the latter as the limit of an operator with conformal dimension $\Delta_1$, for $\Delta_1\rightarrow1-s_1$. This definition has to be understood as a regulatization procedure we adopted in order to properly cancel out divergent $\Gamma$-functions that arise in the calculation at $\Delta_1=1-s_1$. To put  the integral in its final form, it was essential to rewrite the exponent of $\bar z_{43}$ as $1=\lim_{\Delta_1\rightarrow1-s_1} (\Delta_1+s_1)$ in the 3rd step. Given this final expression,} 
we can now
use \eqref{KGR} and the conformal integral identity
\cite{Dolan:2011dv}
\be
\f{1}{2\pi}\int_{S_3}
\f{1}{z_{13}^{2h}}
\f{1}{z_{23}^{2-2h}}
\f{1}{\bar z_{13}^{\bar h}}
\f{1}{\bar z_{23}^{2-\bar h}}
=2\pi \delta^2(z_1,z_2)
\f{\Gamma(2\bar h-1)\Gamma(1-2\bar h)}{\Gamma(2-2h)\Gamma(2h)}
=2\pi \delta^2(z_1,z_2)
\f{\Gamma(2 h-1)\Gamma(1-2 h)}{\Gamma(2-2\bar h)\Gamma(2\bar h)}\,,
\la{intid-2}
\ee
to arrive at
\be
S[N_{s_1}](z_1)S[N^-_{\Delta_2}](z_2)
\sim&-\f{i\k}{8 {s_1}!}\lim_{\Delta_1\rightarrow 1-s_1}\f{K_{\Delta_2}}{2\pi}(-)^{3-\Delta_1}\sum_{n=0}^{s_1}\f{1}{(s_1-n)!}(-)^{n}(\Delta_2-2)_{n}({s_1})_{s_1-n}
\cr
&\times\f{1}{(2-\Delta_1)_n}\f{1}{(-\Delta_1-2)_{s_1-2}}\f{\Gamma(3-\Delta_1)}{\Gamma(2+\Delta_1)}
\bar\partial_{z_1}^{s_1-2}\partial_{z_1}^n\left(\f{\bar z_{21}^{\Delta_2}}{z_{21}^{4-\Delta_2}}\partial_{z_1}^{s_1-n}N^-_{\Delta_2+1-{s_1}}(z_1)\right)
\cr\sim&-\f{i\k}{8 {s_1}!}\lim_{\Delta_1\rightarrow 1-s_1}\f{K_{\Delta_2}}{2\pi}(-)^{3-\Delta_1}\sum_{n=0}^{s_1}\f{1}{(s_1-n)!}(-)^{n+s_1}(\Delta_2-2)_{n}({s_1})_{s_1-n}
\cr
&\times\f{\Gamma(3-\Delta_1-n)}{\Gamma(s_1+\Delta_1)}
\bar\partial_{z_1}^{s_1-2}\partial_{z_1}^n\left(\f{\bar z_{21}^{\Delta_2}}{z_{21}^{4-\Delta_2}}\partial_{z_1}^{s_1-n}N^-_{\Delta_2+1-{s_1}}(z_1)\right)
\cr
\sim&-\f{i\k}{8 {s_1}!}\f{K_{\Delta_2}}{2\pi}\sum_{n=0}^{s_1}(-)^{s_1-n}(\Delta_2-2)_{s_1-n}({s_1})_{n}(n+1)
\bar\partial_{z_1}^{s_1-2}\partial_{z_1}^{s_1-n}\left(\f{\bar z_{12}^{\Delta_2}}{z_{12}^{4-\Delta_2}}\partial_{z_1}^{n}N^-_{\Delta_2+1-{s_1}}(z_1)\right)\,,
\cr
\ee
which is identical to \eqref{tNSG}.


\subsection{$\bN_{s_1} \tN_{s_2}$}\la{App:bNtN}

Starting from the OPE \eqref{sbng} and using the diamond in Fig. \ref{fig:CD-GR-1}, we can compute first the OPE of a positive helicity soft graviton and a shadow charge. Supposing initially $ s_2 \ge s_1+4$, we have
\be
\bar N_{s_1}(z_1)S[ q^1_{s_2}](z_2)
\sim&\f{\k}{8i}\sum_{n=0}^{s_1}\f{(s_1+s_2-n-4)_{s_1-n}}{(s_1-n)!}\f{1}{n!}\bar\partial^{s_2-2}_{z_2}\bigg(\f{\bar z_{12}^{n+1}}{z_{12}}\bar \partial_{z_2}^nN_{s_1+s_2-1}(z_2)\bigg)
\cr
\sim&\f{\k}{8i}\sum_{n=0}^{s_1}\sum_{k=0}^{s_2-2}\f{(s_1+s_2-n-4)_{s_1-n}}{(s_1-n)!}\f{1}{n!}\f{(s_2-2)_k}{k!}\bar\partial^{k}_{z_2}\bigg(\f{\bar z_{12}^{n+1}}{z_{12}}\bigg)\bar \partial_{z_2}^{s_2-2-k+n}N_{s_1+s_2-1}(z_2)
\cr
\sim&\f{\k}{8i}\bigg[2\pi\sum_{n=0}^{s_1}\sum_{k=n+2}^{s_2-2}(-)^n(n+1)\f{(s_1+s_2-n-4)_{s_1-n}}{(s_1-n)!}\f{(s_2-2)_k}{k!}
\cr&\times\bar\partial^{k-2-n}_{z_2}\delta^2(z_1,z_2)\bar \partial_{z_2}^{s_2-2-k+n}N_{s_1+s_2-1}(z_2)
\cr&
+\sum_{n=0}^{s_1}\sum_{k=0}^{n+1}\f{(n+1)}{(n+1-k)!}\f{(s_1+s_2-n-4)_{s_1-n}}{(s_1-n)!}\f{(s_2-2)_k}{k!}(-)^k\f{\bar z_{12}^{n+1-k}}{z_{12}}\bar \partial_{z_2}^{s_2-2-k+n}N_{s_1+s_2-1}(z_2)\bigg]
\cr
\sim&\f{\k}{8i}\bigg[2\pi\sum_{k=2}^{s_2-2}\sum_{n=0}^{k-2}(-)^n(n+1)\f{(s_1+s_2-n-4)_{s_2-4}}{(s_2-4)!}\f{(s_2-2)_k}{k!}
\cr&\times\bar\partial^{k-2-n}_{z_2}\delta^2(z_1,z_2)\bar \partial_{z_2}^{s_2-2-k+n}N_{s_1+s_2-1}(z_2)
\cr
&+\sum_{k=0}^{s_1+1}\left(\sum_{n=0}^{s_1+1-k}(-)^{s_1-k+1}\f{(n+k)}{k!}\f{(3-s_2)_{s_1-n-k+1}}{(s_1-n-k+1)!}\f{(s_2-2)_{n}}{n!}\right)\f{\bar z_{12}^{k}}{z_{12}}\bar \partial_{z_2}^{s_2-3+k}N_{s_1+s_2-1}(z_2)\bigg]
\,.
\cr
\ee
Let us now compute the two double sums above separately. We start with
\be
&\sum_{k=2}^{s_2-2}\sum_{n=0}^{k-2}(-)^n(n+1)\f{(s_1+s_2-n-4)_{s_1-n}}{(s_1-n)!}\f{(s_2-2)_k}{k!}
\cr&\times\bar\partial^{k-2-n}_{z_2}\delta^2(z_1,z_2)\bar \partial_{z_2}^{s_2-2-k+n}N_{s_1+s_2-1}(z_2)
\cr
&=\sum_{n=0}^{s_2-4}
\left(\sum_{k=n}^{s_2-4}(-)^{n+k}(k-n+1)\f{(s_1+s_2-k+n-4)_{s_1-k+n}}{(s_1-k+n)!}\f{(s_2-2)_{k+2}}{(k+2)!}
\right)
\cr
&\times\bar\partial^{n}_{z_2}\delta^2(z_1,z_2)\bar \partial_{z_2}^{s_2-4-n}N_{s_1+s_2-1}(z_2)
\cr
&=\sum_{n=0}^{s_2-4}
\f{1}{(s_1+n+1)(s_1+n+2)}
\f{(s_1+s_2-2)!}{s_1! n!(s_2-n-4)!}
\bar\partial^{n}_{z_2}\delta^2(z_1,z_2)\bar \partial_{z_2}^{s_2-4-n}N_{s_1+s_2-1}(z_2)\,.
\ee
The second double sum gives
\be
&\sum_{k=0}^{s_1+1}\left(\sum_{n=0}^{s_1+1-k}(-)^{s_1-k+1}\f{(n+k)}{k!}\f{(3-s_2)_{s_1-n-k+1}}{(s_1-n-k+1)!}\f{(s_2-2)_{n}}{n!}\right)\f{\bar z_{12}^{k}}{z_{12}}\bar \partial_{z_2}^{s_2-3+k}N_{s_1+s_2-1}(z_2)
\cr
&=\sum_{k=0}^{s_1+1}\f{(s_2-2)}{k!}(-)^{s_1-k+1}\f{(0)_{s_1-k}}{(s_1-k)!}\f{\bar z_{12}^{k}}{z_{12}}\bar \partial_{z_2}^{s_2-3+k}N_{s_1+s_2-1}(z_2)
\cr
&+\sum_{k=1}^{s_1+1}\f{1}{(k-1)!}(-)^{s_1-k+1}\f{(1)_{s_1-k+1}}{(s_1-k+1)!}\f{\bar z_{12}^{k}}{z_{12}}\bar \partial_{z_2}^{s_2-3+k}N_{s_1+s_2-1}(z_2)\,,
\ee
where, on the RHS above, in the first sum only the $k=s_1$ contribution survives, while in the second sum only $k=s_1+1,s_1$.
We thus finally arrive at
\be
\bar N_{s_1}(z_1)S[ q^1_{s_2}](z_2)
\sim&\f{\k}{8i}\bigg[2\pi(s_1+s_2-2)_2\binom{s_1+s_2-4}{s_1}\sum_{n=0}^{s_2-4}
\f{1}{(s_1+n+2)_2}
\binom{s_2-4}{n}
\bar\partial^{n}_{z_2}\delta^2(z_1,z_2)
\cr&\times\bar \partial_{z_2}^{s_2-4-n}N_{s_1+s_2-1}(z_2)
+\f{1}{s_1!}\f{\bar z_{12}^{s_1}}{z_{12}}(2-s_2-s_1+\bar z_{12}\bar\partial_{z_2})S[ q^1_{s_1+s_2-1}](z_2)
\bigg]
.
\la{ssbngbis-app}
\ee

For the case $ s_1+3\geq s_2$, one can instead split the sums in the second passage as $\sum_{k=0}^{s_2-2}\sum_{n=0}^{s_1}=\sum_{k=2}^{s_2-2}\sum_{n=0}^{k-2} +\sum_{k=0}^{s_2-2}\sum_{n=k-1}^{s_1}$ and, with similar manipulations, arrive at the same final expression. 
Taking $\partial_{z_2}^{s_2+2}$ derivatives of \eqref{ssbngbis-app}, we arrive at the OPE 
\be
\bar N_{s_1}(z_1,\bz_1)\tilde N_{s_2}(z_2,\bz_2)
&\sim
\f{\k}{8i}\bigg[2\pi(s_1+s_2-2)_2\binom{s_1+s_2-4}{s_1}\sum_{n=0}^{s_2-4}
\f{1}{(s_1+n+2)_2}
\binom{s_2-4}{n}
\cr
&\times\partial_{2}^{s_2+2}\left(\bar\partial^{n}_{2}\delta^2(z_1,z_2)\bar \partial_{2}^{s_2-4-n}N_{s_1+s_2-1}(z_2,\bz_2)\right)
\cr
&
+\f{1}{s_1!}\partial_{z_2}^{s_2+2}\left(\f{\bar z_{12}^{s_1}}{z_{12}}(2-s_2-s_1+\bar z_{12}\bar\partial_{2})S[q^1_{s_1+s_2-1}](z_2,\bz_2)\right)
\bigg]\,.
\la{bNtN-app}
\ee

If we restrict to the case $s_1=0$ in \eqref{bNtN-app}, we get
\be
\bar N_{0}(z_1)\tilde N_{s_2}(z_2)\sim&\f{\k}{8i}\bigg[
c.t.
+\partial_{z_2}^{s_2+2}\left(\f{1}{z_{12}}(2-s_2+\bar z_{12}\bar\partial_{z_2})S[ q^1_{s_2-1}](z_2)\right)
\bigg]
\cr\sim&\f{\k}{8i}\bigg[c.t.
+\sum_{k=2}^{s_2+2}\binom{s_2+2}{k}\partial_{z_2}^{k}\f{1}{z_{12}}(2-s_2+\bar z_{12}\bar\partial_{z_2})\partial_{z_2}^{s_2+2-k}S[q^1_{s_2-1}](z_2)
\cr&
+z_{12}^{-2}(s_2+2+z_{12}\partial_{z_2})(2-s_2+\bar z_{12}\bar\partial_{z_2})\tilde N_{s_2-1}(z_2)
\bigg]
\cr\sim&\f{\k}{8i}\bigg[c.t.
-\f{1}{2\pi}(s_2+2)_2z_{12}^{-3-s_2}\int_{S_3}\left(\bar z_{12}+(s_2-2)\bar z_{23}\right)\f{z_{13}^{s_2}}{\bar z_{23}^2}\tilde N_{s_2-1}(z_3)
\cr&
+z_{12}^{-2}(s_2+2+z_{12}\partial_{z_2})(2-s_2+\bar z_{12}\bar\partial_{z_2})\tilde N_{s_2-1}(z_2)
\bigg],
\la{nsdgen-app}
\ee
where $c.t.$ stands for the contact terms given by the sum contribution on the RHS of \eqref{bNtN-app}, and we manipulated the $S[q^1_{s_2-1}]$ terms as follows
\be
&\f{1}{2\pi}\sum_{k=2}^{s_2+2}\binom{s_2+2}{k}k!z_{12}^{-(1+k)}(2-s_2+\bar z_{12}\bar\partial_{z_2})\partial_{z_2}^{s_2+2-k}\f{1}{s_2!}\int_{S_3}\f{z_{23}^{s_2}}{\bar z_{23}}\tilde N_{s_2-1}(z_3)
\cr=&\f{1}{2\pi}(s_2+2)_2\sum_{k=0}^{s_2}\f{s_2!}{k!(s_2-k)!}z_{12}^{-(3+k)}(2-s_2+\bar z_{12}\bar\partial_{z_2})\int_{S_3}\f{z_{23}^{k}}{\bar z_{23}}\tilde N_{s_2-1}(z_3)
\cr=&-\f{1}{2\pi}(s_2+2)_2z_{12}^{-3-s_2}\int_{S_3}\left(\bar z_{12}+(s_2-2)\bar z_{23}\right)\f{z_{13}^{s_2}}{\bar z_{23}^2}\tilde N_{s_2-1}(z_3),
\ee
where in the last passage we used
\be
z_{13}^{s_2}=(z_{12}+z_{23})^{s_2}=\sum_{k=0}^{s_2}\binom{s_2}{k}z_{12}^{s_2-k}z_{23}^k.
\ee

\subsection{$\bN_{s_1} S[N^-_{s_2}]$}\la{App:bNSN}

Starting from the OPE \eqref{bng} and taking the shadow of the second entry for a general $\Delta_2$, we get
\be
\bar N_{s_1}(z_1)S[N^-_{\Delta_2}](z_2)
\sim&\f{\k}{8i{s_1}!}\f{K_{\Delta_2}}{2\pi}\f{\tilde K_{2-(\Delta_2+1-{s_1})}}{2\pi}\sum_{n=0}^{s_1}(-)^{n-{s_1}}(\Delta_2+2)_{{s_1}-n}({s_1})_n\f{1}{n!}\int_{S_3}\f{\bar z_{23}^{\Delta_2}}{z_{23}^{4-\Delta_2}}\f{\bar z_{13}^{n+1}}{z_{13}}
\cr
&\times\int_{S_4}\bigg[\sum_{k=0}^n\binom{n}{k}\bar \partial_{z_3}^k\left(\f{1}{z_{34}^{\Delta_2-1-{s_1}}}\right)\bar \partial_{z_3}^{n-k}\left(\f{1}{\bar z_{34}^{3+\Delta_2-{s_1}}}\right)\bigg]S[N^-_{\Delta_2+1-{s_1}}](z_4)
\cr\sim&\f{\k}{8i{s_1}!}\f{K_{\Delta_2}}{2\pi}\f{\tilde K_{2-(\Delta_2+1-{s_1})}}{2\pi}\sum_{n=0}^{s_1}(-)^{n-{s_1}}(\Delta_2+2)_{{s_1}-n}({s_1})_n\f{1}{n!}\int_{S_3}\f{\bar z_{23}^{\Delta_2}}{z_{23}^{4-\Delta_2}}\f{\bar z_{13}^{n+1}}{z_{13}}
\cr
&\times\int_{S_4}\bigg[\sum_{k=0}^n\binom{n}{k}\bar \partial_{z_3}^k\left(\f{1}{z_{34}^{\Delta_2-1-{s_1}}}\right)({s_1}-\Delta_2-3)_{n-k}\left(\f{1}{\bar z_{34}^{3+\Delta_2-{s_1}+n-k}}\right)\bigg]S[N^-_{\Delta_2+1-{s_1}}](z_4)
\cr
\sim&\f{\k}{8i{s_1}!}\f{K_{\Delta_2}}{2\pi}\f{\tilde K_{2-(\Delta_2+1-{s_1})}}{2\pi}\sum_{n=0}^{s_1}\f{1}{n!}(-)^{n-{s_1}}(\Delta_2+2)_{{s_1}-n}({s_1})_n({s_1}-\Delta_2-3)_n
\cr&\times\int_{S_3}\int_{S_4}\f{1}{z_{13}}\f{1}{z_{23}^{4-\Delta_2}}\f{1}{z_{34}^{-{s_1}+\Delta_2-1}}\f{1}{\bar z_{13}^{-n-1}}\f{1}{\bar z_{23}^{-\Delta_2}}
\f{1}{\bar z_{34}^{3+\Delta_2-{s_1}+n}}S[N^-_{\Delta_2+1-{s_1}}](z_4)
\cr
&+\f{\k}{8i{s_1}!}\f{K_{\Delta_2}}{2\pi}\f{\tilde K_{2-(\Delta_2+1-{s_1})}}{2\pi}\sum_{n=0}^{s_1}(-)^{n-{s_1}}(\Delta_2+2)_{{s_1}-n}({s_1})_n\f{1}{n!}\sum_{k=1}^n({s_1}-\Delta_2-3)_{n-k}\binom{n}{k}
\cr
&\times\int_{S_3}\int_{S_4}\f{1}{z_{13}}\f{1}{z_{23}^{4-\Delta_2}}\bar \partial_{z_3}^k\left(\f{1}{z_{34}^{\Delta_2-1-{s_1}}}\right)\f{1}{\bar z_{13}^{-n-1}}\f{1}{\bar z_{23}^{-\Delta_2}}
\f{1}{\bar z_{34}^{3+\Delta_2-{s_1}+n-k}}S[N^-_{\Delta_2+1-{s_1}}](z_4).
\la{bNSN-app}
\ee
We see that the second contribution in the last passage above would vanish for $\Re \Delta_2\leq 1$.  
{When taking the second soft limit, this condition is always fulfilled (since we have $\Delta_2=1-s_2\le1$), so the extra term always drops out}. In fact, we have
\be
\bar N_{s_1}(z_1)S[N_{{s_2}}](z_2)\sim&\f{\k}{8i{s_1}!}\f{K_{1-{s_2}}}{2\pi}\f{\tilde K_{2-(2-{s_2}-{s_1})}}{2\pi}\sum_{n=0}^{s_1}\f{1}{n!}(-)^{n-{s_1}}(3-{s_2})_{{s_1}-n}(s_1)_n({s_1}+{s_2}-4)_n
\cr&\times\int_{S_3}\int_{S_4}\f{1}{z_{13}}\f{1}{z_{23}^{3+{s_2}}}\f{1}{z_{34}^{-{s_1}-{s_2}}}\f{1}{\bar z_{13}^{-n-1}}\f{1}{\bar z_{23}^{-\Delta_2}}
\f{1}{\bar z_{34}^{4-{s_2}-{s_1}+n}}S[N_{{s_1}+{s_2}-1}](z_4)
\cr&+\f{\k}{8i{s_1}!}\f{K_{1-{s_2}}}{2\pi}\f{\tilde K_{2-(2-{s_2}-{s_1})}}{2\pi}\sum_{n=0}^{s_1}(-)^{n-{s_1}}(3-{s_2})_{{s_1}-n}({s_1})_n\f{1}{n!}\sum_{k=1}^n({s_1}+{s_2}-4)_{n-k}\binom{n}{k}
\cr&\times\int_{S_3}\int_{S_4}\f{1}{z_{13}}\f{1}{z_{23}^{3+{s_2}}}\bar \partial_{z_3}^k\left(\f{1}{z_{34}^{-{s_2}-{s_1}}}\right)\f{1}{\bar z_{13}^{-n-1}}\f{1}{\bar z_{23}^{-1+{s_2}}}
\f{1}{\bar z_{34}^{4-{s_2}-{s_1}+n-k}}S[N_{{s_1}+{s_2}-1}](z_4)\,,
\ee
and for all $s_1,s_2\geq0$ 
 of our interest, the terms in the third and fourth rows above vanish. Therefore, restricting to the $s_1=0$ case, we have
 \be
\bar N_0(z_1)S[N_{s_2}](z_2)
\sim&\f{\k}{8i}
\bigg[
-(s_2+2)_2\f{1}{2\pi}z_{12}^{-s_2-3}\bar z_{12}^{3-s_2}\int_{S_4}z_{41}^{s_2}\bar z_{24}^{-2}\bar z_{41}^{-2+s_2}S[N_{s_2-1}](z_4)
\cr&+z_{12}^{-2}\bar z_{12}^{3-s_2}\bigg(s_2+2+z_{12}\partial_{z_2}\bigg)\bar\partial_{z_2}\left(\bar z_{12}^{-2+s_2}S[N_{s_2-1}](z_2)\right)
\bigg]
\cr
\sim&\f{\k}{8i}
\bigg[
-(s_2+2)_2\f{1}{2\pi}z_{12}^{-s_2-3}\bar z_{12}^{3-s_2}\int_{S_4}z_{41}^{s_2}\bar z_{24}^{-2}\bar z_{41}^{-2+s_2}S[N_{s_2-1}](z_4)
\cr&+z_{12}^{-2}(s_2+2+z_{12}\partial_{z_2})(2-s_2+\bar z_{12}\bar \partial_{z_2})\left( S[N_{s_2-1}](z_2)\right)
\bigg].
\la{nssgen}
\ee

 Let us manipulate the non-local term
\be
&-(s_2+2)_2\f{1}{2\pi}z_{12}^{-s_2-3}\bar z_{12}^{3-s_2}\int_{S_4}z_{14}^{s_2}\bar z_{24}^{-2}\bar z_{14}^{-2+s_2}S[N_{s_2-1}](z_4)
\cr=&-(s_2+2)_2\f{1}{2\pi}z_{12}^{-s_2-3}\bar z_{12}^{3-s_2}\sum_{k=0}^{s_2-2}\binom{s_2-2}{k}\int_{S_4}z_{14}^{s_2}\bar z_{24}^{-2}\bar z_{12}^{k}\bar z_{24}^{s_2-2-k}S[N_{s_2-1}](z_4)
\cr=&-(s_2+2)_2\f{1}{2\pi}z_{12}^{-s_2-3}\int_{S_4}(\bar z_{12}+(s_2-2)\bar z_{24}) \f{z_{41}^{s_2}}{\bar z_{24}^{2}}S[N_{s_2-1}](z_4)
\cr&-(s_2+2)_2\f{1}{2\pi}z_{12}^{-s_2-3}\sum_{k=0}^{s_2-4}\binom{s_2-2}{k}\bar z_{12}^{3-s_2+k}\int_{S_4}z_{41}^{s_2}\bar z_{24}^{s_2-4-k}S[N_{s_2-1}](z_4).
\la{nlnssgen}
\ee

The last row above can be written as
\be
&-(s_2+2)_2\f{1}{2\pi}z_{12}^{-s_2-3}\sum_{k=0}^{s_2-4}\binom{s_2-2}{k}\bar z_{12}^{3-s_2+k}\int_{S_4}z_{41}^{s_2}\bar z_{24}^{s_2-4-k}S[N_{s_2-1}](z_4)
\cr
=&-(s_2+2)_2\f{1}{2\pi}z_{12}^{-s_2-3}\sum_{k=0}^{s_2-4}\sum_{n=0}^{s_2-4-k}(-)^{s_2+n}\binom{s_2-2}{k}\binom{s_2-4-k}{n}\bar z_{12}^{3-s_2+k+n}\int_{S_4}z_{14}^{s_2}\bar z_{14}^{s_2-4-k-n}S[N_{s_2-1}](z_4)
\cr
=&-(s_2+2)_2\f{1}{2\pi}z_{12}^{-s_2-3}\sum_{k=0}^{s_2-4}\sum_{n=0}^{s_2-4-k}(-)^{s_2+n}\binom{s_2-2}{k}\f{(s_2-4-k)!}{n!(s_2-4)!}\bar z_{12}^{3-s_2+k+n}
\bar\partial_{z_1}^{k+n}\int_{S_4}z_{14}^{s_2}\bar z_{14}^{s_2-4}S[N_{s_2-1}](z_4)
\cr
=&(s_2+2)_2\f{1}{2\pi}z_{12}^{-s_2-3}\sum_{k=0}^{s_2-4}\sum_{n=0}^{s_2-4-k}(-)^{n}\binom{s_2-2}{k}\f{(s_2-4-k)!}{n!(s_2-4)!}\bar z_{12}^{3-s_2+k+n}
\bar\partial_{z_1}^{k+n}\left(\lim_{\Delta\rightarrow 2-s_2}\f{\Gamma(3-\Delta)}{\Gamma(2+\Delta)}N_{s_2-1}(z_1)\right)
\cr
=&(s_2+2)_2\f{1}{2\pi}z_{12}^{-s_2-3}\sum_{k=0}^{s_2-4}\sum_{n=0}^{s_2-4-k}(-)^{n}\binom{s_2-2}{k}\f{(s_2-4-k)!}{n!(s_2-4)!}\bar z_{12}^{3-s_2+k+n}
\left(\lim_{w\rightarrow 0}\f{\Gamma(1-w+s_2)}{\Gamma(w-(s_2-4))} \right)\bar\partial_{z_1}^{k+n} N_{s_2-1}(z_1)
\cr
=&\, 0\,,
\ee
where we used the definitions \eqref{scg2}, \eqref{tKGR} and, in the last passage, we used Euler's reflection formula as follows
\be
\lim_{w\rightarrow 0}\f{\Gamma(1-w+s_2)}{\Gamma(w-(s_2-4))}=&\lim_{w\rightarrow 0}(-)^{s_2-1}\f{\Gamma(1-w+s_2)\Gamma(s_2-3-w)}{\Gamma(-w)\Gamma(1+w)}
\cr=&\lim_{w\rightarrow 0}(-)^{s_2-1}(-w)\f{\Gamma(1+s_2)\Gamma(s_2-3)}{\Gamma(1)}=0.
\ee


\section{Shadow OPE in Yang-Mills}\la{App:YM-OPE}

\subsection{$F_{{s_1}} S[F_{s_2}]$}\la{App:FsSFs}

From the OPE \eqref{FsFD}, assuming a general $\Delta_2\in \C$ e taking the shadow of the second operator, we have
\be
F_{{s_1}}^{a}(z_1)S[F_{\Delta_2}^{-b}](z_2)\sim& -i\f{g_{YM}^2}{4\pi}f^{ab}{}_c\f{K_{\Delta_2}}{2\pi}\f{1}{{s_1}!}\sum_{n=0}^{s_1}(-)^{{s_1}-n}  \binom{{s_1}}{n}(\Delta_2-2)_{{s_1}-n}
\int_{S_3}\f{1}{z_{23}^{3-\Delta_2}}\f{1}{\bar z_{23}^{1-\Delta_2}}\f{z_{13}^n}{\bar  z_{13}}  \partial^n F_{\Delta_2-s_1}^{-c}(z_3)
\cr\sim& -i\f{g_{YM}^2}{4\pi}f^{ab}{}_c\f{K_{\Delta_2}}{2\pi}\f{\tilde K_{2-\Delta_2+s_1}}{2\pi}\f{1}{{s_1}!}\sum_{n=0}^{s_1}(-)^{{s_1}-n}  \binom{{s_1}}{n}(\Delta_2-2)_{{s_1}-n}
\cr
&\times\int_{S_3}\int_{S_4}\f{1}{z_{23}^{3-\Delta_2}}\f{1}{\bar z_{23}^{1-\Delta_2}}\f{z_{13}^n}{\bar  z_{13}}  \partial^n_{z_3} \left(\f{1}{z_{34}^{-1+\Delta_2-s_1}}\f{1}{\bar z_{34}^{1+\Delta_2-s_1}}\right)S[F^{-c}_{\Delta_2-s_1}](z_4)
\cr\sim& -i\f{g_{YM}^2}{4\pi}f^{ab}{}_c\f{K_{\Delta_2}}{2\pi}\f{\tilde K_{2-\Delta_2+s_1}}{2\pi}\f{1}{{s_1}!}\sum_{n=0}^{s_1}(-)^{{s_1}-n}  \binom{{s_1}}{n}(\Delta_2-2)_{{s_1}-n}\bigg[
\cr
&\times(1-\Delta_2+s_1)_n\int_{S_3}\int_{S_4}\f{1}{z_{13}^{-n}}\f{1}{z_{23}^{3-\Delta_2}}\f{1}{z_{34}^{-1+\Delta_2-s_1+n}}\f{1}{\bar  z_{13}} \f{1}{\bar z_{23}^{1-\Delta_2}} \f{1}{\bar z_{34}^{1+\Delta_2-s_1}}S[F^{-c}_{\Delta_2-s_1}](z_4)
\cr
&+\sum_{k=1}^{n}\binom{n}{k}(1-\Delta_2+s_1)_{n-k}\int_{S_3}\int_{S_4}\f{1}{z_{13}^{-n}}\f{1}{z_{23}^{3-\Delta_2}}\f{1}{z_{34}^{-1+\Delta_2-s_1+n-k}}\f{1}{\bar z_{23}^{1-\Delta_2}}
\cr&\times\partial_{z_3}^k\left(\f{1}{\bar z_{34}^{1+\Delta_2-s_1}}\right)S[F^{-c}_{\Delta_2-s_1}](z_4)\bigg]\,.
\la{FsSFD-app}
\ee
If we take the second conformally soft limit, we get
\be
F_{{s_1}}^{a}(z_1) S[F^{b}_{s_2}](z_2)\sim& -i\f{g_{YM}^2}{4\pi}f^{ab}{}_c\f{K_{1-s_2}}{2\pi}\f{\tilde K_{1+s_2+s_1}}{2\pi}\f{1}{{s_1}!}\sum_{n=0}^{s_1}(-)^{{s_1}-n}  \binom{{s_1}}{n}(-s_2-1)_{{s_1}-n}\bigg[
\cr
&\times (s_1+s_2)_n\int_{S_3}\int_{S_4}\f{1}{z_{13}^{-n}}\f{1}{z_{23}^{2+s_2}}\f{1}{z_{34}^{-s_2-s_1+n}}\f{1}{\bar  z_{13}}\f{1}{\bar z_{23}^{s_2}}  \f{1}{\bar z_{34}^{2-s_2-s_1}}S[F^{c}_{s_1+s_2}](z_4)
\cr
&+\sum_{k=1}^{n}\binom{n}{k}(s_1+s_2)_{n-k}\int_{S_3}\int_{S_4}\f{1}{z_{13}^{-n}}\f{1}{z_{23}^{2+s_2}}\f{1}{z_{34}^{-s_2-s_1+n-k}}\f{1}{\bar  z_{13}}\f{1}{\bar z_{23}^{s_2}}   
\cr
&\times\partial_{z_3}^k\left(\f{1}{\bar z_{34}^{2-s_2-s_1}}\right) S[F^{c}_{s_1+s_2}](z_4)\bigg]
\cr\sim& -i\f{g_{YM}^2}{4\pi}f^{ab}{}_c\f{K_{1-s_2}}{2\pi}\f{\tilde K_{1+s_2+s_1}}{2\pi}\f{1}{{s_1}!}\sum_{n=0}^{s_1}(-)^{{s_1}-n}  \binom{{s_1}}{n}(-s_2-1)_{{s_1}-n}\bigg[
\cr
&\times (s_1+s_2)_n\int_{S_3}\int_{S_4}\f{1}{z_{13}^{-n}}\f{1}{z_{23}^{2+s_2}}\f{1}{z_{34}^{-s_2-s_1+n}}\f{1}{\bar  z_{13}}\f{1}{\bar z_{23}^{s_2}}  \f{1}{\bar z_{34}^{2-s_2-s_1}}S[F^{c}_{s_1+s_2}](z_4)
\cr
&+\theta(1-s_1-s_2)\sum_{k=1}^{n}\binom{n}{k}(s_1+s_2)_{n-k}\int_{S_3}\int_{S_4}\f{1}{z_{13}^{-n}}\f{1}{z_{23}^{2+s_2}}\f{1}{z_{34}^{-s_2-s_1+n-k}}\f{1}{\bar  z_{13}}\f{1}{\bar z_{23}^{s_2}}   
\cr
&\times\f{2\pi}{(1-s_1-s_2)!}(-)^{k-1}\partial_{z_4}^{k-1}\bar \partial_{z_4}^{1-s_1-s_2}\left(\delta^2(z_3,z_4)\right) S[F^{c}_{s_1+s_2}](z_4)\bigg]
\cr
\sim& -i\f{g_{YM}^2}{4\pi}f^{ab}{}_c\f{K_{1-s_2}}{2\pi}\f{\tilde K_{1+s_2+s_1}}{2\pi}\f{1}{{s_1}!}\sum_{n=0}^{s_1}(-)^{{s_1}-n}  \binom{{s_1}}{n}(-s_2-1)_{{s_1}-n}\bigg[
\cr
&\times (s_1+s_2)_n\int_{S_3}\int_{S_4}\f{1}{z_{13}^{-n}}\f{1}{z_{23}^{2+s_2}}\f{1}{z_{34}^{-s_2-s_1+n}}\f{1}{\bar  z_{13}}\f{1}{\bar z_{23}^{s_2}}  \f{1}{\bar z_{34}^{2-s_2-s_1}}S[F^{c}_{s_1+s_2}](z_4)
\cr
&+\theta(1-s_1-s_2)\sum_{k=1}^{n}\binom{n}{k}(s_1+s_2)_{n-k}\sum_{p=0}^{k-1}\binom{k-1}{p}(-)^{k-1-p}(s_1+s_2+k-n)_{k-1+p}\cr&\times\int_{S_3}\int_{S_4}\f{1}{z_{13}^{-n}}\f{1}{z_{23}^{2+s_2}}z_{34}^{s_2+s_1-n+1+p}
\f{1}{\bar  z_{13}}\f{1}{\bar z_{23}^{s_2}}   
\f{2\pi}{(1-s_1-s_2)!}
\cr
&\times(-)^{1-s_1-s_2}\delta^2(z_3,z_4) \partial_{z_4}^{p}\bar \partial_{z_4}^{1-s_1-s_2}S[F^{c}_{s_1+s_2}](z_4)\bigg]
\cr
\sim& -i\f{g_{YM}^2}{4\pi}f^{ab}{}_c\f{K_{1-s_2}}{2\pi}\f{\tilde K_{1+s_2+s_1}}{2\pi}\f{1}{{s_1}!}\sum_{n=0}^{s_1}(-)^{{s_1}-n}  \binom{{s_1}}{n}(-s_2-1)_{{s_1}-n}(s_1+s_2)_n
\cr
&\times \int_{S_3}\int_{S_4}\f{1}{z_{13}^{-n}}\f{1}{z_{23}^{2+s_2}}\f{1}{z_{34}^{-s_2-s_1+n}}\f{1}{\bar  z_{13}}\f{1}{\bar z_{23}^{s_2}}  \f{1}{\bar z_{34}^{2-s_2-s_1}}S[F^{c}_{s_1+s_2}](z_4).
\ee
In the last step, we used that the power of $z_{34}$ in the $\theta$-terms is such that $s_1+s_2+1+p-n\ge s_2+1\ge1$, so integrating the Dirac delta against it always yields zero.


\subsection{$S[F_{0}] S[F^-_{\Delta_2}]$}\la{App:SF0SF}

We want to take the shadow of the second entry of the OPE \eqref{SF0FD}; we consider first the case  $\Re\Delta_2\le1$, which gives
\be
S[F_{0}^{a}](z_1) S[F_{\Delta_2}^{-b}](z_2)\sim& -i\f{g^2_{YM}}{4\pi}f^{ab}{}_c\f{K_{\Delta_2}}{2\pi}\int_{S_3}\f{1}{z_{23}^{3-\Delta_2}}\f{1}{\bar z_{23}^{1-\Delta_2}}\f{1}{ z_{13}} F_{\Delta_2}^{-c}(z_3)
\cr\sim& i\f{g^2_{YM}}{4\pi}f^{ab}{}_c\f{K_{\Delta_2}}{2\pi}\f{\tilde K_{2-\Delta_2}}{2\pi}\int_{S_4}\bigg[\int_{S_3}\f{1}{ z_{31}}\f{1}{z_{32}^{3-\Delta_2}}\f{1}{z_{34}^{-1+\Delta_2}}\f{1}{\bar z_{32}^{1-\Delta_2}} \f{1}{\bar z_{34}^{1+\Delta_2}}\bigg]S[F_{\Delta_2}^{-c}](z_4)
\cr\sim& -i\f{g^2_{YM}}{4\pi}f^{ab}{}_c\f{1}{(2-\Delta_2)}\f{K_{\Delta_2}}{2\pi}\f{\tilde K_{2-\Delta_2}}{2\pi}\int_{S_4}\partial_{z_4}\bigg[\int_{S_3}\f{1}{ z_{31}}\f{1}{z_{32}^{3-\Delta_2}}\f{1}{z_{34}^{\Delta_2-2}}\f{1}{\bar z_{32}^{1-\Delta_2}} \f{1}{\bar z_{34}^{1+\Delta_2}}\bigg]S[F_{\Delta_2}^{-c}](z_4)
\cr\sim& -i\f{g^2_{YM}}{4\pi}f^{ab}{}_c\f{1}{(2-\Delta_2)}\f{K_{\Delta_2}\tilde K_{2-\Delta_2}C_{124}}{2\pi}z_{12}^{-3+\Delta_2}\bar z_{12}^{\Delta_2}\int_{S_4}\partial_{z_4}\bigg[z_{41}^{2-\Delta_2}\bar z_{24}^{-1}\bar z_{41}^{-\Delta_2}\bigg]S[F_{\Delta_2}^{-c}](z_4)
\cr\sim& -i\f{g^2_{YM}}{4\pi}f^{ab}{}_c\bigg[-\f{1}{2\pi}(2-\Delta_2)z_{12}^{\Delta_2-3}\bar z_{12}^{\Delta_2}\int_{S_4}z_{41}^{1-\Delta_2}\bar z_{24}^{-1}\bar z_{41}^{-\Delta_2}S[F_{\Delta_2}^{-c}](z_4)+ z_{12}^{-1}S[F_{\Delta_2}^{-c}](z_2)\bigg]\,,
\cr
\la{SF0SFD-app}
\ee
where we used the constant expressions \eqref{KYM}, \eqref{tKYM}, \eqref{C124}.
If we take $\Re\Delta_2>1$, we have
\be
S[ F_{0}^{a}](z_1) S[F_{\Delta_2}^{-b}](z_2)\sim& -i\f{g^2_{YM}}{4\pi}f^{ab}{}_c\f{K_{\Delta_2}}{2\pi}\int_{S_3}\f{1}{z_{23}^{3-\Delta_2}}\f{1}{\bar z_{23}^{1-\Delta_2}}\f{1}{ z_{13}} F_{\Delta_2}^{-c}(z_3)
\cr\sim& i\f{g^2_{YM}}{4\pi}f^{ab}{}_c\f{K_{\Delta_2}}{2\pi}\f{\tilde K_{2-\Delta_2}}{2\pi}\int_{S_4}\bigg[\int_{S_3}\f{1}{ z_{31}}\f{1}{z_{32}^{3-\Delta_2}}\f{1}{z_{34}^{-1+\Delta_2}}\f{1}{\bar z_{32}^{1-\Delta_2}} \f{1}{\bar z_{34}^{1+\Delta_2}}\bigg]S[F_{\Delta_2}^{-c}](z_4)
\cr\sim& i\f{g^2_{YM}}{4\pi}f^{ab}{}_c\f{K_{\Delta_2}}{2\pi}\f{\tilde K_{2-\Delta_2}}{2\pi}\f{1}{(2-\Delta_2)}\partial_{z_2}\int_{S_4}\bigg[\int_{S_3}\f{1}{ z_{31}}\f{1}{z_{32}^{2-\Delta_2}}\f{1}{z_{34}^{-1+\Delta_2}}\f{1}{\bar z_{32}^{1-\Delta_2}} \f{1}{\bar z_{34}^{1+\Delta_2}}\bigg]S[F_{\Delta_2}^{-c}](z_4)
\cr\sim& i\f{g^2_{YM}}{4\pi}f^{ab}{}_c\f{K_{\Delta_2}\tilde K_{2-\Delta_2}C_{124}}{2\pi}\f{1}{(2-\Delta_2)}\partial_{z_2}\int_{S_4}\bigg[z_{12}^{-2+\Delta_2}z_{41}^{1-\Delta_2}\bar z_{12}^{\Delta_2}\bar z_{24}^{-1}\bar z_{41}^{-\Delta_2}\bigg]S[F_{\Delta_2}^{-c}](z_4)
\cr\sim& -i\f{g^2_{YM}}{4\pi}f^{ab}{}_c\bigg[-\f{1}{2\pi}(2-\Delta_2)z_{12}^{-3+\Delta_2}\bar z_{12}^{\Delta_2}\int_{S_4}z_{41}^{1-\Delta_2}\bar z_{24}^{-1}\bar z_{41}^{-\Delta_2}S[F_{\Delta_2}^{-c}](z_4)
+ z_{12}^{-1}S[F_{\Delta_2}^{-c}](z_2)\bigg].
\cr
\ee
We thus obtain the same result as in \eqref{SF0SFD-app}.
If we take the conformally soft limit $\Delta_2\to 1$ of the second argument, we obtain
\be
S[ F_{0}^{a}](z_1)S[F_{0}^{b}](z_2)\sim& -i\f{g^2_{YM}}{4\pi}f^{ab}{}_c\bigg[
\f{1}{z_{12}}S[F_0^c](z_2)
-\f{1}{2\pi}\f{\bar z_{12}}{z_{12}^{2}}\int_{S_4}\bar z_{24}^{-1}\bar z_{41}^{-1}S[ F_0^{c}](z_4)\bigg]
\cr
\sim&-i\f{g^2_{YM}}{4\pi}f^{ab}{}_c \left[
\f{1}{z_{12}}S[F_0^c](z_2)
+\f{1}{2\pi z_{12}^2}
\int_{S_4}\left(\f1{\bar z_{41}}+\f1{\bar z_{24}}\right)S[F_0^c](z_4)
\right]
\cr\sim&-i\f{g^2_{YM}}{4\pi}f^{ab}{}_c \left[
\f{1}{z_{12}}S[F_0^c](z_2)
+\f{1}{ z_{12}^2}
(\tilde r_0^{1c}(z_2)-\tilde r_0^{1c}(z_1))
\right]\,.
\ee

\subsection{$S[F_{{s_1}}]S[F_{\Delta_2}^{-}]$}\la{App:SFsSFD}

Following similar steps as in Appendix \ref{App:SNsSND}, the OPE
\eqref{SFsSFD} can be obtained starting from the first line in \eqref{FsSFD-app}, we can write
\be
S[F_{{s_1}}^{a}](z_1)S[F_{\Delta_2}^{-b}](z_2)
\sim& -i\f{g_{YM}^2}{4\pi}f^{ab}{}_c\lim_{\Delta_1\rightarrow 1-s_1}\f{K_{\Delta_1}}{2\pi}\f{K_{\Delta_2}}{2\pi}\f{1}{{s_1}!}\sum_{n=0}^{s_1}(-)^{n}  \binom{{s_1}}{s_1-n}(\Delta_2-2)_{n}
\cr
&\times
\int_{S_4}\int_{S_3}\f{1}{z_{14}^{3-\Delta_1}}\f{1}{\bar z_{14}^{1-\Delta_1}}\f{1}{z_{43}^{-1+\Delta_1+n}}\f{1}{\bar  z_{43}^{1+\Delta_1-1+s_1}}\f{1}{z_{23}^{3-\Delta_2}}\f{1}{\bar z_{23}^{1-\Delta_2}}  \partial^{s_1-n} F_{\Delta_2-s_1}^{-c}(z_3)
\cr
\sim& -i\f{g_{YM}^2}{4\pi}f^{ab}{}_c\lim_{\Delta_1\rightarrow 1-s_1}\f{K_{\Delta_1}}{2\pi}\f{K_{\Delta_2}}{2\pi}\f{1}{{s_1}!}\sum_{n=0}^{s_1}(-)^{n}  \binom{{s_1}}{s_1-n}(\Delta_2-2)_{n}\f{(-)^n}{(1-\Delta_1)_n}\f{(-)^{s_1-1}}{(-1-\Delta_1)_{s_1-1}}
\cr
&\times
\int_{S_4}\int_{S_3}\f{1}{z_{14}^{3-\Delta_1}}\f{1}{\bar z_{14}^{1-\Delta_1}}\partial_{z_3}^{n}\f{1}{z_{43}^{-1+\Delta_1}}\bar \partial_{z_3}^{s_1-1}\f{1}{\bar  z_{43}^{1+\Delta_1}}\f{1}{z_{23}^{3-\Delta_2}}\f{1}{\bar z_{23}^{1-\Delta_2}}  \partial^{s_1-n} F_{\Delta_2-s_1}^{-c}(z_3)
\cr
\sim& -i\f{g_{YM}^2}{4\pi}f^{ab}{}_c\lim_{\Delta_1\rightarrow 1-s_1}\f{K_{\Delta_1}}{2\pi}\f{K_{\Delta_2}}{2\pi}\f{1}{{s_1}!}\sum_{n=0}^{s_1}(-)^{n}  \binom{{s_1}}{s_1-n}(\Delta_2-2)_{n}\f{1}{(1-\Delta_1)_n}\f{1}{(-1-\Delta_1)_{s_1-1}}
\cr
&\times
\int_{S_4}\int_{S_3}\f{1}{z_{14}^{3-\Delta_1}}\f{1}{\bar z_{14}^{1-\Delta_1}}\f{1}{z_{43}^{-1+\Delta_1}}\f{1}{\bar  z_{43}^{1+\Delta_1}}\bar \partial_{z_3}^{s_1-1}\partial_{z_3}^{n}\left(\f{1}{z_{23}^{3-\Delta_2}}\f{1}{\bar z_{23}^{1-\Delta_2}}  \partial^{s_1-n} F_{\Delta_2-s_1}^{-c}(z_3)\right)
\cr
\sim& -i\f{g_{YM}^2}{2}f^{ab}{}_c\lim_{\Delta_1\rightarrow 1-s_1}(-)^{\Delta_1}\f{K_{\Delta_2}}{2\pi}\f{1}{{s_1}!}\sum_{n=0}^{s_1}(-)^{n}  \binom{{s_1}}{s_1-n}(\Delta_2-2)_{n}\f{1}{(1-\Delta_1)_n}\f{1}{(-1-\Delta_1)_{s_1-1}}
\cr
&\times\f{\Gamma(-\Delta_1)}{\Gamma(\Delta_1-1)}
\bar \partial_{z_1}^{s_1-1}\partial_{z_1}^{n}\left(\f{1}{z_{21}^{3-\Delta_2}}\f{1}{\bar z_{21}^{1-\Delta_2}}  \partial^{s_1-n} F_{\Delta_2-s_1}^{-c}(z_1)\right)
\cr
\sim& -i\f{g_{YM}^2}{4\pi s_1!} K_{\Delta_2}
f^{ab}{}_c
\sum_{n=0}^{s_1}(-)^{s_1-n}  (s_1)_{n}(\Delta_2-2)_{s_1-n}
\bar \partial_{z_1}^{s_1-1}\partial_{z_1}^{s_1-n}\left(\f{1}{z_{12}^{3-\Delta_2}}\f{1}{\bar z_{12}^{1-\Delta_2}}  \partial^{n} F_{\Delta_2-s_1}^{-c}(z_1)\right)
\,.
\cr
\la{SFsSFD-app}
\ee

\subsection{$\bF_{0} S[F_{s_2}]$}\la{App:bF0SFs}

Starting from the OPE \eqref{bF0SFs}, {for $s_2\ge1$}, we can  use the identity
\be
\f1{\bz_{23}}\left(\f{\bz_{13}}{\bz_{12}}\right)^{s_2-1}
=\f{1}{\bz_{23}} \sum_{k=0}^{s_2-1}
\f{(s_2-1)_k}{k!}\left(\f{\bz_{23}}{\bz_{12}}\right)^k\,,
\ee
following from the binomial formula, to rewrite  it  as
\be
\bar F_{0}^{a}(z_1,\bz_1)  S[F_{s_2}^{b}](z_2,\bz_2)
&\sim
 -i\f{g^2_{YM}}{4\pi}f^{ab}{}_{c}\bigg[
 \f1{z_{12}}S[F^c_{s_2}](z_2,\bz_2)
 +\f{(s_2+1)}{2\pi}\f{1}{z_{12}^{s_2+2}}
 \int_{S_3}
 \f{z_{13}^{s_2}}{\bar z_{23}}
 S[F^c_{s_2}](z_3,\bz_3)
 \cr
& +\f{(s_2+1)}{2\pi}\f{1}{z_{12}^{s_2+2}}
 \sum_{k=1}^{s_2-1}
 \f{(s_2-1)_k}{k!}
 \f1{\bz_{12}^k}
 \int_{S_3}
 z_{13}^{s_2}
\bz_{23}^{k-1}
 S[F^c_{s_2}](z_3,\bz_3)\bigg]\,.
 \la{bF0SFs-2-app}
\ee
Next, we want to show that the second line contribution on the RHS of  \eqref{bF0SFs-2-app} vanishes. In order to do this, we use the manipulations
\be
&\f{(s_2+1)}{2\pi}\f{1}{z_{12}^{s_2+2}}
 \sum_{k=1}^{s_2-1}
 \f{(s_2-1)_k}{k!}
 \f1{\bz_{12}^k}
 \int_{S_3}
 z_{13}^{s_2}
\bz_{23}^{k-1}
 S[F^c_{s_2}](z_3,\bz_3)
\cr=&\f{(s_2+1)}{2\pi}\f{1}{z_{12}^{s_2+2}}
 \sum_{k=0}^{s_2-2}
 \f{(s_2-1)_{k+1}}{(k+1)!}
 \f1{\bz_{12}^{s_2-2-k+1}}
 \int_{S_3}
 z_{13}^{s_2}
\bz_{23}^{s_2-2-k}
 S[F^c_{s_2}](z_3,\bz_3)
\cr=&\f{(s_2+1)}{2\pi}\f{1}{z_{12}^{s_2+2}}
 \sum_{k=0}^{s_2-2}
 \f{(s_2-1)}{(k+1)!}
 \f1{\bz_{12}^{s_2-2-k+1}}
\bar\partial_{2}^k\left(\lim_{\Delta\rightarrow 1-s_2}(-)^\Delta\f{\Gamma(2-\Delta)}{\Gamma(1+\Delta)}
 F^c_{s_2}(z_2,\bz_2)\right)
 \cr
 =&\,0\,,
\ee
where  we used the definitions \eqref{FD}, \eqref{tKYM}, and Euler's reflection formula as follows
\be
\lim_{\Delta\rightarrow 1-s_2}(-)^\Delta\f{\Gamma(2-\Delta)}{\Gamma(1+\Delta)}=&\lim_{w\rightarrow 0}(-)^{w+1-s_2}\f{\Gamma(1-w+s_2)}{\Gamma(2+w-s_2)}
\cr=&\lim_{w\rightarrow 0}(-)^{w}\f{\Gamma(1-w+s_2)\Gamma(s_2-1-w)}{\Gamma(-w)\Gamma(1+w)}
\cr=&\lim_{w\rightarrow 0}(-)^{w}(-w)\f{\Gamma(1+s_2)\Gamma(s_2-1)}{\Gamma(1)}=0.
\ee

\subsection{$\bF_{s_1} S[r^1_{s_2}]$}\la{App:bFstrs}

In order to derive the OPE \eqref{ftilder}, let us consider first the case $s_1\ge s_2-1$. In this case we have
\be
\bar F_{s_1}^{a}(z_1) S[r_{s_2}^{1b}](z_2)&
\sim -i\f{g^2_{YM}}{4\pi}f^{ab}{}_{c} \sum_{n=0}^{s_1} \f{1}{n!} \frac{(s_1+s_2-n-2)_{s_1-n}}{(s_1-n)!}\bar\partial_{z_2}^{s_2-1}\left(\f{\bar z_{12}^n}{z_{12}}\bar \partial^n F_{s_1+s_2}^{c}(z_2)\right)
\cr
&\sim -i\f{g^2_{YM}}{4\pi}f^{ab}{}_{c} \bigg[\sum_{k=1}^{s_2-1}\sum_{n=0}^{k-1} \f{1}{n!} \frac{(s_1+s_2-n-2)_{s_1-n}}{(s_1-n)!}\f{(s_2-1)_k}{k!}\bar\partial_{z_2}^{k}\left(\f{\bar z_{12}^n}{z_{12}}\right)\bar \partial^{s_2-1-k+n} F_{s_1+s_2}^{c}(z_2)
\cr
&+\sum_{k=0}^{s_2-1}\sum_{n=k}^{s_1} \f{1}{n!} \frac{(s_1+s_2-n-2)_{s_1-n}}{(s_1-n)!}\f{(s_2-1)_k}{k!}\bar\partial_{z_2}^{k}\left(\f{\bar z_{12}^n}{z_{12}}\right)\bar \partial^{s_2-1-k+n} F_{s_1+s_2}^{c}(z_2)\bigg]
\cr
&\sim -i\f{g^2_{YM}}{4\pi}f^{ab}{}_{c} \bigg[2\pi\sum_{k=1}^{s_2-1}\sum_{n=0}^{k-1} \frac{(s_1+s_2-n-2)_{s_1-n}}{(s_1-n)!}\f{(s_2-1)_k}{k!}(-)^{n+1}\cr&\times\bar\partial_{z_2}^{k-1-n}\delta^2(z_1,z_2)\bar \partial^{s_2-1-k+n} F_{s_1+s_2}^{c}(z_2)
\cr
&+\sum_{k=0}^{s_2-1}\sum_{n=k}^{s_1} \f{1}{(n-k)!} \frac{(s_1+s_2-n-2)_{s_1-n}}{(s_1-n)!}\f{(s_2-1)_k}{k!}(-)^{k}\f{\bar z_{12}^{n-k}}{z_{12}}\bar \partial^{s_2-1-k+n} F_{s_1+s_2}^{c}(z_2)\bigg]
\cr
&\sim -i\f{g^2_{YM}}{4\pi}f^{ab}{}_{c} \bigg[2\pi\sum_{n=0}^{s_2-2}\left(\sum_{k=1+n}^{s_2-1}(-)^{n+k} \frac{(s_1+s_2-k+n-1)_{s_2-2}}{(s_2-2)!}\f{(s_2-1)_k}{k!}\right)\cr&\times\bar\partial_{z_2}^{n}\delta^2(z_1,z_2)\bar \partial^{s_2-2-n} F_{s_1+s_2}^{c}(z_2)
\cr&+\sum_{k=0}^{s_1}\sum_{n=k}^{s_1} \f{1}{(n-k)!} \frac{(s_1+s_2-n-2)_{s_1-n}}{(s_1-n)!}\f{(s_2-1)_k}{k!}(-)^{k}\f{\bar z_{12}^{n-k}}{z_{12}}\bar \partial^{s_2-1-k+n} F_{s_1+s_2}^{c}(z_2)\bigg]
,
\ee
where in the last double sum above we can take the upper bound for $k$ to be $ s_1$, as $(s_2-1)_{k}=0$ $\forall k>s_2-1$.

Let us now compute the two double sums above separately. The first double sum gives
\be
&\sum_{n=0}^{s_2-2}\left(\sum_{k=1+n}^{s_2-1}(-)^{n+k} \frac{(s_1+s_2-k+n-1)_{s_2-2}}{(s_2-2)!}\f{(s_2-1)_k}{k!}\right)\cr&\times\bar\partial_{z_2}^{n}\delta^2(z_1,z_2)\bar \partial^{s_2-2-n} F_{s_1+s_2}^{c}(z_2)
\cr&=-(s_2+s_1-1)
\left(\begin{matrix}
s_2+s_1-2\\
s_1
\end{matrix}\right)
\sum_{n=0}^{s_2-2}\f1{(s_1+n+1)}
\left(\begin{matrix}
s_2-2\\
n
\end{matrix}\right)
\cr
&\times\bar\partial_{z_2}^{n}\delta^2(z_1,z_2)\bar \partial^{s_2-2-n} F_{s_1+s_2}^{c}(z_2).
\ee
The second double sum gives
\be
&\sum_{k=0}^{s_1}\sum_{n=k}^{s_1} \f{1}{(n-k)!} \frac{(s_1+s_2-n-2)_{s_1-n}}{(s_1-n)!}\f{(s_2-1)_k}{k!}(-)^{k}\f{\bar z_{12}^{n-k}}{z_{12}}\bar \partial^{s_2-1-k+n} F_{s_1+s_2}^{c}(z_2)
\cr&=\sum_{k=0}^{s_1}\sum_{n=0}^{s_1-k} \f{1}{n!} \frac{(s_1+s_2-n-k-2)_{s_1-n-k}}{(s_1-n-k)!}\f{(s_2-1)_k}{k!}(-)^{k}\f{\bar z_{12}^{n}}{z_{12}}\bar \partial^{s_2-1+n} F_{s_1+s_2}^{c}(z_2)
\cr&=\sum_{n=0}^{s_1}\left(\sum_{k=0}^{s_1-n} \f{1}{n!} \frac{(s_1+s_2-k-n-2)_{s_1-n-k}}{(s_1-n-k)!}\f{(s_2-1)_k}{k!}(-)^{k}\right)\f{\bar z_{12}^{n}}{z_{12}}\bar \partial^{s_2-1+n} F_{s_1+s_2}^{c}(z_2)
\cr&=\sum_{n=0}^{s_1}\f{(-)^{s_1-n}}{n!}\f{(0)_{s_1-n}}{(s_1-n)!}\f{\bar z_{12}^{n}}{z_{12}}\bar \partial^{s_2-1+n} F_{s_1+s_2}^{c}(z_2),
\ee
where, on the RHS above, only the $n=s_1$ contribution survives. 
We thus finally arrive at
\be
\bar F_{s_1}^{a}(z_1) S[r_{s_2}^{1b}](z_2)\sim& -i\f{g^2_{YM}}{4\pi}f^{ab}{}_{c} \bigg[-2\pi (s_2+s_1-1)
\left(\begin{matrix}
s_2+s_1-2\\
s
\end{matrix}\right)
\sum_{n=0}^{s_2-2}\f1{(s_1+n+1)}
\left(\begin{matrix}
s_2-2\\
n
\end{matrix}\right)
\cr
&\times\bar\partial_{z_2}^{n}\delta^2(z_1,z_2)\bar \partial^{s_2-2-n} F_{s_1+s_2}^{c}(z_2)
+\f{1}{s_1!}\f{\bar z_{12}^{s_1}}{z_{12}} S[r_{s_1+s_2}^{1c}](z_2)\bigg].
\la{ftilder-app}
\ee
In the case $s_1\le s_2-1$, one can instead split the sums as
$
\sum_{k=0}^{s_2-1}\sum_{n=0}^{s_1}=\sum_{n=0}^{s_1}\sum_{k=n+1}^{s_2-1}+\sum_{n=0}^{s_1}\sum_{k=0}^{n}
$
and, with similar manipulations, arrive at the same expression \eqref{ftilder-app}.


\bibliographystyle{bib-style2.bst}
\bibliography{biblio-w.bib}

\end{document}